\documentclass{cup-elements}

\usepackage{blindtext}
\usepackage{multirow}
\usepackage{graphicx}

\usepackage[natbibapa]{apacite}
\usepackage{xfrac}
\usepackage{bbm}
\usepackage{slashed}
\usepackage{mathtools}
\usepackage{dsfont}

\usepackage{amsthm, amsmath, array}

\newcommand{\defeq}{\vcentcolon=}
\newcommand{\rdefeq}{=\vcentcolon}
\renewcommand\P{\mathcal{P}}
\newcommand\M{\mathcal{M}}

\newcommand\RR{\mathbb{R}}
\newcommand\CC{\mathbb{C}}
\newcommand\jone{\textbf{1}}

\newcommand\G{\mathcal{G}}
\renewcommand\H{\mathcal{H}}
\newcommand\A{\mathcal{A}}

\newcommand\C{\mathcal{C}}
\newcommand\SU{\mathcal{SU}}
\newcommand\U{\mathcal{U}}

\newcommand\K{\mathcal{K}}
\newcommand\J{\mathcal{J}}
\renewcommand\O{\mathcal{O}}
\newcommand\GL{\mathcal{GL}}
\newcommand\D{\mathcal{D}}

\newcommand\vphi{\varphi}

\renewcommand\u{\text{u}}

\renewcommand\epsilon{\varepsilon}

\newcommand\rarrow{\rightarrow}

\newcommand*\vect[1]{\begin{pmatrix}#1 \end{pmatrix}}

\renewcommand\t{\tilde}

\renewcommand\b{\bar }
\newcommand\w{\wedge}

\newcommand\s{\sigma}
\newcommand\bs{\boldsymbol}

\renewcommand\-{^{-1}}
\newcommand\Ad{\text{Ad}}

\renewcommand\jone{\mathds{1}}

\DeclareMathOperator{\Diff}{Diff}
\DeclareMathOperator{\Aut}{Aut}
\DeclareMathOperator{\Tr}{Tr}
\DeclareMathOperator{\vol}{vol}

\DeclareMathOperator{\re}{Re}
\DeclareMathOperator{\im}{Im}

\newcommand{\mh}{m_{\mathrm{h}}}

\newcommand{\Schrodinger}{Schr\"{o}dinger}
\newcommand{\Poincare}{Poincar\'{e}}
\newcommand{\Lagr}{\mathcal{L}}

\newtheorem{thm}{Theorem}

\newtheorem{prop}[thm]{Proposition}
\theoremstyle{definition}
\newtheorem{definition}[thm]{Definition}

\begin{document}

\CUPseries{}
\CUPelements{}

\title{Gauge Symmetries, Symmetry Breaking, and Gauge-Invariant Approaches}

\author{Philipp Berghofer\footnote{philipp.berghofer@uni-graz.at}}
\affil{University of Graz, Department for Philosophy, Heinrichstraße 26/5, 8010 Graz, Austria}

\author{Jordan François\footnote{jordan.francois@umons.ac.be}}
\affil{University of Mons, UMONS, 20 Place du parc, 7000 Mons, Belgium}

\author{Simon Friederich\footnote{s.m.friederich@rug.nl}}
\affil{University of Groningen, University College Groningen, Hoendiepskade 23/24, 9718BG Groningen, The Netherlands}

\author{Henrique Gomes\footnote{gomes.ha@gmail.com}}
\affil{University of Cambridge, Cambridge, UK}

\author{Guy Hetzroni\footnote{guyhe@openu.ac.il}}
\affil{Faculty of Philosophy, University of Oxford, Oxford, UK}
\affil{Department of Natural Sciences, The Open University of Israel, 1 University Road, Raanana 43107, Israel}
\affil{Freudenthal Institute, Utrecht University, Utrecht, The Netherlands}

\author{Axel Maas\footnote{axel.maas@uni-graz.at}}
\affil{University of Graz, NAWI Graz, Universit\"atsplatz 5, 8010 Graz, Austria}

\author{Ren\'e Sondenheimer\footnote{rene.sondenheimer@iof.fraunhofer.de}}
\affil{University of Graz, NAWI Graz, Universit\"atsplatz 5, 8010 Graz, Austria}
\affil{Fraunhofer Institute for Applied Optics and Precision Engineering, Albert-Einstein-Strasse 7, 07745 Jena, Germany}

\frontmatter  
\maketitle

\begin{abstract}
Gauge symmetries play a central role, both in the mathematical foundations as well as the conceptual construction of modern (particle) physics theories. However, it is yet unclear whether they form a necessary component of theories, or whether they can be eliminated. It is also unclear whether they are merely an auxiliary tool to simplify (and possibly localize) calculations or whether they contain independent information. Therefore their status, both in physics and philosophy of physics, remains to be fully clarified.

In this overview we review the current state of affairs on both the philosophy and the physics side. In particular, we focus on the circumstances in which the restriction of gauge theories to gauge invariant information on an observable level is warranted, using the Brout-Englert-Higgs theory as an example of particular current importance. Finally, we determine a set of yet to be answered questions to clarify the status of gauge symmetries.
\end{abstract}




\mainmatter  

\section{Introduction}\label{sec:intro}

Gauge symmetry is a central concept in essentially all of modern fundamental physics. The framework of theories in which gauge symmetries play a central role -- gauge theories -- is very general, and many physicists expect that any future discoveries will be accommodated within it. However, there are unresolved issues in the foundations of gauge theories, notably, concerning which features of gauge theories are descriptively redundant, and which are crucial for empirical adequacy. The aim of this Element is is to present precisely what is known on gauge symmetries and the possibility of gauge symmetry breaking, stressing the relevance of foundational and philosophical issues to current scientific practice and open questions, and to further outline what we take to be the most promising avenues forward. This article is thus an invitation to anyone interested in understanding the conceptual foundations of gauge theories, and a reflection upon how these features shape the way we think about elementary fields and particles.

Most results on gauge theories stem from approaches which make some drastic simplifications. Gauge theories with weak interactions are often treated using perturbative approximations. In these approximations, many of the geometric properties of non-Abelian gauge theories, like their non-trivial topological features, play little to no role. (Lattice) Simulations, suitable especially for strongly-interacting theories, can be formulated in such a way that the gauge symmetry plays essentially no role in practice. Thus, the conceptual questions concerning the gauge symmetries themselves do not usually arise as problems in practice. 

However, already a little conceptual reflection shows that the central implicit and explicit foundational assumptions on gauge symmetries are not always consistent with one another. Gauge-dependent objects depend on the choice of gauge fixing, which is made on pragmatic grounds, not dictated by any choice of gauge made on ``nature's'' behalf. This is one among several reasons why it is commonly stated that gauge-dependent objects cannot directly correspond to anything physically real. This assertion, however, casts doubt on the physical reality of elementary particles such as electrons and quarks, together with the fields that represent them. This is in sharp tension with the common discourse and with aspects of the scientific practice in which these gauge-dependent fields  are taken to be physically real in the same sense as, say, atoms are usually taken to be physically real. This tension already  highlights why properly understanding gauge symmetries is important from an ontological point of view.

There are three standard ways to avoid  gauge-dependent objects in the treatment of the  gauge interactions that form part of the Standard Model of elementary particle physics [\cite{Maas:2017wzi}]: i) in QED a so-called "photon cloud dressing" reestablishes gauge-invariance by including, in the description of the electron, what one might characterize as its "Coulomb tail" [\cite{Haag:1992hx}], ii) in QCD, the resolution, or rather the irrelevance of gauge dependence, is due to confinement, which requires that only uncharged (with respect to the non-Abelian color charge), and thereby gauge-invariant, objects appear at distances at or beyond the radius of hadrons [\cite{McMullan-Lavelle97,BeiglboCk:2006lfa}], and iii) in the electroweak sector, though much less known, it is due to the Fr\"ohlich-Morchio-Strocchi mechanism [\cite{Frohlich:1980gj,Frohlich:1981yi}]. While these three mechanisms will appear quite different at first sight, they eventually all boil down to canceling gauge-dependency by either eliminating the gauge degrees of freedom, or, at least, ensuring they do not appear in the empirically accessible range.

However, the fact that gauge-dependent objects can be eliminated in an unobtrusive way in the gauge theories just mentioned seems to depend on features that are specific to the theories combined in the Standard Model and may not hold in extensions of it. A more systematic strategy for eliminating gauge dependence may be necessary for future progress in search for physics beyond the standard model. A ``literal interpretation'' of gauge fields that regards different gauge symmetry-related field configurations as physically distinct, in contrast, may well be an obstacle to such progress. Thus it is necessary to establish whether a manifestly gauge-invariant approach to gauge theories, replacing the current way of thinking about elementary particles, is compelling or perhaps even necessary for further progress. This need, as we shall see, mirrors themes in the recent philosophical discourse on gauge symmetries. 

The structure of this Element is as follows: It starts out with a review of general features of (gauge) symmetries in Section \ref{Sec:Symmetry}. Many conceptual and technical complications surrounding gauge dependence arise in connection with the spontaneous breaking of gauge symmetry. The understanding of gauge symmetry breaking is particularly central in the context of the Brout-Englert-Higgs effect, and this is discussed in Section \ref{Sec:Breaking}. Based on this discussion, we motivate the search for gauge-invariant approaches in Section \ref{Sec:Why}, and their implementation, given at various level of details, in Sections \ref{The dressing field method} and \ref{Sec:FMS approach}. In Section \ref{Sec:Reflection} we conclude with some reflections about the ultimate consequences of the results presented, and which key steps would have to be taken to answer all substantive open questions about gauge symmetries.

\section{State of the Art: The Interpretation of Gauge Symmetries}\label{Sec:Symmetry}

\subsection{Introduction}

\subsubsection{Symmetries}
\label{SYM}
It has almost become a cliché to emphasize that symmetry plays a central role in modern physics. Weyl declared that “[a]s far as I see, all a priori statements in physics have their origin in symmetry” \cite[126]{Weyl52}, Yang argued that “symmetry dictates interactions” \cite[42]{Yang1980}, Weinberg famously said that “[s]ymmetry principles have moved to a new level of importance in this century and especially in the last few decades: there are symmetry principles that dictate the very existence of all the known forces of nature” \cite[142]{Weinberg92}, and philosopher Christopher Martin called 20\textsuperscript{th} century physics the “Century of Symmetry” \citep{martin2003continuous}. Cliché or not, it is simply a fact that our currently best physical theories exhibit symmetries. In some cases, symmetry considerations played an important heuristic role in formulating the respective physical theories (e.g., the theory of relativity). In other cases, symmetries were found retrospectively (e.g., classical electrodynamics). Due to the omnipresence of symmetries and the important heuristic role symmetry considerations play in modern physics, it is fair to say that understanding and interpreting symmetries is crucial for understanding our physical theories and what they tell us about the world. Accordingly, reflecting on the mechanism, nature, and heuristic significance of symmetries has become a central task of physicists and philosophers (see, e.g., \cite{Weyl52}, \cite{Wigner1967symmetry}, \cite{Yang1996}, \cite{BradingCastellani2003}).

In particular, here we are interested in questions concerning the \textit{ontology} of symmetries. Are certain symmetries mere mathematical artifacts or are they physically real transformations? Here we put a special focus on \textit{gauge symmetries} which are at the very heart of modern physics. Importantly, we do not only argue that understanding the nature of gauge symmetries helps us to better understand our physical theories. We also argue that critical conceptual reflection on the nature of gauge symmetries indicates that textbook accounts of the BEH mechanism are misleading\footnote{It is important to point out that the treatment of the BEH mechanism in the quantum field theoretical setting given by most of these textbooks is based on perturbation theory. In this approach it is stipulated, actually even necessary, that there is a smooth transition from an interacting (non-Abelian) gauge theory to a non-interacting (non-gauge or at most Abelian) theory asymptotically. While this not only violates basic field-theoretical statements like Haag's theorem, this necessarily blurs the line between types of symmetries.}. As we will elaborate below, this line of reasoning is not unfamiliar in the philosophy of physics community. It is one of our aims to present it in its strongest form and show how such considerations naturally lead to gauge-invariant approaches to the BEH mechanism. In this context, as mentioned in Section \ref{sec:intro}, we investigate some available methods for reducing the theories to its gauge-invariant syntax in Sections \ref{The dressing field method} and \ref{Sec:FMS approach}. 

Before we turn to gauge symmetries, we shall begin with some general remarks about symmetries in physics. Symmetries can concern physical objects and states or physical theories and laws. We say that an object or theory possesses a \textit{symmetry} if there are \textit{transformations} that leave certain features of the objects or the theory to which the transformations are applied \textit{preserved} or \textit{unchanged}. With respect to these features, the object or theory is invariant concerning the respective transformations. In this sense, one can say that “[s]ymmetry is invariance under transformation” \cite[83]{Kosso}. Transformations that leave certain aspects unchanged are referred to as \textit{symmetry transformations}. Groups of such transformations are referred to as \textit{symmetry groups}. Types of symmetries can be distinguished according to types of transformations. In physics, we often find the following distinctions: continuous vs. discrete symmetries, external vs. internal symmetries, and global vs. local symmetries.

Continuous symmetry is invariance under continuous transformation. An example of a continuous transformation would be the rotation of a circle. Since the appearance of a circle does not change under continuous rotations, we say that circles possess a continuous symmetry. The appearance of snowflakes, on the other hand, remains unchanged by rotations by sixty degrees but not, say, fifty degrees. This would be an example of a discrete transformation and thus snowflakes possess a discrete symmetry. We are particularly concerned with continuous symmetries. Mathematically, continuous symmetries can be described by Lie groups.

External symmetry is invariance under transformations that involve a change of the spacetime coordinates. Examples of external transformations are spatial rotations. Accordingly, circles possess a continuous \textit{external} symmetry and snowflakes a discrete \textit{external} symmetry. Internal symmetries are symmetries in which the respective transformations do \textit{not} involve a change of the spacetime coordinates. Examples of internal transformations would be permutations of particles or phase transformations. The symmetries we are interested in, i.e., the gauge symmetries in quantum field theory, are \textit{internal} symmetries.

For our purposes, the most important distinction is the one between \textit{global} and \textit{local} symmetries. It is standard (but slightly inaccurate) to characterize global transformations  as transformations that are performed identically at each point in spacetime. Similarly, local transformations are ones that are performed arbitrarily at each point in spacetime. More accurately, we follow Brading and Brown in making a “distinction between symmetries that depend on constant parameters (global symmetries) and symmetries that depend on arbitrary smooth functions of space and time (local symmetries)” \cite[649]{BradingBrown2004}.\footnote{The naive formulation is inappropriate since, in the appropriate mathematical formalism for gauge theories, namely, principal fiber bundles, constant gauge transformations have no meaning without the introduction of a trivialization, which can only be introduced patchwise in spacetimes. So, more accurately, one defines a group of global symmetries to have a finite number of generators---and therefore this group acts `rigidly' on spacetime, since transformations at different spacetime points are not independent. On the other hand, a group of local symmetries acts `malleably', in the sense that transformations at distant points \textit{are} independent \cite{Gomes2021}. Since we are considering in this paper only simple spacetimes, that are covered by a single coordinate patch, the difference between the naive and the precise definition is immaterial. }

Let us exemplify the difference between a global and a local transformation by considering a field $\Psi$ undergoing the following phase transformation:
$$\Psi(x) \rightarrow \Psi'(x) = e^{i\theta}\Psi(x). $$

This is a (continuous, internal) \textit{global} transformation because the phase change $\theta$ is independent of the spacetime point ($\theta$ does not depend on x). Now consider the phase transformation
$$\Psi(x) \rightarrow \Psi'(x) = e^{i\theta(x)}\Psi(x). $$

This is a (continuous, internal) \textit{local} transformation because the phase change $\theta(x)$ depends on the position in the field. The space of theories exhibiting local symmetries is much smaller---i.e. more constrained---than the space of theories only exhibiting a global symmetry (since global symmetries, when they exist, are subgroups of local symmetries).

A prominent example of an external global transformation is the Lorentz transformation of special relativity. Accordingly, special relativity is based on an external global symmetry. Prominent examples of external \textit{local} transformations are the arbitrary differentiable coordinate transformations we find in general relativity. Due to its general covariance, i.e., the invariance with respect to such transformations, general relativity is based on an external local symmetry.

The symmetries that underlie modern particle physics are internal local symmetries. The Standard Model of particle physics, describing three of the four known fundamental interactions, is a non-abelian gauge theory with an internal local U(1) × SU(2) × SU(3) symmetry group. Concerning our terminology, it is to be noted that we use the terms “local symmetry” and “gauge symmetry” synonymously. Accordingly, any field theory in which the Lagrangian remains invariant under local transformations is a gauge theory. This means that all four fundamental interactions are described by gauge theories (Standard Model + general relativity). Only the interactions between the Higgs particle and fermions as well as Higgs self-interactions are not of this type.

Because gauge symmetries play such a prominent role in modern physics, they more than deserve due conceptual reflection. The central topic of this monograph is born out of such reflection, namely, the ontological status of gauge symmetries. Should they be interpreted as mere mathematical structure of our \textit{descriptions} of reality or do they \textit{represent the structure} of reality? To approach this question discussed in more detail in the following section, it is instructive to consider one of the finest examples of synergies between mathematics and physics: Noether’s results concerning the relationship between mathematical symmetries and physical theories.

The famous Noether theorem, also known as Noether’s first theorem, relates continuous global symmetries to conserved quantities. Stated informally, the theorem says that to every continuous global symmetry there corresponds a conservation law (see \cite{BradingBrown2003}). Conversely, every conserved quantity corresponds to a continuous global symmetry. Accordingly, global symmetries seem to be physical symmetries, symmetries of nature. And, indeed, there is some consensus that global symmetries are observable and that they have direct empirical significance (\cite{Kosso}, \cite{BradingBrown2004}, \cite{Healey2009}, \cite{Friederich2015}, \cite{Gomes2021}).

The situation is very different with respect to local symmetries. The difference can be illustrated by turning from Noether’s first theorem to what is sometimes called Noether’s second theorem (see \cite{BradingBrown2003}, \cite{Earman2002, Earman2004}, \citealt[55f.]{Rickles2008}).

It has been pointed out that Noether’s results imply that local symmetries impose “powerful restrictions on the possible form a theory can take” \cite[105]{BradingBrown2003}. Specifically, the second Noether theorem ensures, through a so-called Gauss law, that the dynamics of the force fields are compatible with the dynamics of the charges that are their sources \citep{GomesRobertsNoether}. More broadly, “Noether’s second theorem tells us that in any theory with a local Noether symmetry there is always a prima facie case of underdetermination: more unknowns than there are independent equations of motion” \cite[104]{BradingBrown2003}. As will be discussed in detail below, this underdetermination inherent to gauge theories implies “an apparent violation of determinism” \cite[212]{Earman2002}.

\subsubsection{Interpreting Gauge Symmetries: A First Look}
\label{INT}

\citet{Wigner1967symmetry} compared the gauge invariance of electromagnetism to a theoretical ghost:\footnote{This type of literal `ghost' should not be confused with the actual Fadeev-Popov ghosts that \textit{are} important in gauge theory (see \citealp{Bertlmann} and also Section \ref{quantization}).} ``This invariance is, of course, an artificial one, similar to that which we could obtain by introducing into our equations the location of a ghost. The equations then must be invariant with respect to changes of the coordinate of that ghost. One does not see, in fact, what good the introduction of the coordinate of the ghost does.''  

This metaphor seems to describe the accepted view that is also, at least officially and explicitly, reflected in physics textbooks as well as that of prominent voices in the philosophy of physics community, declaring gauge theories to contain ``surplus structure'' \citep{redhead2002gauge}, ``formal redundancy'' \citep{martin2003continuous}, or ``descriptive fluff'' \cite[1239]{Earman2004}. On one widespread understanding, the surplus structure of many theories is manifested in multiplicity of mathematical representation for each physical state of affairs. Under this definition, surplus structure is ubiquitous in physics: the simple use of coordinates in spacetime physics would count as surplus. In some cases, such a representational multiplicity may be simply understood as the capacity of the theory to accommodate the viewpoints of different observers, that is what symmetries are often understood to be about (e.g. in special relativity).  Therefore, a more useful definition of surplus structure should not encompass the kind of representational multiplicity that is epistemically unavoidable, or even a theoretical virtue.

 \citet{redhead2002gauge} provides a more precise characterization of surplus structure. Given the mathematical structure $M$ of a theory used to represent a physical structure $P$, if $P$ actually maps isomorphically only onto a substructure $M'$ of $M$, then by definition the surplus structure of the theory is the complement of $M'$ in $M$. Situations in which the mathematical structure that is understood as correlating to a physical structure is embedded within a larger mathematical structure are not uncommon in modern physics. It is therefore not hard to find examples that fit this description and present no special interpretive difficulties --- the use of complex number in classical wave mechanics or in circuit analysis comes to mind. 

But obviously there are instances in modern physics where capturing the notion of surplus structure is harder, as the boundary between surplus and essential theoretical structures is hard to draw and may evolve as more knowledge is gained. Generally, which of the mathematical structures of a given theory (if any) correlates to a physical structure is a matter of an interpretation of the theory, that would generally have to balance different theoretical virtues.

We take the relation of mathematical formalism to reality to have three layers: i) the measurable, ii) the ontological (or `real', or physical), and iii) the mathematical. With these layers in mind,  we can stipulate three desiderata concerning interpretations of gauge theories:

\begin{enumerate}[label=(D{{\arabic*}})]
	\item To avoid ontological indeterminism.
	\item To avoid ontological commitments to quantities that are  not measurable even in principle.
	\item To avoid surplus mathematical structure that has no direct ontological correspondence.  
\end{enumerate}

The first two desiderata motivate an interpretation of unobservable or underdetermined theoretical concepts as  structure that has no bearing on the ontology and is in that sense `surplus'. On the other hand, considerations of locality\footnote{See Section \ref{ABeffect} on the Aharonov-Bohm effect.} and explanatory capacity would often push in the opposite direction, supporting the indispensability of the surplus mathematical structure. In the context of spacetime theories, the desiderata (D2) and (D3) are intimately related to symmetry principles aiming to bring together the symmetries of spacetime and those of the dynamics \citep{earman1989ST}. These principles can be applied in interpreting physical theories as well as in constructing them. A possible way of applying analogous principles in the context of gauge theories is by requiring that the symmetries of the dynamics would coincide with the kinematical symmetries, i.e. with the automorphisms of the mathematical structure taken to represent the possible physical states \citep{hetzroni2020ghosts}.  

Broadly speaking, interpretations of gauge theories that take gauge transformations to be physically real---i.e. to relate physically distinct objects---support a realist commitment to gauge-dependent quantities. Let us call these T1 interpretations. T1  is clearly in tension with D1 and D2, but  not in conflict with D3 because, for T1, gauge transformations have  ontological correspondence. In contrast, interpretations that take gauge symmetries to be manifestation of surplus mathematical structure may no longer be in conflict with either D1 or D2, since they  restrict ontological commitments to gauge-invariant quantities  whose evolution is deterministic (and which may even be measurable). Let us call these interpretations T2. According to T2, gauge theories, in their standard formulations, have mathematical surplus structure, in conflict with D3.\footnote{There are interpretations that lie between these two main options T1 and T2. For instance, one could argue that gauge variant quantities can be real but that within each gauge orbit there is only one phase point that represents a physically possible state \cite[58]{Rickles2008}. With respect to the vector potential, which is a gauge variant quantity, one might hold “that the vector potential was real, and that there is ONE TRUE GAUGE which describes it at any time” \cite[367]{maudlin1998healey}. Such an interpretation makes ontological commitments to gauge-variant quantities but may still fulfill D1. The main problem with this interpretation is that “no amount of observation could reveal the ONE TRUE GAUGE” (\citealp[367]{maudlin1998healey}; see also \citealp[49]{martin2003continuous}). Thus, Redhead calls it “a highly \textit{ad hoc} way of proceeding as a remedy for restoring determinism” \cite[292]{redhead2002gauge}. Technically, choosing one true gauge may be achieved via the procedure of gauge fixing, but this procedure faces the problem that the choice faces an obstruction to locality referred to as Gribov-Singer ambiguity (see \citealp[58f.]{Rickles2008}; \citealp[378]{Attard_et_al2017}; \citealp{Gribov78}; \citealp{Singer78}; \citealp{McMullan-Lavelle97}).}

And in fact, D3, as it stands, is a matter of degree: all physical theories harbor \textit{some} amount of surplus structure that is not directly measurable. After all, physics is not formulated solely in terms of---as brackets containing long conjunctions and disjunctions of---directly observable phenomena, like positions of dials and so on. And even if we weaken `measurable' to `ontological', some amount of surplus structure---for instance, a choice of coordinates, units, etc---will usually remain in our description.

Thus, these issues do not pertain merely to the interpretation of gauge theories, they can also motivate the reformulation and extension of existing theories that on the one hand remove superfluous structure to obtain a more parsimonious representation, or on the other hand, promote what initially seems like surplus structure to physical structure. 

Therefore, to proceed, we need to separate the chaff from the wheat with regards to surplus structure, and this requires a more refined notion of the term, identifying it with theoretical or formal features which can be excised from a theory without incurring any detriment to its explanatory and pragmatic virtues. Of course, such criteria still leave open what should be counted as explanatory and pragmatic virtues, an issue that depends on one's viewpoint and goals. Yet, on occasions the criteria are rather clear-cut, and even in more complicated realistic situations, we suggest that some consensus should be pursued so as to make the criteria effective. 

Here we will exemplify this by the usage of one more criterion: locality. Thus, redundancy of representation in a theory will be counted as surplus structure if eliminating it still allows us to describe physical states via locally determined quantities, or local variables. This will be the motivation for some of the approaches presented in Section \ref{The dressing field method}. The issue at  hand is therefore not limited to whether gauge symmetries manifest surplus structure or not; the refined question is how to distinguish between genuinely surplus structure and that having physical signature related to non-locality or non-separability of gauge physics. Answering this refined question provides a hard and fast criterion for a theory to meet all the desiderata D1-D3.

\subsection{The Development of Gauge Theories}
\subsubsection{Gauge Invariance in Classical Electromagnetism}
\label{CEM}

The formal property known today as gauge invariance already appeared in Maxwell’s 1856 \textit{On Faraday’s Lines of Force}, in which he showed, \textit{inter alia},  that the magnetic vector potential, introduced a few years earlier by William Thomson, can give rise to a unified mathematical description of different phenomena described by Faraday. This gauge invariance later allowed for the elimination of the vector potential from the equations, in the modern formulation of Maxwell’s equations by Hertz and Heaviside. In classical electromagnetism the equations of motion of the fields are Maxwell's equations:
\begin{align}
\nabla \cdot \vec{B} &= 0 \label{maxwell1} \\ 
\nabla \times \vec{E} + \frac{\partial \vec{B}}{\partial t} &= 0 \label{maxwell2} \\ \nabla \cdot \vec{E} &= \rho \label{maxwell3} \\
\nabla \times \vec{B} - \frac{\partial \vec{E}}{\partial t} &= \vec{j}. \label{maxwell4}
\end{align}
The equations of motion of the particles include the Lorentz force $q\left(\vec{E} + \vec{v}\times B\right)$ derived from the fields. 

This description appears to have a straightforward interpretation; the electric field $\vec{E}$ and the magnetic field $\vec{B}$ constitute the basic field ontology, Maxwell’s equations determine their behavior. The local values of the field can be found empirically based on the action of Lorentz force on particles.  This interpretation seems to yield a local understanding of the interactions, and a picture of a continuous flow of energy in space through electromagnetic radiation. Yet, this theory is not free of conceptual problems.\footnote{Primarily the divergence of the energy density and the total energy in the vicinity of charged point particles, and the self-interaction problem, i.e.\ the question of the influence of the field produced by a particle on the motion of the particle itself  (see technical and foundational discussion in \citet{rohrlich2007particles}, and \citet{frisch2005electromagnetism} for a philosophical point of view).} During the first half of the 20th century these problems raised the question of whether the electric and magnetic fields are real mediators of the interaction, or merely a mathematical tool which helps physicists to keep track of it. The central alternative to the field ontology was a picture of point particles directly interacting with each other at a distance. This kind of theory was famously advocated by \citet{wheeler1949action} based on earlier theories. 
The gauge freedom of the theory is related to a third mathematical representation, based on the electric potential $V$ and the magnetic vector potential $\vec{A}$. This representation is particularly convenient in various kinds of physical situations (such as those involving conductors). It also has the advantage of having \eqref{maxwell1}, \eqref{maxwell2} follow as identities from the kinematics rather than as additional dynamical equations. The potentials are defined such that the fields satisfy. $\vec{E} = -\nabla V - \frac{\partial \vec{A}}{\partial t}$ and $\vec{B} = \nabla \times \vec{A}$. Yet, the potentials are underdetermined by the fields: the same magnetic field can be represented by many mathematically distinct potentials. For given potentials $\vec{A}$ and $V$, the potentials $\vec{A}'=\vec{A} + \nabla f$ and $V'=V - \frac{\partial f}{\partial t}$ represent the same values of the fields $\vec{E}$ and $\vec{B}$ for any arbitrary smooth function of space and time $f$. 

The following transformation is therefore considered as the \emph{gauge transformation} of the theory:
\begin{equation}
\label{gaugetransEM}
\begin{aligned}
\vec{A} \rightarrow \vec{A} + \nabla f, \\
V \rightarrow V - \frac{\partial f}{\partial t}.
\end{aligned}
\end{equation}
Under this transformation the field values do not change, and Maxwell's equations therefore remain invariant;  Maxwell’s equations possess a gauge symmetry.

According to the field interpretation mentioned above the electric and magnetic potentials are devoid of ontological importance. Gauge invariance is therefore naturally interpreted as a manifestation of a redundancy in the way the potential represents the physical situation, rendering them as mere mathematical auxiliaries. 
However, future developments in quantum theory and particle physics gave three reasons to think that things may be more complicated. The first is the formal indispensability of the potentials in the theory. The second is the Aharonov-Bohm effect (both described in the next subsection). The third reason is that the property of gauge invariance can be promoted to a \emph{gauge principle} using which the equations that govern the interaction can be derived without appealing to prior knowledge of the classical limit. This method is the basis for the symmetry dictates interaction conception of contemporary field theories (Subsection \ref{principle}). This profound significance of gauge invariance appears to many to stand in conflict to the view that regards this invariance as mere matter of mathematical redundancy (Section \ref{The Gauge Principle Meets Philosophy}). This tension is further sharpened in the context of spontaneous gauge symmetry breaking (see Sections \ref{Sec:Breaking} and \ref{Sec:Why}).

\subsubsection{The Aharonov-Bohm Effect}
\label{ABeffect}

In classical electromagnetism the potentials are non-measurable  quantities whose local values are not well defined, and do not form an essential part of the mathematical description of the dynamics, that can be expressed in terms of the fields using Maxwell's equations and the Lorenz force equation. Yet, there is a significant place in which they become indispensable, at least from a formal point of view, and this is the Hamiltonian (and also the Lagrangian) formulation of the theory. In quantum mechanics the Hamiltonian formulation gains fundamental significance as the generator of the temporal dynamics. The theory does not include force as a fundamental entity, but only as a derived phenomenon at a classical limit. Accordingly, it is the electric and magnetic magnetic potentials, and not the fields, which appear in the Hamiltonian and thus in the \Schrodinger\ equation. 

\begin{figure}[h]
  \caption{The Aharonov-Bohm experiment. Figure taken from \citet{aharonovbohm1959}.}
  \label{ABfigure}
  \includegraphics[width=0.8\textwidth]{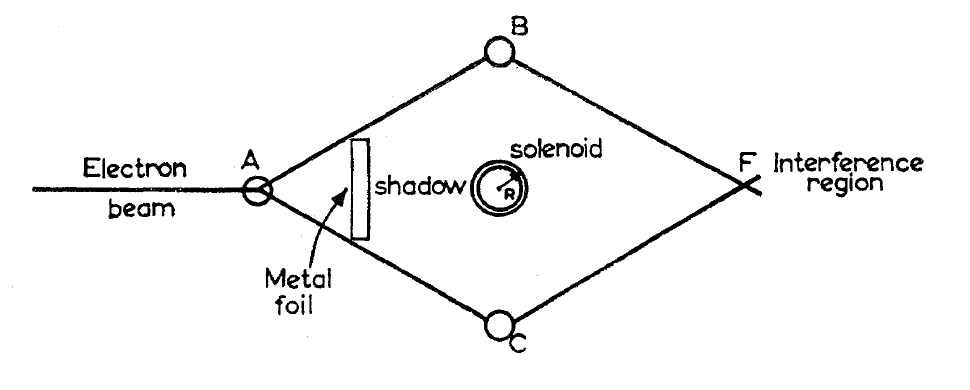}
\end{figure}

Aharonov and Bohm were intrigued by the question of whether this theoretical difference between the quantum and the classical can make an observational difference, and proposed an experiment in which it does. The proposed experiment is an electron interference experiment, in which a beam is split into two branches which are brought back together to form an interference pattern (Fig. \ref{ABfigure})\footnote{\citeauthor{aharonovbohm1959}'s work was conducted independently of the work by \citet{ehrenberg1949refractive} who proposed the same experiment with a different framing in a work that did not receive much attention at the time. In addition, according to \citet{hiley2013ABeffect}, Walter Franz described a similar situation in a talk in 1939.}. The electric and magnetic field along the possible trajectories is zero. In addition, a conducting solenoid is placed between the two branches, in an area of space that is shielded from the beam (the wave function at the neighborhood of the solenoid is zero). The current in the solenoid induces a magnetic field inside it, but the foil insures that there is no overlap between the support of the wave function in space-time and the magnetic field. The surprising fact is that in this scenario the interference pattern would depend on the magnetic field! 

The dependence of the interference pattern on the field can be expressed in terms of a phase difference between the two branches that is acquired in the process. The following calculation of the phase factor emphasizes the role of gauge invariance. Let us begin with the case $\vec{A}=0$ everywhere and at all times, which, in a particular gauge, describes the experiment conducted with a zero magnetic field (we shall also use $\phi=0$ in all cases). Let us denote the solutions of the \Schrodinger\ equation for the first branch ABF and for the second branch ACF by $\psi_1^0\left(\vec{r},t\right)$ and $\psi_2^0\left(\vec{r},t\right)$ respectively. An interference pattern is obtained in the interference region, the overlap of the support of these two wave functions, and in this case its form is given by $\left|\psi_1^0\left(\vec{r},t\right)+\psi_2^0\left(\vec{r},t\right)\right|^2$. 

The experiment is described by the Hamiltonian: \begin{equation}
\label{lorentzhamiltonian}
H=\frac{1}{2m}\left(\vec{p}-\frac{q}{c}\vec{A}\right)^2+q\phi.
\end{equation}
It is invariant under the gauge transformation: 
\begin{equation}
\label{magneticgaugetrans}
\vec{A}(\vec{r},t)\rightarrow\vec{A}(\vec{r},t)-\nabla\Lambda(\vec{r},t). 
\end{equation}
For arbitrary smooth $\Lambda$. This invariance is helpful for the calculation of the effect of the magnetic field. Let us denote by $\vec{A}^*\left(\vec{r},t\right)$ the magnetic vector potential that describes the situation with a given non-zero magnetic flux $\Phi_B$. For simplicity, the choice of gauge is such that $A^*$ is time independent and that at point A we have $\vec{A}^*=0$. The magnetic field along the first branch is zero. That means that there is a certain gauge transformation $\Lambda_1\left(\vec{r}\right)$ such that the potential $\vec{A}_1\equiv \vec{\nabla}\Lambda_1$ corresponds to a situation with zero magnetic field everywhere (i.e. $\vec{\nabla}\times \vec{A}_1=0$ at all places), but along the first branch $\vec{A}_1=\vec{A}^*$. Similarly, since at any point along the second branch the magnetic field is zero and $\vec{\nabla}\times \vec{A}^*=0$, there exist a gauge transformation $\Lambda_2\left(\vec{r}\right)$ and a corresponding potential $\vec{A}_2\equiv \vec{\nabla}\Lambda_2 $ such that $\vec{\nabla}\times \vec{A}_2=0$ at all places, but along the second branch $\vec{A}_2=\vec{A}^*$.

Now, any electromagnetic gauge transformation \eqref{magneticgaugetrans} is a symmetry of the Hamiltonian \eqref{lorentzhamiltonian} when it is accompanied by the local phase transformation $\psi\rightarrow e^{iq\Lambda\left(\vec{r},t\right)/\hbar c}\psi$. As a consequence, if we start from $\vec{A}=0$ everywhere and apply the gauge transformation defined by $\Lambda_1\left(\vec{r}\right)$ which transforms the vector potential into $\vec{A}_1$, the wave function $\psi_1^0\left(\vec{r},t\right)$ would be transformed into $e^{i\Lambda_1\left(\vec{r}\right)}\psi_1^0\left(\vec{r},t\right)$. But since $\vec{A}^*=\vec{A}_1$ along the first branch, the wave function $e^{i\Lambda_1\left(\vec{r}\right)}\psi_1^0\left(\vec{r},t\right)$ would also be the wave function of the wave-packet that travels along the first branch for the case of nonzero magnetic field. Similarly, if we start from $\vec{A}=0$ everywhere and apply the gauge transformation defined by $\Lambda_2\left(r\right)$, the wave function $\psi_2^0\left(\vec{r},t\right)$ would be transformed into $e^{iq\Lambda_2\left(\vec{r}\right)/\hbar c}\psi_2^0\left(\vec{r},t\right)$. This expression would also hold for the wave function that travels along the second branch when the magnetic field is non-zero. Therefore, in this case the interference pattern would be given by $\left|e^{iq\Lambda_1\left(\vec{r}\right)/\hbar c}\psi_1^0\left(\vec{r},t\right)+e^{iq\Lambda_2\left(\vec{r}\right)/\hbar c}\psi_2^0\left(\vec{r},t\right)\right|^2=\left|\psi_1^0\left(\vec{r},t\right)+e^{iq\left[\Lambda_2\left(\vec{r}\right)-\Lambda_1\left(\vec{r}\right)\right]/\hbar c}\psi_2^0\left(\vec{r},t\right)\right|^2$. 

The definitions above imply that at any point along the first branch $\Lambda_1\left(\vec{r}\right)=\int_A^r{A_1\left(\vec{r'}\right)\cdot\vec{dr'}}$, where the integral is taken from point $A$ along the path of the particle in the first branch. But since $\vec{A}_1$ equals $\vec{A}^*$ in that region, we get $\Lambda_1\left(\vec{r}\right)=\int_A^r{A^*\left(\vec{r}'\right)\cdot\vec{dr}'}$. Similarly, along the second branch, $\Lambda_2\left(\vec{r}\right)=\int_A^r{A^*\left(\vec{r}'\right)\cdot\vec{dr}'}$. In the interference region, the phase difference which is responsible for the change in the interference pattern is given by the difference between these two phase factors which is exactly
\begin{equation}
\label{ABphase}
\Delta\varphi_{AB}=\frac{q}{\hbar c}\oint{A^*\left(\vec{r}\right)\cdot\vec{dr}}, 
\end{equation}
where the loop integral is over the closed loop ABFCA. 

While this calculation was performed in a specific gauge, its final outcome is gauge independent. According to Stokes' theorem the loop integral equals the magnetic flux, such that the phase difference is $\Delta\varphi_{AB}=q\Phi_B/\hbar c$.

The effect gave rise to new controversies concerning both the foundations of quantum theory and of gauge theories. One possible conclusion is that locality and gauge invariant ontology cannot be reconciled. We discuss the issue in more detail in Section \ref{gaugeontology} (and see also Section \ref{AVS}). 

A closely related issue is the role of topology in physics, which is considered by some physicists and philosophers as the main issue raised by the Aharonov-Bohm effect \citetext{e.g. \citealp{ryder1996qft, Nounou2003AB}}. The point is that it is possible to gauge away the potential in the domain of each paths ABF and ACF, but not in union of the domains, due to the non-trivial topology of this union. In the coordinate-free formulation of classical electromagnetism\footnote{See, for example,  \citet{baez1994gauge}.} the vector potential 1-form $A$ and the magnetic field 2-form $B$ satisfy $B=dA$. In any point within the domain accessible by the particle it holds that $dA=B=0$. The vector potential is therefore a \emph{closed} 1-form. According to the \Poincare\ lemma, on any contractible sub-domain of the configuration space, the differential form $A$ is also \emph{exact}, i.e. in the sub-domain there exists a form $\Lambda$ such that $A=d\Lambda$. Each of the two possible paths lies on such a contractible sub-domain. All holonomies in such a sub-domain would therefore equal zero. But this is not true for the entire configuration space of the particle, whose topology is non-trivial, and therefore the premises of the \Poincare\ lemma are not satisfied. In the minimal coupling scheme the vector potential 1-form takes the role of the connection of the  $U\left(1\right)$ principle bundle over the spacetime manifold that correspond to the region available to the particle. The connection determines the phase-change that occurs in infinitesimal displacements, thus determining the holonomies. Due to the non-trivial topology of the configuration space of the particle, the connection form is not necessarily exact, and holonomies may appear. 

\subsubsection{The Gauge Principle}
\label{principle}

In the early 20th century the gauge invariance of classical electromagnetism was not considered as bearing any fundamental significance. An original attempt to rethink the issue was made by  \citet{weyl1918gravitation}\footnote{And also in other writings from the same year, see \citet{scholz2004weyl}.}, who aimed for a geometrical unification of gravitation and electromagnetism based on an extension of Einstein's general theory of relativity, introducing  the term `gauge transformations' for the first time.\footnote{For more details on the philosophical considerations that motivated Weyl’s gauge principle, particularly the influence of Edmund Husserl’s phenomenology, see \cite{Ryckman2005, Ryckman2020}.} Weyl saw great importance in the notion of locality that is expressed in general relativity in the dependence of the metric on spacetime points. However, the geometry of the theory does not fully manifest the desired form of locality, due to the invariance of the inner product under parallel transport, which defines a global standard of length. Relaxing this condition introduces (in a modern terminology) a connection $\phi$ of the local scaling group, and a corresponding local scale factor $\lambda$ for the transformation of the metric $g(x)\rightarrow\Tilde{g}(x)=\lambda(x)g(x)$. Weyl then identified the components of the connection 1-form $\phi$ with the components of the electromagnetic four-potential. Thus, the curvature of the scaling connection was identified with electromagnetism in the same way the curvature of the Levi-Civita connection is associated with gravity. The empirical adequacy of this identification was soon criticized by Einstein and others.

In Weyl's theory the length scale acquires a non-integrable measure factor along a trajectory in spacetime. Just after the introduction of quantum mechanics, \citet{london1927gauge} suggested to reinterpret Weyl's theory based on the observation that the the quantum phase factor can be seen as an imaginary version of Weyl's measure factor. 

This discrepancy suggested that the relationship revealed through the concept of gauge between electromagnetism and gravity is that of an analogy, rather than a straightforward unification. This analogy was the basis of Weyl's gauge \emph{principle} \citet{weyl1929elektron,weyl1929gravitation} . In these papers Weyl used the rejection of global parallelism to motivate a formulation of general relativity using tetrads (which had been recently introduced by Einstein), and used this formulation to emphasize similarities between the structure of geometry and electromagnetism. Weyl noted that the metric does not fully determine the tetrad: there is a freedom of local Lorentz transformations. Weyl's theory is based on 2-spinors, whose rotation (in internal space) can be regarded as a representation of the same Lorentz group. The tetrads, that define distant-parallelism, thus determine not only  spacetime curvature and the connection, but also those of the  spinors' internal space. The important point is that the choice of tetrad does not fully determine the state of the spinors, as there remains a freedom of a phase factor. By analogy to the gravitational case, this freedom should be manifested as an invariance under local phase transformations.
\begin{quote}
``The transformation of the $\psi$ induced by the rotation of the tetrad is determined only up to such a factor. In special relativity we must regard this gauge-factor as a constant because here we have only a single point-independent tetrad. Not so in general relativity; every point has its own tetrad and hence its own arbitrary gauge factor; because by the removal of a rigid connection between tetrads at different points the gauge-factor necessarily becomes an arbitrary function of position.'' \footnote{English translation in \citet{ORaifeartaigh1997gauge}, pp. 139-140.}
\end{quote}

This desired local invariance motivated the introduction of a covariant derivative that includes a local quantity $f$ (the connection term in the covariant derivative) such that the action is invariant under the transformation:
\begin{align}
\psi\rightarrow e^{i\lambda(x)}\psi && f_p\rightarrow f_p-\frac{\partial\lambda}{\partial x_p}.
\end{align}
Weyl then notes that the resultant $f$ term in the action is identical to ``the manner [..] that the electromagnetic potential interacts with matter according to experiment. This justifies the identification of the quantities $f_p$ introduced here with the electromagnetic potentials''.  Weyl then identifies the electromagnetic field $f_{pq}=\frac{\partial f_q}{\partial x_p}-\frac{\partial f_p}{\partial x_q}$.  He further notes the connection of the transformation to conservation of electric charge, and stresses the analogy to the connection between conservation of momentum and angular momentum in general relativity and the invariance under local Lorentz transformations (rotation of the tetrads in spacetime). Weyl thus regards the gravitational interaction and the electromagnetic interaction as manifesting the same new principle of gauge invariance (see also \citealp{weyl1929gravitation}). In developing this principle, and later in reflecting on it, Weyl appealed to philosophical considerations, mainly ones that originate in an idealist or Kantian tradition \citep{ryckman2003gauge, hetzroni2021analogies}.

Weyl's gauge terminology  with respect to the electromagnetic interaction was embraced by \citet{pauli1933handbuch,pauli1941particles} in his influential writings on quantum physics. It allowed for a reconstruction of the electromagnetic interaction in a quantum context from simple principles that do not appeal to classical electromagnetism. For example, the electromagnetic interaction term is introduced into the free Dirac equation $i\gamma^\mu\partial_\mu\psi-m\psi=0$ using the gauge principle by replacing the derivative $\partial_\mu$ with a gauge covariant derivative $D_\mu=\partial_\mu +ieA_\mu$. The resulting equation $i\gamma^\mu\left(\partial_\mu+i e A_\mu\right)\psi-m\psi=0 $ is invariant under local gauge transformations 
\begin{align}
\psi\rightarrow e^{-i\lambda(x)}\psi && A_\mu \rightarrow A_\mu+\frac{1}{e}\partial_\mu \lambda(x). 
\end{align}

The analogy with the electromagnetic case  motivated the successful attempt of  \citet{yangmills1954} to localize the $SU(2)$ isospin symmetry. The theory involves two Dirac spinors of equal mass, that can be described using the Lagrangian $\Lagr=i\Bar{\psi}\gamma^\mu\partial_\mu\psi-\Bar{\psi}m\psi$ with $\psi\equiv\begin{pmatrix}\psi_1\\\psi_2\end{pmatrix}$. It is initially invariant under a global $SU(2)$ isospin symmetry. The Yang-Mills field $B_\mu$ (a $2\times 2 $ matrix) is similarly introduced by replacing the derivative with a corresponding covariant derivative that renders the theory invariant under local SU(2) transformations.

Gauge theories of Yang-Mills type were soon recognized as easy to renormalize. This provided a central motivation to pursue this line  \citep{kibble2015electroweak,brown1993renormalization}. Thus, while Yang and Mills's original theory failed to describe the strong nuclear interaction it was try to account for, the method it presented soon became a template used in the construction  of the theory of electroweak interactions (see Chapter \ref{Sec:Breaking}) as a gauge theory of the group $SU(2)\otimes U(1)$, and later also in the construction of the strong interaction. The  development of the standard model was in this sense based on applying the same pattern of achieving local invariance by introducing gauge fields \citep{mills1989fields}. A retrospective description of these developments is given by \citeauthor{ORaifeartaigh1997gauge} (\citeyear{ORaifeartaigh1997gauge}, emphasis in original):
\begin{quotation}
invariance with respect to the local symmetry \emph{forces the introduction of the vector fields $A_\mu(x)$ and determines the manner in which these fields interact with themselves and with matter}. The fields $A_\mu(x)$ turn out to be just the well-known radiation fields of particle physics, namely, the gravitational field, the electromagnetic field, the massive vector meson fields $Z^0$, $W^\pm$ of the weak interactions and the coloured gluon fields $A^c_\mu$ of the strong interactions. Thus gauge symmetry introduces all the physical radiation fields in a natural way and determines the form of their interactions, up to a few coupling constants. 
It is remarkable that this variety of physical fields, which play such different roles at the phenomenological level, are all manifestations of the same simple principle and even more remarkable that the way in which they interact with matter is prescribed in advance. It is not surprising, therefore, to find that the covariant derivative has a deep geometrical significance. [...] the [local gauge groups] $G(x)$ are identified as sections of principal fiber-bundles and the radiation fields $A_\mu(x)$ are mathematical \emph{connections}.
\end{quotation}

In addition to the gauge symmetries in particle physics, the analogy with gravitation continued to play a role in various works aiming to go beyond general relativity in the description of gravity, or to provide a unified framework for all interactions. In 1956 Ryoyu \citet{Utiyama1956} fully articulated the modern understanding of the gauge argument in terms of  ``gauging'' a global symmetry using a Lie group to obtain a gauge theory.\footnote{His work, completed in mid-1954, is independent of the $\SU(2)$ Yang-Mills paper issued the same year and is much more  general, the $\SU(2)$ case being yet again independently worked out by R. Shaw - a student of A. Salam - in 1954 and published in his PhD thesis \citep{Shaw1955}. It is arguably an unjust historical oversight that gauge theories are associated mainly to the names of Yang and Mills, while Utiyama's far reaching contribution is seldom remembered. See \citep{ORaifeartaigh1997gauge} for an historical account of the events.}
Utiyama applied his approach also to general relativity, showing for the first time that General Relativity can be recovered from a gauge principle applied to the rigid Lorentz group. 

The theory of \citet{brans1961dicke} generalized general relativity based on conformal transformations, similar to the ones presented in \citet{weyl1918gravitation}. After the success of gauge theoretic approaches to the nuclear interactions in the mid to late 60s and early 70s, and after the inception  of supersymmetry, explorations of gauge theories of gravity and super-gravity blossomed in the second half of the 70s 
 (see  \citealp{scholz2020metric} for a historical review and  \citealp{Blagojevi-et-al2013} for a detailed review of the theories). These theories describe gravity using different non-Riemannian geometries introduced through the process of gauging different groups larger than the Lorentz group. The aim is to bridge the language gap with the description of the other interactions, so as to  facilitate either their unification or the quantization of gravity. 

The classical theory of non-gravitational gauge interactions were understood in terms of the geometry of Ehresmann connections on fiber bundles (see section \ref{Geometric background}). For a long time, the gauge potentials of gravity were thought about in the same  terms. Only quite recently was it recognised that gauge gravity is better understood in terms of the geometry of Cartan connections on principal bundles, see e.g.\citep{Wise09, Wise10}. Cartan geometry is indeed the natural generalization of (pseudo) Riemannian geometry, as introduced by {\'E}. Cartan in the 20s, and the direct precursor of the notion of connection introduced in the late 40s by C. Ehresmann (who was a pupil of {\'E}. Cartan). See  \citep{Sharpe, Cap-Slovak09} for modern introductions. 

Interpretive issues surrounding gauge symmetries are thus pressing for gravitational and non-gravitational theories alike. 

\subsubsection{Quantization}
\label{quantization}

The advent of quantum field theory required to reshape the understanding of gauge symmetries substantially. Especially, the role of gauge transformations moves from a transformation of solutions of the equations of motion to an integral part of the definition of the quantum theory. This also modifies, to some extent, how gauge transformations are perceived in a quantum field theory, as will be outlined here.

In principle, quantizing a gauge theory is performed similarly as with non-gauge theories, but for a few subtleties. For the purpose of illustration, this will be done here using a path-integral approach. We select  the vector potential as the integration variable\footnote{Other choices will lead in general to a nonequivalent path integral. While this is mathematically perfectly feasible, these are not empirically adequate. Why this is so is a very good, and unsolved, question.}.

A way [\cite{Bohm:2001yx}] to understand the origin of the ensuing subtleties when integrating over the vector potential is to note that any gauge transformation leaves the action invariant, and, as a shift, also does not influence the measure. Hence, there are flat\footnote{A flat direction is one where under the variation of the (path) integration variable the (path) integral kernel does not change, and thus one adds up a constant. This is what happens in a gauge theory. Along the gauge orbit the integrand does not change, thus amassing the infinity. The often quoted non-invertebility of the Hessian is only an issue if one attempts a saddle-point approximation, like in perturbation theory.} directions of the path integral, and thus the integral diverges when integrating along these directions.

There are only few possibilities to deal with these divergencies. One is to perform the quantization on a discrete space-time grid in a finite volume. In this way the divergences become controllable, and can be removed before taking the limit to the original theory [\cite{Montvay:1994cy}]. 

Another one is to transform to manifestly gauge-invariant variables. This would be achieved by a variable transformation to the dressed fields of section \ref{The dressing field method}. However, in most cases, this has been thus far found to be practically impossible.

The third option is to remove the divergences by fixing a gauge [\cite{Bohm:2001yx}]. This is achieved by sampling every gauge orbit only partially in such a way that the result is finite while gauge-invariant quantities are not altered. Even though gauge-variant information is removed, this is not equivalent to introducing a gauge-invariant formulation. Any gauge condition will define one distinct way of removing the superfluous degrees of freedom, but what is removed and what is left differs for every choice of gauge. 

As gauge fixing plays a central role in contemporary particle physics, as well as in section \ref{Sec:FMS approach}, it is worthwhile to detail it further. Gauge-fixing proceeds by selecting a gauge condition $C_\Lambda$, which may involve the gauge field as well as any other fields, and which can be parametrized by some quantity or function $\Lambda$ in an arbitrary way.  Necessary conditions for this gauge condition are that every gauge orbit has at least one representative fulfilling this condition. We will furthermore assume, for reasons of practicality, that there is only one such representative. In non-Abelian gauge theories the Gribov-Singer ambiguity makes this requirement very involved [\cite{Gribov78,Singer78}] to implement and formulate, even to the point of practical impossibility [\cite{McMullan-Lavelle97,Vandersickel:2012tg,Maas2011}]. Conceptually, however, this will not be an issue for now.

The procedure for determining the amplitude of any gauge-invariant observable $f$ goes as follows:
\begin{align}
&\frac{1}{{\cal N}}\int_\Omega{\cal D}A_\mu f(A_\mu)\exp(iS[A_\mu])\label{q1}\\
&=\frac{1}{{\cal N}}\int_{\Omega/\Omega_C}{\cal D}g\int_{\Omega_c}{\cal D}A_\mu\Delta[A_\mu]\delta(C_\Lambda)f(A_\mu)\exp(iS[A_\mu])\label{q2}\\
&=\frac{1}{{\cal N'}}\int_{\Omega}{\cal D}A_\mu\Delta[A_\mu] f(A_\mu)\delta(C_\Lambda)\exp(iS[A_\mu])\label{q3}\\
&=\frac{1}{{\cal N^{''}}}\int_{\Omega_C}{\cal D}A_\mu\Delta[A_\mu]f(A_\mu)\exp(iS[A_\mu])\label{q4},
\end{align}
in which possible further fields are suppressed. The original expression (\ref{q1}) is an integral over the whole set $\Omega$ of all gauge orbits and all representatives on every gauge orbit. In (\ref{q2}) this set is split into the set $\Omega_C$, which contains for all orbits only the representatives fulfilling $C_\Lambda$, and the remainder $\Omega/\Omega_C$. This allows us to write the integration along gauge orbits $g$ and over gauge orbits separately. This separation requires the introduction of a $\delta$-function on the gauge condition and an additional weight factor, the Faddeev-Popov determinant $\Delta$. The latter ensures that the weight of the representative is the same for all gauge orbits, and thus gauge-invariant, ensuring $\langle 1\rangle=1$. Then the integrals along the gauge orbits are orbit-independent, and can be absorbed in the normalization in (\ref{q3}). Finally, resolving the $\delta$-function in (\ref{q4}) yields the gauge-fixed path integral, yielding yet another normalization. This approach is standard for perturbation theory\footnote{It is often convenient to integrate over the parametrization of the gauge condition, which is e.\ g.\ done for linear covariant gauges like the Feynman gauge.} [\cite{Bohm:2001yx}]. As noted above, beyond perturbation theory the Gribov-Singer ambiguity makes this procedure in practice cumbersome. This manifests itself in the form and properties of both the Faddeev-Popov determinant $\Delta$ and the gauge condition $C_\Lambda$ [\cite{McMullan-Lavelle97,Vandersickel:2012tg,Maas2011}].

  In principle, the Fadeev-Popov determinant is a simple functional extension of the following argument: for a function with a single root, $f(x_o)=0$, the Dirac delta function obeys the identity: 
   \begin{equation}\label{delta_identity}\delta(f(x))=\frac{\delta(x-x_o)}{|\det{f\rq{}(x_o)}|}\end{equation} Since the integral of $\delta(x-x_o)$ over $x$ gives unity, then
$$|\det{f\rq{}(x_o)}|\int dx \, \delta(f(x))=1$$
The functional setting works in the same way, 
with \begin{equation}\label{dirac_FP}
|\det{J\rq{}(\varphi_o)}|\int \mathcal{D}\varphi\, \delta(F(\varphi))=1
\end{equation} Usually the field $\varphi$ is a gauge-parameter, with \eqref{dirac_FP} determining the factor for a gauge-fixing. 

 Geometrically, the Fadeev-Popov determinant emerges as a functional Jacobian for a  change of variables, since one now decomposes, as in Fubini's theorem, a functional integration over all variables into an integration over a gauge-fixing section and an integration over the orbits (see e.g. \cite{Babelon:1979, MottolaFP}). Of course, such a decomposition is also vulnerable to the Gribov ambiguity, and thus would not be available non-perturbatively. Nonetheless, this interpretation leads straightforwardly to the Fadeev-Popov determinant as follows:
 For the transformation $A_\mu\rightarrow A_\mu^{Cg}$, where $A^C_\mu$ is the gauge-fixed connection, we obtain the respective Jacobian for the measure ${\cal D}A\rightarrow \det{(J)}{\cal D}A^C{\cal D}g$.\footnote{
     But note that establishing the Jacobian requires a normalisation of the measures. This is usually defined by a unit Gaussian measure. We then combine that with the variable transformation. For instance, for Landau gauge (in Euclidean signature, and for the Abelian theory), the gauge-fixed potential is a projection of the potential onto its transverse-free part: 
     \begin{equation}
         A^C_\mu(A)=(\delta^\nu_\mu-\partial_\mu\square^{-2}\partial^\nu)A_\nu.
     \end{equation}   Of course, this case rather trivially furnishes only a field-independent Jacobian, since $J$ does not depend on $A$.}

It is an interesting question to ask what happens if one tries to calculate a gauge-dependent amplitude.  In fact, if done so by putting a gauge-dependent $f$ in (\ref{q1}), the answer is always zero, up to some $\delta$-functions at coinciding arguments. Of course, such a calculation requires a method like the lattice-regularization. The reason is that for any gauge field configuration with some value $A_\mu(x_0)$ at the fixed position $x_0$, there exists a gauge transformation, which is only non-vanishing at $x_0$, such that the value of the gauge transformed gauge field is $-A_\mu(x_0)$. In this way, any integration over the full gauge group yields zero. The only exception can happen if arguments coincide, yielding squares of the fields. On the other hand, evaluating a gauge-dependent quantity $f$ using (\ref{q4}) yields very non-trivial results.

The reason for this apparent disagreement is the step from (\ref{q2}) to (\ref{q3}). Here, the integral $\int{\cal D}g$ was absorbed in the normalization, because none of the remaining expressions depended on it, because all were gauge-invariant. This is no longer true, if $f$ is gauge-dependent. Then the integral over gauge transformations can no longer be separated as a factor and be removed.

Thus, from a purely mathematical point of view, the expressions (\ref{q1}-\ref{q2}) and (\ref{q3}-\ref{q4}) are distinct theories. From the point of view of physics, there is just an infinite number of equivalent quantum theories, the one without gauge fixing and the infinitely many choices of $\Omega_c$, or equivalently $C_\Lambda$, which all yield the same gauge-invariant observables, but differ for gauge-dependent ones.

Alternatively, this can also be taken to imply that any choice of theory with the action $S'=S-i\ln\Delta$ gives the same gauge-invariant quantities, provided they are integrated over the corresponding set $\Omega_C$, either directly implemented as integration range or by a $\delta$-function. In either way, this leads ultimately to the expression (\ref{q4}).

This infinite degeneracy of quantum theories is a consequence of working with gauge freedom. This problem of infinite degeneracy would vanish if a transformation to gauge-invariant variables could be performed, as is aimed at in the dressing-field approach in section \ref{The dressing field method}. Alternatively, an approach like the one in section \ref{Sec:FMS approach} will take all these theories to be equivalent by defining the (gauge-invariant) observables to be only those, which are the same for all. This yields the same observable result as the lattice regularization.

\subsection{Interpreting Gauge Theories}
\label{Interpreting Gauge Theorie}
\subsubsection{The Gauge Principle Meets Philosophy}
\label{The Gauge Principle Meets Philosophy}

In the late 20th century gauge symmetries began to attract the attention of philosophers. Many of them identified the gauge principle as a prominent example of a fundamental shift in the methodology of theoretical physics, intertwined with the elevation of abstract mathematics. \citeauthor{steiner1998applicability} (\citeyear{steiner1989application,steiner1998applicability}) raised the issue in the context of the question of the applicability of mathematics to natural science. He regarded the gauge argument (especially in the version of Yang and Mills) as a Pythagorean analogy, i.e. one that can only be expressed in a mathematical language and is not based on a physical similarity. Such mathematical analogies, according to Steiner, are motivated by human values (such as aesthetics) that guide the development of mathematics, and their repeated success in physics puts physics at odds with naturalism.  

A closely related issue is the apparent conflict between the standard understanding of gauge symmetries as a matter of a choice of convention and their great methodological significance.  \citeauthor{yangmills1954} described gauge freedom in terms of local conventions:
\begin{quote}
``The difference between a neutron and a proton is then a purely arbitrary process. As usually conceived, however, this arbitrariness is subject to the following limitation: once one chooses what to call a proton, what a neutron, at one space-time point, one is then not free to make any choices at other space-time points. It seems that this is not consistent with the localized field concept that underlies the usual physical theories.'' 
(p.192)
\end{quote} 
This view of gauge invariance became part of the received view, in part due to \citet{Wigner1964invariance} who emphasized the distinction between dynamical symmetries, such as gauge symmetries, and other symmetries such as Lorentz transformations. A philosophical view that adopts this approach was presented by \citet{auyang1995qft} in her book that aimed to present quantum field theory in a Kantian categorical framework of objective knowledge. \citet{teller1997metaphysics,teller2000gauge} criticized this view, raising the question of  ``[h]ow can an apparently substantive conclusion follow from a fact about conventions?'' (2000, p. 469). 

This tension between the significance of gauge invariance as an essential part of the basis for the formulation of successful theories, and the appearance of gauge as essentially a manifestation of mathematical redundancy was dubbed by \citet{redhead2002gauge} as ``the most pressing problem in current philosophy of physics'' (p. 299) and has been the subject of extensive philosophical inquiry. Many \citep{martin2003continuous, norton2003gauge} have pointed the similarity of the question to the controversial issue of the role of general covariance in general relativity \citep{norton1993dispute}, in which a mathematical constraint on the formulation of the theory is seen as fundamental to the construction of the theory. 

One possible way to approach this question is to adopt a deflationary approach towards  the gauge principle. \citet{brown1998objectivity} noted that neither the curvature of the fiber bundle nor the back-reaction of the matter on the gauge field can be explained by the principle. \citet{martin2003continuous} noted that in  the gauge argument, the gauge principle is applied in conjuction with other principles such as Lorentz invariance, renormalizability and simplicity. The requirement for a local symmetry does not by itself determine the form of the interaction, nor does it dictate its existence. While this criticism does amend misconceptions that appear in some textbooks, it does not claim to fully resolve the foundational and philosophical questions. 

A different approach appeals to a notion of direct observability of gauge symmetries, in which the symmetry is restricted to observations performed on a certain subsystem (Section \ref{DES}). A third approach \citep{mack1981gauge,lyre2000equivalence,lyre2001gauging,lyre2003phase} acknowledges gauge covariance as a formal requirement by itself, and aims to account for its applicability by supporting it with an additional equivalence principle, formulated by analogy with Einstein's presentation of general relativity based on general covariance and equivalence principle. The principle roughly states that there is always a choice of gauge such that the gauge-field vanishes at a given point. 

Applying a generalized version of the equivalence principle can be beneficial in understanding gauge not only at the interpretational level, but already at the level of heuristics. The ``methodological equivalence principle''  prescribes the introduction of an interaction based on an the way the dynamical law of a theory violates invariance requirement. It provides a unified framework of understanding standard gauge theories, the use of general covariance in general relativity and tangent-space symmetries in gauge theories of gravity \citep{hetzroni2020equivalence,hetzroni2021analogies,hetzroni2021gravity}. This approach can motivate an interpretation of gauge theories as grounded in ontology of relational quantities, in harmony with a recent account by  \citet{rovelli2014gauge}, and stressing a structural similarity between the gauge argument and Mach's principle \citep{hetzroni2020ghosts}. The origin of the gap between gauge independent physical phenomena and gauge dependent theoretical representation is that while fundamental physical degrees of freedom amount to relations between pairs of physical entities, our theories use variables that refer to single objects, and not their relations. From this perspective gauge invariance naturally arises as a formal property. What calls for an explanation is actually the non-invariance of interaction-free theories (special relativity under diffeomorphisms, Dirac equation under change of local phase convention etc.). The explanation is based on the existence of a gauge field not taken into account by those theories. 

\subsubsection{Interpretation of Gauge Invariance}
\label{gaugeinterpretation}

The appearance of a gauge symmetry in a given theory may have crucial implications on its possible interpretations and its ontology. The question of whether gauge symmetries reflect descriptive  mathematical redundancy or some physical property is deeply entwined  with issues of locality, separability, and determinism. Reconciling such theoretical virtues with each other, and in particular recall (D1)-(D3) introduced in Section \ref{INT}, is a non-trivial matter. 

Let us now introduce some terminology. We shall do so using the case study of electromagnetism (see \citealp[48]{Rickles2008}). Potentials related by the gauge transformation \eqref{gaugetransEM} (i.e., potentials that lead to the same fields) are \textit{gauge equivalent}. Gauge equivalent configurations lie on the same \textit{gauge orbit}. Accordingly, all potentials related by a gauge transformation are on the same gauge orbit. Our fields $\vec{E}$ and $\vec{B}$ remain constant under the gauge transformation \eqref{gaugetransEM}, they are \textit{gauge invariant}. The 4-vector potential $A^\mu$, i.e. the gauge field, is \textit{gauge variant}. Many mathematically distinct potentials lie on the same gauge orbit and lead to the same field. The fact that it does not matter which vector potential from a given gauge orbit we choose to express the fields is referred to as \textit{gauge freedom}.

Gauge freedom implies that our gauge theory is mathematically underdetermined: the values of the gauge field throughout space in a given moment of time together with the Maxwell equations do not determine the potentials in different times. In some interpretations, this underdetermination corresponds to a form of indeterminism. If we have a point $p_0$ on phase space\footnote{Technically, the phase space of the theory can be constructed from the magnetic vector potential $\vec{A}$ and the electric field $\vec{E}$, see \citet{belot1998EM}.} as our starting point, the system does not evolve into a state represented by a unique phase point $p_t$, instead ``we now have an indeterministic time-evolution, with a unique $p_t$ replaced by a gauge orbit'' \cite[289]{redhead2002gauge}. This is to say that if we take a gauge variant quantity such as the vector potential to be physically real, and mathematically distinct vector potentials to be physically distinct, then not only is there  no way to determine the state empirically, the state is also underdetermined by the equations of motion. 

Of course, this problem of indeterminism vanishes if one’s realism is constrained to gauge-invariant quantities, i.e. by identifying all gauge equivalent states with one physical state, an assumption that is often referred to as \textit{Leibniz Equivalence} (see for example \citealp{saunders2003leibniz}). We may therefore make a very broad distinction between two kinds of interpretations of gauge theories. On the one hand those that deny Leibniz equivalence, therefore taking gauge transformations to be physical transformations, allowing for realist commitments to gauge variant quantities, and interpretations that adopt Leibniz equivalence. Such interpretations would usually regard gauge freedom as an expression of  surplus mathematical structure, and restrict realist commitments to gauge-invariant quantities.

Proponents of the former type of interpretations may argue for a one-to-one correspondence between points on phase space and physical states such that each point on phase space represents a distinct physical state. This would mean that different points within the same gauge orbit would correspond to physically distinct states. In our case of classical electrodynamics, this would mean that gauge equivalent vector potentials would correspond to physically distinct states. These states would be empirically equivalent in the sense that no possible observation or measurement could reveal which gauge equivalent potential is currently in effect. As Rickles puts it, “these physically distinct possibilities are qualitatively indistinguishable” \cite[62]{Rickles2008}. Accordingly, such interpretations that allow gauge variant quantities to be physically real come with the great disadvantage of an awkward indeterminism that allows for physical quantities that are in principle unobservable. However, it is a virtue of such interpretations to avoid the existence of surplus mathematical structure that is typically considered a characteristic feature of gauge theories. Since mathematically distinct descriptions are related to physically distinct states, the problem of explaining surplus mathematical structure vanishes.

Proponents of the latter type of interpretations may argue for a many-to-one correspondence between points on phase space and physical states such that points within the same gauge orbit would correspond to one and the same physical state. In our case of classical electrodynamics, this would mean that vector potentials are unphysical and gauge equivalent vector potentials would correspond to the same physical state. Accordingly, such interpretations that restrict realist commitments to gauge-invariant quantities come with the virtue of avoiding indeterminism and restricting realist commitments to quantities that are in principle measurable. However, such interpretations that opt for a many-to-one correspondence between mathematical descriptions and physical states still face the problem of dealing with all the mathematical redundancy. Many physicists and philosophers share the sentiment of Zee that ``gauge theories are also deeply disturbing and unsatisfying in some sense: They are built on a redundancy of description'' \cite[187]{Zee2010}. Instead, many would desire a more direct correspondence between the mathematics and the physics, such that every mathematical degree of freedom has observable effects.

\subsubsection{Ontology of Gauge Theories}
\label{gaugeontology}

Interpretations of gauge theories are often presented as a response to the tension between gauge invariance and  locality that is revealed in the Aharonov-Bohm effect (Section \ref{ABeffect}). 

In their original paper Aharonov and Bohm argued that 
\begin{quotation}
in quantum mechanics, the fundamental physical entities are the potentials, while the fields are derived from them by differentiations[...] 
Of course, our discussion does not bring into question the gauge invariance of the theory. But it does show that in a theory involving only local interactions (e.g., \Schrodinger's or Dirac's equation, and current quantum-mechanical field theories), the potentials must, in certain cases, be considered as physically effective, even when there are no fields acting on the charged particles. [...] we are led to regard $A_\mu(x)$ as a physical variable. This means that we must be able to define the physical difference between two quantum states which differ only by gauge transformation. 
\end{quotation}

The radical suggestion that the actual physics behind electromagnetism is not gauge invariant is supposed to maintain locality. \citetext{A straightforward way to interpret classical electromagnetism in this way is to revive the concept of the aether, and consider the vector potential as its velocity field, and the electric field as its acceleration. See \citealp{belot1998EM}}. 

The idea has several significant disadvantages. One problem is that the interpretation leads to the radical indeterminism discussed in the previous subsection. 
Such an interpretation is not necessarily local; it might also lead to a form of action at a distance that is more explicit than the nonlocality of the Aharonov-Bohm effect. An electric current which starts to flow, for example, can (depending on the actual gauge) immediately change the values of the vector potential throughout space. Furthermore, it is controversial whether the potentials can provide a truly local explanation for the Aharonov-Bohm effect, due to the issue of separability \citep{healey1997nonlocalityAB, maudlin1998healey, healey1999maudlin, eynck2001AvsB}. The potential's approach was criticized by Aharonov himself, due to the unexplained gap it contains between the gauge independent observed phenomena and their gauge dependent theoretical description \citep{aharonovrohrlich2008QP}.

In light of the evident downsides of potentials' ontology, it may be worthwhile to reconsider the standard field interpretation of electromagnetism and see if it can be adapted to account for the Aharonov-Bohm effect. Such an approach was presented by \citet{dewitt1962AB}, who concluded that 
\begin{quotation}
Nonrelativistic  particle mechanics as well as relativistic quantum field theories with an externally imposed electromagnetic field can therefore be formulated solely in terms of field strengths, at the expense, however, of having the field strengths appear nonlocally in line integrals.
\end{quotation}
A different account of a nonlocal influence of fields is described by \citet{aharonovrohrlich2008QP} (mainly Chapters 4-5) in terms of modular variables. Nevertheless, even in this nonlocal approach the potentials are an essential part of the theoretical description of the interaction.\footnote{
\citet{strocchi1973superselection} suggest a local gauge invariant explanation by writing \Schrodinger's equation as an equation for the current density $j$ expressed in terms of the fields. The locality depends on the existence of a ``tail'' of a non-vanshing wave-function in overlap with the source. A very different local and gauge invariant explanation was offered by \citet{vaidman2012AB} based on a transient entanglement induced by the influence of the field generated by the particle on the source. See discussion in \citet{aharonov2015comment,vaidman2015reply} concerning the question of the formal indispensability of the potentials. } 

A third explanation was suggested by \citet{wu1975factors}. It closes the gap between the observable properties and the postulated physical properties in a straightforward way. \citeauthor{wu1975factors} noted that the actual observable quantity is not the AB phase difference $\Delta\varphi_{AB}$ itself (Equation \eqref{ABphase}), but rather, the phase factor (or holonomy, or anholonomy):
\begin{equation}
\label{ABphasefactor}
\Phi_{AB}=\exp\left(i\Delta\varphi_{AB}\right)=\exp\left(i\frac{q}{\hbar c}\oint{A^*\left(\vec{r}\right)\cdot\vec{dr}}\right). 
\end{equation}
Their approach, known today as the \emph{holonomies} approach to the Aharonov-Bohm effect, is to promote this factor (defined for any non-intersecting closed curve in space-time) to a fundamental variable from which all other electromagnetic quantities can be derived. 	The description of electromagnetism based on holonomies is gauge invariant and nonlocal. This kind of nonlocality is not a dynamical action at a distance, but a kinematic non-separability. This interpretation was advocated by \citet{belot1998EM} as a fruitful interpretation and by \citet{eynck2001AvsB} because of the local action and measurability of the holonomies. 

Forming an analogy with the ontology of space debate, \citet{arntzenius2014space} relates to  holonomies interpretation  as \textit{gauge relationism}, which he contrasts with \textit{fibre-bundle substantivalism} he presents. The ontology in this approach is a literal reading of the fibre bundle representation of gauge theories. It consists of fibre bundle representing the possible states of a gauge field over spacetime manifold. The structure of the fibre (vector space, Abelian\textbackslash non-Abelian group, etc.) determines the property of the interaction represented by the gauge fields, understood in terms of connections over the bundle. In the most literal reading, gauge transformations in these interpretations do reflect physical change (as in \citealp{maudlin1998healey}). It is possible, however, to formulate ``sophisticated'' version of this substantivalism, which denies the existence of possible worlds that are only connected by a gauge transformation. The main reason to favour substantivalist approaches to gauge, according to Arntzenius, is the availability and simplicity of theories formulated in terms of gauge dependent variable, in contrast to the difficulties in constructing a dynamical theory  in which holonomies are a fundamental variable.

\subsubsection{Direct Empirical Significance}
\label{DES}

The debate about the ontological correlates of symmetries is sometimes linked to the question of which symmetries have ``direct empirical significance'' and which do not. The idea here is that if a symmetry is a mere descriptive redundancy, it cannot be said to have any direct empirical significance. (Though it may have indirect empirical significance: the very fact that the theory can be formulated in terms of these symmetries has specific further implications, e.g. conservation laws.) In contrast, symmetries that connect states that are physically distinct yet empirically equivalent are said to have direct empirical significance.

This question is intimately related to the distinction between systems and subsystems. There is no physical point of view from which a state of the entire universe and a symmetry-related state could be distinguished empirically, and it is therefore common to regard symmetries of the entire universe as not carrying any empirical significance. For symmetries applied only to proper subsystems of the universe, the situation is different. Here the standard example is the thought experiment known as Galileo's ship:
the inertial state of motion of a ship is immaterial to how events unfold in the cabin, but is registered in the values of relational quantities such as the distance and velocity of the ship relative to the shore.

The question that is in the centre of the debate about direct empirical significance is whether the same holds for local symmetries as in gauge theories. The orthodox view, that local gauge symmetries do \textit{not} carry empirical significance, has recently been argued to be derivable from three simple assumptions about direct empirical significance and physical identity of subsystem states \citep{Friederich2015,Friederich2017}, see \citep{Murgueitio} for criticism of the assumptions used there. However, recently, dissonant voices have become more prominent e.g. \cite{GreavesWallace,Teh_emp, Wallace2019a, Wallace2019b}. The unorthodox view reproduces, for any subsystem, a common interpretation of asymptotic symmetries within physics: that any gauge transformation that preserves the state at the boundary and yet is not asymptotically the identity, acquires empirical significance. 

But this transposition of asymptotic conclusions to compact subsystems has come under attack from many directions.  For instance, in \cite{Gomes2021} counter-examples were shown to exist in vacuum electromagnetism; follow-up work gauge-fixes the local symmetry and asserts that only global symmetries can carry empirical significance as subsystem transformations, and this occurs only when such global symmetries are associated with covariantly conserved quasi-local charges \cite{GomesNoether}. Moreover, the common view of asymptotic symmetries \textit{can} be recovered within such a gauge-fixed formalism (cf. \cite{RielloSoft}).

\subsection{Artificial vs. Substantial Gauge Symmetries}
\label{AVS}

Depending on the physicist, and the time at which they write, the appearance of gauge symmetries in our most successful theories of fundamental physics is regarded either as a deep insight into the structure of  matter and  fundamental interactions, or as a mere feature of convenience, reflecting a mathematical redundancy of the descriptive formal apparatus (see section \ref{The Gauge Principle Meets Philosophy}). 
Indeed the physical importance of gauge symmetries (if only at an heuristic level) is attested by the widespread acceptance of the \emph{gauge principle} -- constraining the form of interacting theories that generalise the free ones possessing fewer symmetries -- as a fundamental concept on par with the relativity or general covariance principle. Both are sometimes seen as symmetry principles uncovering deep structural aspects of physical reality. Such a position, and endorsement of the spirit of Yang's already mentioned aphorism that ``symmetry dictates interactions" \citep{Yang1980}, is prima facie in tension with the view of gauge symmetries as `surplus' theoretical structure. 

Resolving this tension, we argued in section \ref{INT}, hinges on a refined understanding of what should count as 'surplus structure' in gauge theory. 
 
A first step in the right direction is to notice that not all gauge symmetries are  necessarily on a par. Indeed, it happens that, for technical reasons, physicists are used to routinely `adding' gauge symmetries to theories that initially had none, using a variety of tools (whose forefather is perhaps the Stueckelberg trick \citep{Ruegg-Ruiz}). But if practically any field theory can be turned into a gauge theory, it is hard to understand how gauge symmetries could be fundamental. 

This state of affair is reminiscent of a well-know discussion within the philosophy of general relativistic physics, spurred by the \emph{Kretschmann objection}: In 1917, E. Kretschmann criticised the foundational status that Einstein gave to the principle of general covariance for his General Relativity, observing that any theory could in principle be recast  - perhaps with some cleverness required on the part of the theoretician - in the language of tensor calculus, thus in a generally covariant way. This prompted a recognition that one should distinguish \emph{substantive} general covariance, native of a theory, from \emph{artificial} general covariance, that is formally forced onto a theory, and that the main question is then to specify a demarcation criterion between the two.\footnote{The complete story is somewhat more subtle, of course. See e.g. \citep{Pooley2017}.}

Given the above observation, a \emph{generalised Kretschmann objection} could similarly be leveraged against the gauge principle, suggesting a provisional distinction between  \emph{substantive} and \emph{artificial} gauge symmetries \citep{Pitts2008, Pitts2009}. The question would then be: when presented with some gauge field theory, on what ground could we make the distinction? Presumably, one may look for physical signatures of substantive gauge symmetries that are lacking for artificial ones. 
As 
  discussions in previous sections   
suggests, there is a feature that is usually recognised as characterising `honest' gauge symmetries: a trade-off between locality and gauge-invariance. It is indeed a fact that some gauge theories can be reformulated in a gauge-invariant way without sacrificing the locality of their basic variables (e.g. scalar electrodynamics and the abelian Higgs model), while others can't (e.g. spinorial electrodynamics and pure Yang-Mills theories) -- see section \ref{The dressing field method}. This certainly does not exhaust the physical content of substantive gauge symmetries, but one may take this trade-off to be a robust physical signature: On the position that physical d.o.f. are gauge-invariant, this criterion would classify  `true' gauge theories as ones that  display a certain non-locality (entirely distinct from quantum non-locality).\footnote{On the (difficult) position that gauge-variant variables are physically distinct but in principle empirically indistinguishable, the trade-off, though not interpreted as a physical signature, remains. Thus the distinction between the two classes of gauge symmetries still holds.} The famous Aharonov-Bohm effect, discussed in section \ref{ABeffect}, seems to provide support to the claim. 

Many -- likely all -- interpretive and conceptual issues raised by gauge symmetries, and discussed in this text, concern the substantive kind. Artificial gauge symmetries pose no -- or at least fewer -- mysteries as they can be redefined away, usually by some change of local field variables. 
Thus, presented with a gauge theory one could first seek to identify which, if any, of its gauge symmetries are artificial, and get rid of them so as to work with a minimal (`Ockhamised') version of the theory displaying only substantive residual gauge symmetries, if any. 
 We might call such a minimal, and still local, version of a gauge theory its ``maximally invariant formulation".
 The latter is certainly the proper candidate to be subjected to in-depth philosophical and theoretical analysis. 
 \footnote{Section \ref{The dressing field method} highlights an approach aiming to obtain maximally invariant formulations of  gauge theories, the Dressing Field Method. Examples are given, but as we will see the application of main interest is the electroweak model -- which is also the main concern of the invariant formulation via the FMS approach of section \ref{Sec:FMS approach}. }

According to the definition of `surplus structure' suggested in section \ref{INT}, artificial gauge symmetries are genuine surplus, while substantial gauge symmetries are not since they signal the non-locality -- or non-separability -- of the fundamental d.o.f. of gauge physics. 
Yet we should remark that, as also hinted in  \ref{INT}, it may be that features initially  deemed surplus are promoted if it is eventually found that they have a direct physical signature, or one more indirect in the form of an important theoretical virtue.  
Thus, while the distinction  between substantive and artificial gauge symmetries holds good for classical gauge field theories,
one may be led to reassess this provisional demarcation in quantum gauge field theory. 

Notably, a centrally attractive feature of gauge theories is their good renormalization properties under quantization.  
For a fundamental theory, renormalizability is often deemed an important virtue with deep physical relevance. 
Admitting this, if it appears that a gauge symmetry classically classified as artificial nonetheless plays a significant role in ensuring that the quantum theory is renormalizable, one would then be justified in promoting it to a substantial symmetry.
Let us raise an intriguing possibility though: 
that on the contrary the  distinction  would extend to QFT, so that
a gauge symmetry initially thought to be necessary to ensure good quantum properties, yet classically marked as artificial, would on second analysis turn out to be dispensable to the quantum theory.
\footnote{Case in point, if $\SU(2)$ in the electroweak model is classically artificial -- as it arguably is according to section \ref{Invariant formulation of the electroweak model} -- is it really unavoidable to prove the renormalisability of the quantized model?} This we leave as an open question.

\subsection{Dualities}
\label{DUAL}

When discussing the idea of the possibility to eliminate (gauge) symmetries, another item necessarily appearing is the concept of dualities: The situation that physical systems can have two or more equivalent theoretical formulations. The different formulations may have different symmetries, even to the point that there exist a formulation with no symmetries at all. In general, if multiple theories exist, which agree on all observables, but have degrees of freedom which are not in a one-to-one-relation with each other, they are called dual to each other.

Such relations are more than just variable transformations. A variable transformation is a mathematical process which is unambiguous, and thus invertible. In theories with the same physical content, but different levels of symmetries, there is necessarily an ambiguity in the relation between variables.

A, relatively trivial, example is in classical mechanics the problem of orbital movement in the usual gravitational potential. It is a superintegrable system, which means that out of the three independent degrees of freedom all but one can be eliminated, using the conservation of the Runge-Lenz vector and of orbital angular momentum. But, one can choose to eliminate only one of the two degrees of freedom. This gives an intermediate theory with inequivalent symmetries to the one with only one degree of freedom, and connected to it by a non-invertible variable transformation.

There are a vast number of further examples. Perhaps one of the most extreme cases is two-dimensional Yang-Mills theory in flat space-time [\cite{Dosch:1978jt}]. This theory is trivial in the sense that there are no physical excitations, and the only physical state is the (empty) vacuum. Nonetheless, there is a denumerable infinite number of associated gauge theories, Yang-Mills theory with an arbitrary Lie group, which all create this same single state. These theories all differ in terms of their gauge-dependent Green's functions, and thus in the way cancellations occur to create such a single state.

There are further theories, which are known to be exactly related to each other, and are non-trivial. However, these are either theories, which are not gauge-theories, like the low-temperature/high-temperature duality of the Ising model [\cite{Kogut:1979wt}], or have a gauge theory only on one side, like the gauged O(4) linear sigma model with punctured target space and the corresponding non-gauge theory of massive scalars and vectors [See section \ref{Examples} and \cite{Maas:2017wzi,Fernandez:1992jh,Evertz:1985fc}]. 

While theories with trivial dynamics appear to be quite irrelevant, there is a wide variety of non-trivial theories where similar features are, to some extent, established, motivated, or conjectured. Probably most notorious are dualities between weakly coupled theories and strongly coupled theories. These are especially prolific for supersymmetric gauge theories [\cite{Weinberg:2000cr}], but have also been conjectured for some so-called walking theories [see e\ g.\ \cite{Sannino:2009za}]. In such cases often generalized electric and magnetic degrees of freedom are exchanged. However, the celebrated AdS/CFT conjecture [\cite{Freedman:2012zz}] suggests even the possibility that a quantum gauge theory in a fixed space-time could be equivalent to a classical gravitational theory in a different space-time. Would this be proven, especially for any applicable theory describing our universe, this would imply that even the fundamental structure of space-time is exchangeable.

Dualities extend the ideas of Kretschmann [\cite{Pitts2009}], whose conjecture was that every theory can be covariantized. These considerations lead to a much stronger conjecture: That for every set of physical observables there exist multiple (gauge) theories, which are not related to each other by unambiguously invertible variable transformations.

Finally, we note that very recently the topic of dualities has received some well-deserved attention in the philosophy of physics (see, e.g., \cite{CastellaniRickles} and \cite{HaroButterfield}). Here we find some notable agreement that duality is closely related to symmetry. More precisely, the idea is “that a duality is like a ‘giant symmetry’: a symmetry between theories” \cite[2974]{HaroButterfield}. In section 2.1.1, we have discussed in what sense symmetry is to be understood as “invariance under transformation.” The idea, now, is that while symmetries typically relate different physical states, or, in the case of gauge symmetries, the same physical state to different mathematical descriptions, in the case of dualities theories are related such that different theories describe the same (observable) physics. What makes the interpretation of duality particularly interesting is that dual theories are often associated with radically different ontological commitments. For instance, as it is the case with respect to the AdS/CFT conjecture, if one theory is formulated in D-dimensional space-time and the (allegedly) dual theory in D-1-dimensional space-time, what does this say about the world we live in? When there are two dual theories, describing the world in very different concepts, is one of them true and the other one false? Are both false, indicating that there is a true theory hidden beyond? Or is a unique mathematical description of the physical world impossible and dual theories constitute different (and possibly incomplete) but equally valid perspectives? 

Obviously, it would go beyond the scope of the paper to discuss this in detail. What we do note, however, is that in the literature dualities have been systematically compared to gauge symmetries (\cite{Rickles2017}, \cite{Haro2017}). This is because “dual theories can indeed ‘say the same thing in different words’—which is reminiscent of gauge symmetries” \cite[68]{Haro2017}. Accordingly, we conclude this section on the interpretation of gauge symmetries by emphasizing the significance of this undertaking. However, we note that the methods we discuss in Sections \ref{The dressing field method} and \ref{Sec:FMS approach} can be regarded as examples of how to establish dualities between gauge theories and non-gauge theories
and their important role in gauge-invariant formulations the BEH mechanism.
For example, the dressing field method provides  a reformulation of the Abelian Higgs model or the Glashow-Weinberg-Salam model of electroweak interactions in terms of gauge-invariant fields at the classical level, see Sec.~\ref{The dressing field method}. In addition, the concept of dualities, relating the state space of different theories, allows for a potential reinterpretation of the BEH mechanism which avoids the terminology of spontaneous gauge symmetry breaking as conjectured in [\cite{Sondenheimer:2019idq}]. The framework proposed by Fr\"{o}lich, Morchio, and Strocchi (FMS) to formulate observables of a BEH theory in a gauge-invariant manner, provides a link between some gauge-invariant observables in one gauge theory to invariant observables with respect to another particular gauge group.\footnote{Note, that the latter group can also be trivial.} Thus, the BEH mechanism, i.e., the introduction of a scalar field with a Mexican hat type interaction potential, does not involve a spontaneous breaking of the original gauge group but induces a duality between particular states of two gauge theories in a specific way that is described by the FMS formulation. The seeming spontaneous breaking rather displays the fact that this duality becomes most apparent in a gauge-fixed language which explicitly breaks the original gauge group. We further elaborate on this relation in Sec.~\ref{Sec:FMS approach}.

\section{Symmetry Breaking and the BEH Mechanism}\label{Sec:Breaking}

The term `symmetry breaking' refers to a collection of phenomena and theoretical notions generally characterized by a situation in which the state of a system does not respect the symmetry of the laws governing it. The concept became dominant in theoretical physics around the early 1960s, in an interplay between condensed-matter physics and particle physics. The crowning glory of this framework is widely considered to be the Brout-Englert-Higgs (BEH) mechanism. A fundamental part of the electroweak model in particle physics, the mechanism is commonly presented as a case of dynamical spontaneous symmetry breaking of gauge symmetry. This standard account, however, has  been challenged on various theoretical and philosophical grounds. 

The issue is clearly interconnected with that of the interpretation of gauge symmetries. In this chapter we shall briefly present the standard account of the mechanism, and recent developments in its understanding. We shall then focus on motivating gauge-invariant approaches (which will be the subject of the  following two chapters) in light of problems with the standard account of symmetry breaking on the one hand, and philosophical discourse on gauge presented in the previous chapter on the other hand. 

\subsection{Symmetry Breaking at a Glance}
\label{symbreak}

Historically, one of the first physicists to recognize the systematic significance of the notion of symmetry breaking was Pierre Curie, the 1894 paper `Sur la sym\'{e}trie dans les ph\'{e}nom\`{e}nes physiques' being his principal work on this topic. Symmetry breaking, he noted, is not some obscure, rare phenomenon but is commonplace and may be the reason for the occurrence of phenomena in the first place. According to Curie,
\begin{quotation}
\textit{A phenomenon can exist in a medium which possesses its characteristic symmetry or that of one of the subgroups of its characteristic symmetry}. In other words, certain symmetry elements can coexist with certain phenomena, but they are not necessary. What is necessary is that some symmetry elements be missing. \textit{Asymmetry is what creates a phenomenon}. \cite[312]{Curie2003}
\end{quotation}

This idea that `[\textit{a}]\textit{symmetry is what creates a phenomenon}' underscores the significance of symmetry breaking and is a theme not unfamiliar in modern (philosophy of) physics. Castellani, for instance, declares:
\begin{quotation}
Any symmetry we can perceive (albeit in an approximate way) is indeed the result of a higher order symmetry being broken. This can actually be said of any symmetry which is not the ‘absolute’ one (i.e.\ including all possible symmetry transformations). But we can say even more: in a situation characterized by an absolute symmetry, nothing definite could exist, since absolute symmetry means total lack of differentiation. \cite[p. 322]{Castellani2003}
\end{quotation}

The claim that phenomena exist due to broken symmetries (or the lack of symmetry) is related to, but not equivalent to, what has become to be known as Curie’s principle. The principle says: “\textit{When certain causes produce certain effects, the symmetry elements of the causes must be found in their effects}” \cite[312]{Curie2003}. Conversely but equivalently, put in terms of asymmetry, the principle reads: "\textit{When certain effects show a certain asymmetry, this asymmetry must be found in the causes which gave rise to them}” \cite[312]{Curie2003}. Partly due to its vagueness, Curie’s principle has met opposing reactions (see \citealp{Earman2004b}). Attempts to make it more precise have resulted in a formulation that, according to prominent voices, “makes it virtually analytic” \cite[173]{Earman2004b}. In slogan form, the principle says: “\textit{if no asymmetry goes in, then no asymmetry comes out}” \cite[580]{Roberts2013}.\footnote{Although the principle is considered “an analytical truth” (\citealp{CastellaniIsmael}), Bryan Roberts has argued that the principle fails in certain cases of time reversal symmetry (\citealp{Roberts2013}). For more details on the relationship between Curie’s principle and spontaneous symmetry breaking, see \cite{Earman2004b}.} This leads us to the following question: How is it possible that our world manifests asymmetries although this world, on a fundamental level, seems to be governed by symmetric laws and theories? How exactly does asymmetry enter the picture and what does it mean and how can it be that a symmetry breaks down?

There are two forms of symmetry breaking relevant in modern physics: explicit and spontaneous.\footnote{There are also other options, like breaking due to quantum anomalies. However, for the purposes at hand, such additional options could be lumped into either of the categories, e.\ g.\ anomalies into explicit breaking.} Explicit symmetry breaking is rather straightforward. A symmetry is broken explicitly when  the equations of motion describing the system (as a consequence of the Lagrangian or Hamiltonian of the system not being invariant under the symmetry), are not covariant under the transformations. This is the case when a symmetry-breaking term is added in some situations to an otherwise covariant dynamics (e.g. due to an external field). 

In the case of spontaneous symmetry breaking (SSB) the equations of motion describing the system are covariant concerning the respective symmetry transformations, but the system is in a state that is not invariant under all symmetry transformations. This means that a system that is governed by a symmetric Hamiltonian can evolve into an asymmetric state. 
In particular, SSB appears when the ground state is degenerate, i.e. there are multiple states with the lowest energy. The transformation that connects these states instantiates a symmetry of the dynamics, yet each of the states does not respect the the symmetry, as it transforms it to one of the other states.

The above description relates to classical theories. At the classical level,  the choice of a particular ground state can be provided by part of the initial conditions. At the quantum level, this becomes more involved [\cite{Maas:2017wzi,Sartori:1992ib,Birman:2013gaa}]]. In quantum systems with finitely many degrees of freedom, the actual ground state always respects the symmetry. It is a superposition of non-symmetric states that are connected via quantum tunneling. Thus in a quantum theory SSB can only occur in systems with infinitely many degrees of freedom. this can be thought of in terms of the tunneling barriers becoming infinite.

In such quantum settings, phenomena associated with SSB can also be described in terms of explicit breaking. If a system shows SSB, it exhibits a well-defined behavior if the symmetry is broken by an external source as well. It will then show a broken (asymmetric) ground state even when taking the limit of a vanishing source\footnote{This is usually observed in quantities known as order parameters, which are local quantities not invariant under the symmetry to be broken. However, it is possible to detect SSB also with other quantities [\cite{Maas:2017wzi}]. Furthermore, in gauge theories there are no gauge-invariant local order parameters [\cite{Frohlich:1981yi,Maas:2017wzi,Elitzur1975}], and whether alternatives exist is unclear, cf footnote \ref{nonlocalfootnote}. We therefore refrain here from using this concept.}.

For gauge theories, any such source would itself be gauge-dependent, and thus break the gauge symmetry explicitly. The question thus naturally arises whether external sources are necessary to define spontaneous symmetry breaking. That can already be addressed at the global level. In fact, because the path integral sums over all configurations, there is no possibility without external source for singling out spontaneously a direction and thus break a symmetry. Thus global symmetries are always intact without external sources [\cite{Frohlich:1981yi,Maas:2017wzi}]. Nonetheless, it is still possible to determine how it would react to an external infinitesimal source using suitable observables [\cite{CaudyGreensite2008,Maas:2017wzi}]. If it would react in a non-analytic way, the system is metastable, and this metastability is equivalent to the usual picture of SSB.

Thus, external sources are not necessary to identify SSB. In fact, they can even be misleading for gauge theories, as cases are known in which metastability is signalled in absence of actual SSB [\cite{Maas:2013sca}]. Furthermore, Elitzur's theorem even states that gauge symmetries can never experience SSB [\citet{Elitzur1975}]\footnote{\label{analyticityfootnote}Note that many statements such statements use a lattice regulator to make the theory well-defined. Analyticity and regularization have an often tricky relation. This is even more so for the lattice regulator which introduces additional non-analyticities by its very definition depending on a second-order phase transition. While exact proofs are notoriously difficult in continuum quantum field theories, the existing evidence, as well as special results [\cite{Frohlich:1981yi}], strongly suggest that the situation for the present topic is under sufficient control [\cite{Maas:2017wzi}].}. Hence, the whole concept of SSB cannot be transferred directly from global symmetries to gauge symmetries.

This leads to the central issue: What is meant when, colloquially, physicists speak of SSB of gauge symmetries? The most prominent example for this issue is the Brout-Englert-Higgs effect (BEH), presented in the following subsection and discussed subsequently.

\subsection{Gauge Symmetry Breaking and the BEH Mechanism}
\label{gaugesymbreak}

The standard text-book account of the BEH mechanism sets off from spontaneous breaking of the global symmetry, and then extends the analysis to the corresponding local gauge symmetry. In a spontaneous breaking of a global continuous symmetry, the different lowest-energy states are connected by a transformation of a degree of freedom that, in particle terms, can be interpreted as massless, spinless `Goldstone's bosons', that interact with the matter fields. The introduction of the field associated with Goldstone's bosons can be described using classical fields, by rewriting the Lagrangian, separating degrees of freedom that are transformed from one ground-state to another and `radial` directions, associated with massive particles. In quantum field theories this reformulation gains further significance, as it corresponds to a perturbative analysis around a particular ground-state. 

 The simplest example concerns two scalar fields $\phi_1$, $\phi_2$ with a $U(1)$ symmetry corresponding to rotation in the internal field space (i.e., continuously transforming the one into the other). The potential is taken to be $V(\phi)=-\frac{1}{2}\mu^2 \phi^2+\frac{1}{4}\lambda \phi^4$, a function of the  gauge independent term  $\phi^2\equiv\phi_1^2+\phi_2^2$. Thus, all states satisfying $\phi=\mu^2/\lambda\equiv v^2$ have minimal energy. In this model, $\phi$ can be treated as a radial direction, corresponding to a massive boson. Transformations along the tangential direction generate the symmetry, and are associated with a Goldstone boson. This description is said to break the symmetry since perturbative analysis around any particular vacuum state (e.g. $\phi_1=v^2$; $\phi_2=0$) leads to the Lagrangian ${\cal{L}}=\left[\frac{1}{2}\partial_\nu \eta \partial^\nu \eta -\mu^2\eta^2\right]+\left[\frac{1}{2}\partial_\nu \xi \partial^\nu \xi \right]+[\text{interaction terms}]$ with $\eta$ radial perturbation and $\xi$ tangential ($\eta\equiv\phi_1-v^2$ and $\xi\equiv\phi_2$ for the aforementioned state). 

When applying the gauge principle, the global symmetry is localized, introducing a new massless bosonic field. In the above example the field $A^\nu$ defines a connection over the $U(1)$ bundle. The central point in this standard account is that when the Lagrangian is expressed in terms of $\xi$ and $\eta$, the gauge field $A^\nu$ obtains a non-vanishing mass term (that equals to $\frac{1}{2} \frac{q\mu}{\hbar c \lambda} A_\nu A^\nu$ in the above example). $A^\nu$ is therefore interpreted as massive bosonic field. The breaking of the local symmetry is  associated with a choice of particular gauge. In order to allow for the massive-boson interpretation, the gauge is chosen such that the terms corresponding to the peculiar and unobservable Goldstone bosons disappear. Thus the mechanism, according to this standard presentation, allows for gauge bosons to obtain mass at the price of sacrificing gauge invariance.  

Moving from this toy model to theories with more empirical relevance, the prototypical theory is initially invariant under some local internal symmetry group and would accordingly have some gauge fields $W_\mu^a$, supplemented by one or more scalar fields $\phi$. In some representation of the gauge group the dynamics are defined by the Lagrangian 
\begin{align}
{\cal L}&=-\frac{1}{4}W_{\mu\nu}^aW^{\mu\nu}_a+(D_{\mu\alpha\beta}\phi_\beta)^\dagger D^\mu_{\alpha\gamma}\phi_\gamma-V(\phi),\label{la:hs}\\
W_{\mu\nu}^a&=\partial_\mu W_\nu^a-\partial_\nu W_\mu^a+gf^{abc}W_\mu^b W_\nu^c,\\
D^\mu_{\alpha\beta}&=\partial^\mu\delta_{\alpha\beta}+gT^R_{a\alpha\beta}W_a^\mu.
\end{align}
The $T$ are suitably normalized generators for the representation $R$ of the scalar field, $f^{abc}$ the corresponding strutcure constants, and $g$ is the newly introduced gauge coupling. The most prominent example is the standard model Higgs sector. In this case the scalar is in the fundamental representation of an SU(2) gauge group. In this special case the theory furnishes also, a global SU(2) symmetry acting only on the scalar field. The Abelian Higgs model is recovered for $f^{abc}=0$ and $T^R=1$.

Since the potential is, by construction, a function of gauge-invariant quantities only, so are the minima given by gauge-invariant conditions,
\begin{align}
\phi^\dagger\phi&=v^2\label{minhpot},
\end{align}
where $v\neq 0$ minimizes the potential $V(\phi)$, and is thus a function of the parameters of the potential. These minima are for a polynomial potential necessarily translationally invariant, and therefore the fields satisfying (\ref{minhpot}) need to be constant. The minima need also to be invariant under any intact global symmetry. The condition $v\neq 0$ follows from the structure and parameters of the potential\footnote{This assumes that the quantum effective potential, in a suitable manner, still allows for (\ref{minhpot}) to hold. Furthermore, a suitable renormalization scheme is required to maintain (\ref{minhpot}) literally beyond tree-level [\cite{Bohm:2001yx}]}. This condition is at the root of the BEH effect, and thus of the so-called gauge SSB, and everything else follows from it.

The existence of such a non-trivial minimum is then exploited by setting the length of the space-time averaged Higgs field as a gauge condition to the value $v$. To complete the gauge, it is customary to use a local condition
\begin{align}
\partial^\mu W_\mu^a+ig\zeta\phi_\alpha T^a_{\alpha\beta} v_\beta+\Lambda^a=0\label{thooftg},
\end{align}
where $v_\beta$ is a vector of length $v$ and $\zeta$ is an arbitrary gauge parameter. The direction of $v_\beta$ is arbitrary, but fixed by this gauge choice. It is really the second term in (\ref{thooftg}), which enforces that one direction is made special, and thereby breaks the gauge symmetry in such a way as to ensure a vacuum expectation value for the Higgs field. It thereby establishes the BEH effect in the usual picture of a condensation of the Higgs field. Such a gauge choice would not be possible, if the potential did not support a non-trivial minimum $v^2>0$.

Returning for a second to the classical level, initial conditions for the equations of motion of the gauge-dependent fields would be needed to be selected to comply with this gauge choice, and therefore implement the BEH effect. At the quantum level, there are no initial conditions. Hence, the gauge choice alone can enforce SSB and, in that sense, the BEH effect. After the gauge is implemented is implemented, calculations can be done, especially in form of perturbation theory.

\subsection{Foundational and Interpretational Issues}

Already in the early literature on the BEH mechanism it was noted that the gauge condition is not necessary [\cite{Lee:1974zg}]. The vacuum expectation value (VEV) of the Higgs field, while sometimes identified with observable properties, is in fact gauge dependent. Furthermore, gauges that do not fix it are possible, as well as gauges that set it to zero. This may seem surprising from the point of view of the standard perspective of the BEH as relying crucially on SSB, but it is actually not surprising in the light of Elitzur's theorem \citep{Elitzur1975}, which states that local symmetries cannot be spontaneously broken, concluding that ``breaking of local symmetry such as the Higgs phenomenon, for example, is always explicit, not spontaneous. The local symmetry
must be broken first explicitly by a gauge-fixing term leaving only global symmetry. This remaining global symmetry can be broken spontaneously'' (p.3981)\footnote{Elitzur's theorem, as well as similar proofs, all rely on lattice regularization. The question to which extent they pertain to continuum theories is not an easy one. See \citep{Maas:2017wzi} for an overview on this issue, cf footnote \ref{analyticityfootnote}}. 

Gauges in which the VEV is set to zero disable perturbation theory, and for that reason they did not find widespread use, see [\cite{Lee:1974zg,Maas:2017wzi}]). However, their existence makes the answer to the question whether the gauge symmetry has genuinely been broken appear by itself dependent on the choice of gauge! This hard-coding of SSB by gauge-fixing raises the question of whether it is physical. At least when using a lattice regulator in the standard-model case \cite{Osterwalder:1977pc} and \cite{Fradkin:1978dv} showed that there exists only a single phase, in the sense of an analytic free energy. Hence, the BEH effect cannot be a physical distinction. In fact, explicit calculations showed that the possibility to fix the vacuum expectation value of the Higgs field itself is gauge-dependent, and thus an unphysical distinction \citep{CaudyGreensite2008}\footnote{ \label{nonlocalfootnote}Note that many attempts exist to still identify phases using non-local conditions, see e.g. \cite{Greensite:2017ajx,Greensite:2018mhh}.}. As \citet{Friederich2013} puts it, as it depends on the choice of gauge-fixing, SSB of local symmetries does not qualify as a ``natural phenomenon.'' These issues, together with more basic considerations of gauge invariance raised by philosophers in this context (see Section \ref{Sec:Why}) suggest that the standard account of the BEH mechanicsm as an instance of gauge symmetry breaking is misleading. 

Nonetheless, ignoring these conceptual issues and identifying SSB after gauge fixing with the BEH mechanism, may have played an important heuristic role in the acquisition of a wealth of experimentally confirmed results in particle physics [\cite{Bohm:2001yx,Zyla:2020zbs}]. This is a baffling state of affairs. It strongly suggests the necessity for a reconciliation of both the formal aspects and the phenomenological successes. One way to achieve such a  reconciliation  will be presented and discussed in Section \ref{Sec:FMS approach}.

\section{Motivating Gauge-Invariant Approaches}
\label{Sec:Why}

As stressed in Section \ref{Interpreting Gauge Theorie}, there is a prioi a tension between the view of gauge symmetries as an insight into the inner workings of Nature -- as acknowledged by the wide recognition of the heuristic value of the Gauge Principle in discovering empirically successful mathematical descriptions of the fundamental interactions -- and the view of gauge symmetries as essentially a descriptive redundancy of our theoretical apparatus -- as manifested by the near universal acknowledgement that physical d.o.f. (observables ) should be gauge-invariant. Call this the ``profundity vs. redundancy" conundrum (see \cite{martin2003continuous}). 

While both these views are expressed in the physics literature, their inconsistency doesn't raise much interests or worries there, as it appears to be considered inconsequential to most practical or technical matters. Philosophers of physics on the other hand, whose job it is to worry about such things, have begun to seize on it when turning their attention to the foundations of gauge theories  some twenty years ago. About this tension, Michael Redhead (2002)  states ``In my view its elucidation is the most pressing problem in current philosophy of physics".
\smallskip

The one area where philosophers' and physicists' preoccupations (should) converge, and where 
the problem manifests itself with unique acuity, is when it comes to the notion of gauge SSB, especially given its implementation via the BEH mechanism in the Standard Model (SM) where it is  widely seen as a concept pivotal to our understanding of the electroweak interaction. 
It is indeed a standard textbook presentation, and the conventional wisdom,  that in the early history of the Universe, elementary particles interacting with the Higgs field (weak bosons and (most or all) fermions) gained their masses when the latter spontaneously broke the fundamental $\SU(2)$ gauge symmetry of the electroweak theory. But then how is this account compatible with the ``redundancy" stance on gauge symmetries?
John Earman expresses the tension particularly vividly:
\begin{quotation}
As the semi-popular presentations put it, ‘Particles get their masses by eating the Higgs field.’ Readers of \textit{Scientific American} can be satisfied with these just-so stories. But philosophers of science should not be. For a genuine property like mass cannot be gained by eating descriptive fluff, which is just what gauge is. Philosophers of science should be asking the Nozick question: What is the objective (i.e., gauge invariant) structure of the world corresponding to the gauge theory presented in the Higgs mechanism? \cite[1239]{Earman2004}
\end{quotation}

This is why we say that there is a worrisome tension at the very heart of modern particle physics. On the one hand gauge symmetries are considered unphysical mathematical redundancy but on the other hand the BEH mechanism is explained by gauge SSB. But how could the breaking of unphysical mathematical redundancy have any physical impact on our world? The underlying idea of this joint work is that instead of employing the conceptually dubious notion of gauge SSB one may look for manifestly gauge invariant approaches to the BEH mechanism. This sentiment has found a considerable echo in the philosophy of physics community, where gauge-invariant reformulation of the BEH mechanism is considered a crucial test-bed to clarify the conceptual foundations of gauge field theories (see \cite{Friederich2013, Friederich2014}; \cite{Lyre2008}; \cite{Smeenk2006}; \cite{Struyve2011}). In the physics community, while a narrative about the BEH in terms of SSB remains commons, dissenting views --  such as \cite{Chernodub2008}; \cite{Ilderton-Lavelle-McMullan2010}; \cite{Maas:2017wzi}; \cite{Sondenheimer:2019idq}  --  indicate an emerging  trend in particle physics where invariant formulations of the BEH mechanism are seen as a promising research endeavor.
\footnote{Especially so in the lattice community. See \cite{Maas:2017wzi} for a list of references, including an overview of the development of such views since the inception of the BEH effect.} 
Some physicists familiar with the philosophy literature share this vision, one of us raising the worry that
\begin{quotation}
\[...\] not acknowledging the insights of philosophers of physics would certainly lead to a long-lived misconception at the heart of particle physics to remain uncorrected for still some times, and important ensuing questions regarding the context of justification of the electroweak model to remain unasked, let alone answered. \cite[475]{Francois2018}
\end{quotation}
\medskip
It was indeed a motivation of this joint text to argue for a better awareness and appreciation of the general conceptual and interpretive issues surrounding gauge symmetries, especially as clear-headedness on these could actually have a concrete impact on the (re-)assessment of the foundation of electroweak physics and, perhaps, on future research avenues in particle physics. 
\medskip

In Section  \ref{INT}, we stressed that a satisfying interpretation of gauge theories may fulfil three desiderata that we remind to be 
\begin{enumerate}[label=(D{{\arabic*}})]
	\item To avoid ontological indeterminism.
	\item To avoid ontological commitments to quantities that are  not measurable even in principle.
	\item To avoid surplus mathematical structure that has no direct ontological correspondence.  
\end{enumerate}

We see that restricting one’s realist commitments to gauge-invariant quantities has the crucial advantage of avoiding a problematic indeterminism that allows for quantities that have no observable effects. This speaks in favor of the redundancy interpretation: Gauge symmetries are unphysical, full stop. While this is certainly a viable view some of us might endorse, above we have introduced a more subtle understanding that combines elements of both sides of the “profundity vs. redundancy” conundrum. We suggested that it should be understood as “surplus” any formal structure of a theory whose excision wouldn't be detrimental to its physical content and interpretation nor to its theoretical/pragmatic virtues. We hinted at the fact that for gauge theories, the (field-theoretic) notion of locality provides a robust criterion to detect such structures\footnote{One should note, however, that locality is seen by many, though not all, physicists as convenient but not necessary. Even though genuine non-local theories are much less understood on a technical level.}. On this notion was elaborated in Section  \ref{AVS}  the distinction between \emph{artificial} and \emph{substantial} gauge symmetries: 
only artificial gauge symmetries can be eliminated without sacrificing the locality of elementary field variables of a gauge theory,  making them genuinely surplus. 
This distinction already goes some way toward alleviating the ``profundity vs redundancy" tension: the gauge principle reveals its profundity when it points toward theories with  substantial gauge symmetries.
 The general issue at hand is therefore not to assess whether gauge symmetries are surplus structure or not; the refined question is how to distinguish between genuinely surplus gauge structure, and the non-surplus gauge structure whose physical signature relates to the non-locality or non-separability of gauge d.o.f.
 
Gauge theories undergoing gauge SSB are no exception to this general discussion. 
Given a gauge theory with SSB, two possibilities arise. Either the  gauge symmetry is substantial and its breaking correlates to (potential) physical observables, in which case gauge SSB is a genuine phenomenon.  Or, the supposedly broken symmetry is artificial, in which case there is an invariant local reformulation of the theory in which SSB cannot occur and is revealed as a formal artifact of the original formulation stemming from an inadequate choice of local field variables. 

To elucidate which of these  obtains in any given theory, one needs to reformulate it  in a maximally invariant way -- in the sense of Section \ref{AVS} --  leaving only substantial gauge symmetries. In particular, such a  gauge-invariant reformulation of the electroweak model is what is needed to answer Earman's question quoted above, and assess the real ontological status of electroweak SSB in the SM.  

In the following two sections, we present the state of the art of gauge-invariant approaches to the BEH mechanism. In Section 5, we highlight a tool adapted to such goals, known as the \emph{dressing field method}. We  illustrate it via several instructive examples before the main application to the electroweak model in Section \ref{Invariant formulation of the electroweak model}. As we will see then, the dressing field  method gives arguments to the effect that the $\SU(2)$ gauge symmetry of the model is artificial and that only the $\mathcal{U}(1)$ symmetry is substantial, so that SSB is superfluous to the empirical success of the SM -- this we observe, is consistent with Elitzur’s theorem. 
Then, in Section \ref{Sec:FMS approach}, we describe a closely related gauge-invariant formulation of the eletroweak model known as the FMS approach -- for Fr\"ohlich, Morchio, and Strocchi -- which has been much further developed, especially in lattice simulations, and is closer to being a viable alternative to the standard account regarding confrontation with  high precision collider experiments. 

\section{The Dressing Field Method of Gauge Symmetry reduction}
\label{The dressing field method}

Given a physical theory with certain gauge symmetries, the discussion in the previous section motivates an attempt to identify its physical degrees of freedom by replacing gauge dependent variables with gauge invariant ones. This step, when successful, results in a  
\emph{reduction of gauge symmetries}, i.e. the formulation of a theory with less symmetries. In fact, gauge fixing and SSB, discussed in previous sections, similarly apply reduction of gauge symmetries to achieve certain goals,  such as to allow for massive gauge fields mediating weak interactions or to control the quantization of the classical theory. In the past few years, another item in the physicist's symmetry reduction toolkit has been developed and is now known as  the \emph{dressing field method} (DFM)  (\cite{GaugeInvCompFields, Francois2014, Attard_et_al2017}).
In a nutshell, this approach allows to systematically build gauge-invariant field variables out of the initial gauge-variant fields of a theory \emph{if} a so-called \emph{dressing field} can be identified within it. In contrast to SSB and gauge fixing the DFM does not reduce the symmetry by replacing gauge invariant Lagrangian with one that has to be expressed in a specific gauge, but rather by removing gauge dependence already at the level of the the field variables.

As a matter of fact, the DFM provides a general framework of which one finds various retrospective applications in gauge theory.
The tetrad field in gauge reformulations of General Relativity (GR) is probably the first example of dressing field in the physics literature (in the writings of Einstein and Weyl).
The ``Stueckelberg trick" (\cite{Stueckelberg1938-I, Stueckelberg1938-II-III, Ruegg-Ruiz}) is another noticeable early instance of the method, as is
Dirac's gauge-invariant formulation of QED (\cite{Dirac55}, \cite{Dirac58} section 80).
Let us also mention e.g. the study of anomalies in QFT  (\cite{Stora1984, Manes-Stora-Zumino1985, Garajeu-Grimm-Lazzarini}),  or in QCD the so-called  ``proton spin decomposition controversy'' (\cite{LorceGeomApproach, Leader-Lorce, FLM2015_I}) and the issue of  constructing physical quark states (\cite{McMullan-Lavelle97, Heinzl:2008bu}).

Regarding the interpretive issues about gauge symmetries that we concern ourselves with in this essay, the main takeaway from the DFM is the following.
A dressing field, and therefore the gauge-invariant variables constructed from it, can either be local or non-local (like e.g. holonomies). In keeping with the nomenclature reminded in Section 4, if a local dressing field can be extracted, and thus a theory rewritten in a gauge-invariant yet local way, one may/could infer that the original gauge symmetry of the theory was an \emph{artificial} one: it was a genuine surplus structure, a mathematical artifact with no physical signature. When applied in particular to the electroweak model, the DFM suggests that the $\SU(2)$ symmetry  -- allegedly broken according to the standard narrative -- is actually artificial, and can be removed in the new "dressed" formulation of the model by a local change of field variables.

Such a surprising conclusion is nonetheless in keeping with a body of literature on invariant reformulations of theories undergoing SSB -- see e.g. \cite{Maas:2017wzi} for references -- going all the way back to \cite{Higgs66} and \cite{ Kibble67}. Philosophers of physics, such as \cite{Smeenk2006, Lyre2008, Struyve2011} and \cite{Friederich2013, Friederich2014}), did not fail to appreciate that such reformulations, which they sometimes rediscovered for themselves, should shed new light on these theories and on the electroweak model in particular.

To flesh out this short preview, in this section we propose an introduction to the DFM. It is formalised within the framework of differential geometry of fiber bundles,\footnote{For the acquainted reader, let us notice that it is close in spirit to  the Bundle Reduction Theorem, various presentations of which can be found in \cite{Trautman, Westenholz, Sternberg, Sharpe}.} which is  nowadays widely understood to be the mathematical foundation of classical gauge field theory. So, while the following presentation is fairly technical, we made efforts to make it self-contained and to give it as logically developed a progression as possible. Our aim is for a motivated non-expert to get a firm grasp  of the key  technical and conceptual  notions.

We start by providing background material in differential geometry, so that the DFM can be properly described next. We highlight how it can help rewrite gauge theories in an invariant fashion, and why it can be a tool to detect genuine surplus structure -- see also \cite{Francois2018} for further discussion of this point. The general discussion is illustrated via several examples before we turn to the main application to the electroweak model, which is followed by final short comments on the merits and limits of its reformulation via DFM.

\subsection{Geometric Background}
\label{Geometric background}

Let's take the view that the differential geometry of fiber bundles is the geometrical underpinning of classical gauge field theories. The recipe for one thus consists in a series of geometric ingredients providing the ``kinematics", so to speak, and a physical one providing the dynamics: the Lagrangian. 

Arguably the central ingredient is then a principal bundle $\P$ over spacetime $\M$ with structure group $H$ (the global/rigid symmetry group) and projection $\pi :\P \rarrow \M$, $p \mapsto \pi(p)=x$. A fiber over $x\in \M$ is $\pi\-(x)=\P_{|x} \subset \P$. Each fiber is an orbit of the right action of the structure group, $\P\times H \rarrow \P$, $(p, h) \mapsto ph =: R_h p$, which is free and transitive. The linearisation of this action induces vectors tangent to the fibers:  $\forall X \in$ Lie$H$ corresponds a vertical vector $X^v_{|p}$ at $p$. At $p \in \P$, the span of these vectors  is a sub-vector space $V_p\P$ of the tangent space $T_p\P$. The~collection of all such subspaces $\forall p \in\P$ is  the canonical vertical subbundle $V\P$ of the tangent bundle $T\P$.  We note $\Gamma(VP)$ the space of vertical vector fields $X^v: \P \rarrow V\P$ (i.e. sections of $V\P$).

Given representations $(\rho_i, V_i)$ of $H$, one naturally builds associated  bundles  to $\P$, $E_i\defeq \P \times_{\rho_i}\!V_i$ with typical fiber $V_i$, whose sections $s_i:\M \rarrow E_i$, $s_i \in \Gamma(E_i)$, represents various kind of matter fields. 
It is a standard result of bundle theory that sections are in 1:1 correspondance with representation-valued equivariant functions on $\P$, i.e. $\Gamma(E_i)\simeq \Omega^0_{\text{eq}}(\P, V_i):=\big\{ \vphi:\P \rarrow V_i\ |\  R_h^*\vphi = \rho_i(h\-) \vphi \text{ i.e. } \vphi(ph)= \rho_i(h\-) \vphi(p) \big\}$. 
More generally, one may define the space of representation-valued equivariant forms $\Omega^\bullet_{\text{eq}}(\P, V_i):=\big\{ \alpha \in \Omega^\bullet(\P, V_i)\ |\  R_h^*\alpha= \rho_i(h\-) \alpha) \big\}$, and the important subspace of \emph{tensorial} forms 
$\Omega^\bullet_{\text{tens}}(\P, V_i):=\big\{ \alpha \in \Omega^\bullet_{\text{eq}}(\P, V_i)\ |\  \alpha(X^v, \ldots)= 0, \text{ for } X^v \in \Gamma(VP) ) \big\}$. A form that satisfies this latter condition, without necessarily being equivariant, is said \emph{horizontal}. 
Notice that  $\Omega^0_{\text{eq}}(\P, V_i) =  \Omega^0_{\text{tens}}(\P, V_i)$. 
Tensorial forms with trivial equivariance are called \emph{basic}. The name is justified by the fact that basic forms induce, or come from, forms on the base $\M$. In fact an alternative definition is 
$\Omega^\bullet_{\text{basic}}(\P, V_i):=\big\{ \alpha \in \Omega^\bullet(\P, V_i)\ |\  \exists \, \beta \in \Omega^\bullet (\M, V_i) \text{ s.t. } \alpha=\pi^*\beta \big\}$. 

The exterior derivative $d$ on $\P$ does not preserve the space of tensorial forms, in particular $d \vphi \notin \Omega^1_{\text{eq}}(\P, V_i)$. Hence the introduction on $\P$  of a connection 1-form $\omega \in \Omega^1(\P, \text{Lie}H)$ defined by:

\begin{equation}
  R^*_h \omega = \Ad_{h\-} \omega,\,\,\text{i.e.}\,\,\omega \in  \Omega_\text{eq}^1(\P, \text{Lie}H),\label{eq:omega_equi}  
\end{equation}
\begin{equation}
\omega_p(X^v_{|p})=X \in \text{Lie}(H), \forall X^v_{|p} \in V_p\P.\label{eq:omega_proj}
\end{equation}

\noindent From these properties follows that one can define a \emph{covariant derivative}, $D^\omega := d\ + \rho_{i*}(\omega) : \Omega^\bullet_{\text{tens}}(\P, V_i) \rarrow \Omega^{\bullet+1}_{\text{tens}}(\P, V_i)$, where $\rho_{i*}$ are  representation maps for Lie$H$. So that in particular $D^\omega \vphi=d\vphi + \rho_{i*}(\omega)\vphi \in \Omega^1_{\text{tens}}(\P, V_i)$, i.e. a connection allows for a good notion of derivation on $\Gamma(E_i)$.
The choice of a connection 1-form on $\P$ is non-canonical. 
The space of all connections $\C$ is an affine space modelled on the vector space $\Omega^1_\text{tens}(\P, \text{Lie}H)$, meaning that for $\omega \in \C$ and $\alpha \in \Omega^1_\text{tens}(\P, \text{Lie}H)$, $\omega':=\omega+\alpha \in \C$. Or, as is clear from the above defining properties, for $\omega, \omega' \in \C$, $\omega+\omega' \notin \C$  and $\omega'-\omega \in \Omega^1_\text{tens}(\P, \text{Lie}H)$.

The curvature of a connection is given by $\Omega=d\omega+ \tfrac{1}{2}[\omega, \omega]$. One shows that it is a tensorial form, $\Omega \in \Omega^2_\text{tens}(\P, \text{Lie}H)$. The covariant derivative thus acts on it trivially, which gives the Bianchi identity: $D^\omega \Omega=0$. It is also easily proved that $D^\omega \circ D^\omega = \rho_*(\Omega)$. 
\medskip

The natural maximal group of transformation of $\P$ is its group of automorphisms $\Aut(\P):=\big\{ \Psi \in \Diff(\P)\ |\ \Psi(ph)=\Psi(p)h \big\}$, i.e the subgroup of $\Diff(\P)$ that respects the fibration structure by sending fibers to fibers, and thus projects on $\Diff(M)$. The subgroup of vertical automorphisms is $\Aut_v(\P):=\big\{ \Psi \in \Aut(\P)\ |\ \pi \circ \Psi=\pi\big\}$, i.e. it is those automorphisms that induce the identity transformation on $\M$. The latter group is isomorphic to the gauge group of $\P$, $\H:= \big\{ \gamma:\P \rarrow H\ |\ R^*_h \gamma =h\- \gamma h \big\}$ (itself isomorphic to the space of sections of the bundle $\P\times_\text{Conj} H$), the isomorphism being $\Psi(p)=p\gamma(p)$. 

The gauge transformation of a differential form $\alpha$ on $\P$ is   defined by $\alpha^\gamma := \Psi^* \alpha$, for $\gamma \in\H$ corresponding to $\Psi \in \Aut_v(\P)$.  
Notably, the gauge transformations of tensorial forms are entirely controlled by their equivariance: for $\alpha \in\Omega^\bullet_{\text{tens}}(\P, V_i)$, $\alpha^\gamma = \rho_i(\gamma\-) \alpha$ (hence the name for such forms). 
The $\H$-transformation of a connection also assumes a simple form: for $\omega \in \C$, 
$\omega^\gamma = \gamma\- \omega \gamma + \gamma\- d\gamma$. 
Transformations induced by the action of $\H \simeq \Aut_v(\P)$ are called \emph{active gauge transformations}, as they are analogous to the action of $\Diff(\M)$ in General Relativity (GR).

Notice that as a special case of its action on tensorial forms,   the gauge group  acts on itself as: $\eta^\gamma =\gamma\- \eta \gamma$ for $\eta, \gamma \in \H$.  This ensures that the action of $\H$ on $\Omega^\bullet_{\text{tens}}(\P, V_i)$ and $\C$ is a right action. 
\medskip

A bundle $\P$ is always locally trivial,  meaning that for $U\subset \M$ we have $\P_{|U}\simeq U \times H$. A trivialising section is a map $\s : U \rarrow \P_{|U}$, $x \mapsto \s(x)$. By its means, one can pull-back objects of $\P$ down to $\M$. In particular the local representative of $\omega \in \C$ is $A:=\s^*\omega \in \Omega^1(U, \text{Lie}H)$, which is a gauge (Yang-Mills) potential. The field strength of $A$ is the local representative of the curvature $F:=\s^*\Omega \in \Omega^2(U, \text{Lie}H)$. Then the local representatives $\phi:=\s^*\vphi \in  \Omega^0(U, V_i)$ are various kinds of matter fields, and $D^A\phi:=\s^*(D^\omega \vphi)  \in  \Omega^1(U, V_i)$ their minimal coupling to the gauge field. The last three are special cases of local representatives of tensorial forms: $a:=\s^* \alpha \in \Omega^\bullet(U, V_i)$.

Considering $U$ and $U'$ s.t. $U \cap U' \neq \emptyset$ and  local sections $\s : U \rarrow \P_{|U}$ and $\s' : U' \rarrow \P_{|U'}$ related on the overlap via $\s'=\s g$ where $g : U \cap U'  \rarrow H$, $x \mapsto g(x)$, is a (well named) transition function of $\P$. 
The local representatives on $U$ and $U'$ obtained via $\s$ and $\s'$ satisfy gluing properties on $U \cap U'$ involving $g$'s. For  local representatives  of a connection and of  tensorial forms we have, 
\begin{align}
\label{PassiveGT}
A' = g\- A g+ g\-dg, \qquad  a'=\rho_i(g\-) a.
\end{align}
As special case of the second equation, we have the gluings of the field strength and matter fields: $F'=g\-Fg$ and $\phi'=\rho_i(g\-)\phi$. 
Equations \eqref{PassiveGT} are called \emph{passive gauge transformations}, as they are entirely analogous to coordinate changes, or passive diffeomorphisms, in GR. 

The latter are formally indistinguishable, yet conceptually different, from \emph{ local active gauge transformations} i.e. the local representatives of the global $\H$-transformations seen above, which on $U$ would read,
\begin{align}
\label{local-activeGT}
A^\upgamma = \upgamma\- A \upgamma+ \upgamma\-d\upgamma, \qquad  a^\upgamma=\rho_i(\upgamma\-) a, \qquad \text{for }\  \upgamma \in \H_\text{loc}
\end{align}
and with the local gauge group over $U$ defined as $\H_\text{loc}:=\big\{  \upgamma=\s^*\gamma, \gamma \in \H \ |\ \upeta^\upgamma=\upgamma\- \upeta \upgamma \big\}$. In the subsequent presentation below we will focus on local active gauge transformations. 
\medskip

With this ends our summary of the geometry underlying the kinematics of a gauge theory. Let us note $\A$ the space of gauge potentials (local connections) and, with slight abuse, $\Gamma(E_i)$ the spaces of matter fields. 
A gauge theory is specified by a Lagrangian functional $L: \A \times \Gamma(E_i) \rarrow \Omega^n(U, \RR)$, $(A, \phi) \mapsto L(A, \phi)$, with $n=$ dim$\M$. 
Requiring the passive gauge invariance of $L$, $L(A', \phi')=L(A, \phi)$, amounts to requiring that it has trivial gluings and is well defined across $\M$: $L(A, \phi)  \in \Omega^n(\M, \RR)$ . But it formally is indistinguishable from requiring its local active gauge transformation, $L(A^\upgamma, \phi^\upgamma)=L(A, \phi)$, which implies that it comes from a $\H$-invariant, i.e. basic, form on $\P$: $\b L(\omega, \vphi) = \pi^* L(A, \phi) \in \Omega^n_\text{basic}(\P, \RR)$.

\subsection{Reduction of Gauge Symmetries via Dressing}
\label{Reduction of gauge symmetries via dressing}

We begin by defining the central object of the DFM. We consider   a $\H$-gauge theory based on a bundle $\P(\M, H)$. 
 
\begin{definition}
\label{Def1}
Suppose $\exists$  subgroups $K \subseteq H$ of the structure group, to which corresponds a subgroup $\K \subset \H$ of the gauge group, and $G$  s.t.  $K\subseteq G \subseteq H$. A $K$-\emph{dressing field}  is a map $u:\P \rarrow G$ defined by its $K$-equivariance~$R^*_k u=k\-u$. 
Denote the space of $G$-valued $K$-dressing fields on $\P$ by $\D r[G, K]$. 
 It  follows immediately that the $\K$-gauge transformation of a dressing field is $u^\gamma=\gamma\- u$, for $\gamma \in\K$.
\end{definition} 


Given the existence of a $K$-dressing field, we have the following,
\begin{prop} \normalfont 
\label{Prop1}
From $\omega \in \C$ and  $\alpha \in \Omega^\bullet_\text{tens}(\P, V_i)$, one defines the  \emph{dressed fields}
\begin{align}
\label{dressed-fields}
\omega^u\defeq  u\- \omega u +u\-du, \quad \text{ and } \quad \alpha^u \defeq \rho_i(u)\-\alpha,  
\end{align}
which have trivial $K$-equivariance and are $K$-horizontal, thus are $K$-basic on $\P$. It~follows that they are $\K$-invariant: $(\omega^u)^\gamma=\omega^u$ and $(\alpha^u)^\gamma=\alpha^u$, for $\gamma \in \K$, as is easily checked. 
The dressed curvature is $\Omega^u =d\omega^u+\sfrac{1}{2}\left[\omega^u, \omega^u\right]= u\-\Omega u$, and appears when squaring the dressed covariant derivative defined as $ D^{\omega^u}\defeq d\, + \rho_*(\omega^u)$. 
It satisfies the Bianchi identity $D^{\omega^u}\Omega^u = 0$. 
\end{prop}

In case the equivariance group of $u$ is $K=H$,  $\alpha^u \in \Omega^\bullet_\text{basic}(\P, V_i)$ and  $\omega^u\in \Omega^1_\text{basic}(\P, \text{Lie}H)$ are $\H$-invariant, thus project as  forms on $\M$. 
The above results for $\alpha^u$  make sense for $G \supset H$ if we assume that representations $(V_i, \rho_i)$ of $H$ extend to representations of $G$.

Let us emphasize an important fact: It should be clear from its  definition that $u \notin \K$, so that \eqref{dressed-fields} are \emph{not} gauge transformations, despite the formal resemblance. This means, in particular, that the dressed connection is no more a $H$-connection, $\omega^u \notin \C$, and a fortiori is not a point in the gauge $\K$-orbit $\O_\K[\omega]$ of $\omega$, so that $\omega^u$ must not be confused with a gauge-fixing of $\omega$. 
\medskip

On $U\subset \M$,  a local  $\K_\text{loc}$-dressing field $\u=\s^*u: U \rarrow G \in \D r[G, K]_\text{loc}$ will be defined (or recognised) by its defining gauge transformation property $\u^\upgamma =\upgamma\-\u$ for $\upgamma \in \K_\text{loc} \subseteq \H_\text{loc}$.  
The local version of proposition \ref{Prop1} above is,
\begin{prop} \normalfont 
\label{Prop2}
Given  $A \in \A$ and  $a=\s^*\alpha$ for $\alpha \in \Omega^\bullet_\text{tens}(\P, V_i)$, the following \emph{dressed fields}
\begin{align}
\label{local-dressed-fields}
A^\u\defeq  \u\- A \u + \u\-d\u, \quad \text{ and } \quad a^\u \defeq \rho_i(\u)\-a,  
\end{align}
are $\K_\text{loc}$-invariant (as is easily checked). In particular, so is the dressed curvature $F^\u=dA^\u+\sfrac{1}{2}[A^\u, A^\u]=\u\- F\u$, which satisfies the Bianchi identity $D^{A^\u}F^\u=0$, where the dressed covariant derivative is $D^{A^\u}=d\, +\rho_*(A^\u)$. 
Of course, in case $\u$ is a $\H_\text{loc}$-dressing field, the dressed fields \eqref{local-dressed-fields} are strictly $\H_\text{loc}$-invariant. 
\end{prop}


\subsubsection{Residual Gauge Transformations}
\label{Residual gauge transformations}

Let us indulge in a brief digression that is also a transition. 
In the BRST framework,  infinitesimal gauge transformations are encoded as $\boldsymbol{s} A=-D^A \boldsymbol v$ and $\boldsymbol{s} a = -\rho_*(\boldsymbol{v})a$,  where  $\boldsymbol s$ is the nilpotent BRST operator and $\boldsymbol v$ the ghost field. The latter has values in Lie$\H$ and satisfies $\boldsymbol{sv}+\sfrac{1}{2}[\boldsymbol{v}, \boldsymbol{v}]=0$. 
For~this reason, $\boldsymbol s$ is best interpreted geometrically as the de Rham derivative on $\H$ and $\boldsymbol v$ as its Maurer-Cartan form [\cite{Bonora-Cotta-Ramusino}]. 

One shows that, at a purely formal level, the dressed variables satisfy a modified BRST$^\u$ algebra:  $\boldsymbol{s} A^\u=-D^{A^\u} \boldsymbol{v}^\u$ and $\boldsymbol{s} a^\u = -\rho_*(\boldsymbol{v}^\u)a^u$, where one defines the \emph{dressed ghost} $\boldsymbol{v}^\u\defeq \u\- \boldsymbol{v} \u + \u\- \boldsymbol{s} \u$. 
In the special where case $\u$ is a $\H_\text{loc}$-dressing, its defining gauge transformation translates as $\bs{s}\u=-\bs{v} \u$. Then the dressed ghost is $\bs{v}^\u=0$ and  BRST$^\u$ is trivial, $\bs{s}A^\u=0$ and $\bs{s}a^\u=0$.
In the more general case of a $\K_\text{loc}$-dressing $\u$ achieving only partial gauge reduction, BRST$^\u$ only makes sense if it encodes  residual gauge transformations of the dressed fields~\eqref{local-dressed-fields}.  
\medskip

To  speak meaningfully about these, we must assume that $K$ is a \emph{normal} subgroup, $K \triangleleft H$, so that the $J\defeq H/K$ is indeed a group, to which corresponds the (residual) gauge subgroup $\J \subset \K$. 
Now, the action of $\J$ on the initial variables $A$ and $\alpha$ is known. Therefore what will determine the $\J$-residual gauge transformations of the dressed fields is the action of $\J$ on the dressing field. And this in turn is determined by its $J$-equivariance. In that regard, consider the following propositions,
\begin{prop} \normalfont 
\label{Residual1}
Suppose the dressing field $u$ has $J$-equivariance given by $R^*_j u =j\-uj$. Then 
 the dressing field has $\J$-gauge transformation $u^\eta=\eta\-u\,\eta$ for $\eta \in \J$, and
the residual gauge transformations of the dressed fields are: $(\omega^u)^\eta=\eta\- \omega^u \eta + \eta\-d\eta$ and $(\alpha^u)^\eta=\rho(\eta)\-\alpha^u$. 
So in particular $(\Omega^u)^\eta=\eta\- \Omega^u \eta$. 
\end{prop}

\noindent The local version is, 

\begin{prop} \normalfont 
\label{Residual2}
If the $\K_\text{loc}$-dressing field $\u$ has $\J_\text{loc}$-transformation given by $\u^\upeta=\upeta\- \u \upeta$ for $\upeta \in \J_\text{loc}$,  then 
the residual $\J_\text{loc}$-gauge transformations of the dressed fields are: $(A^\u)^\upeta=\upeta\- A^\u \upeta + \upeta\-d\upeta$ and $(a^\u)^\upeta=\rho(\upeta)\-a^\u$, so in particular $(F^\u)^\upeta=\upeta\- F^\u \upeta$. 
That is, in this case the dressed fields are genuine $\J_\text{loc}$-gauge fields.
\end{prop}

In the BRST language, the normality of $K$ in $H$ implies $\bs{v}=\bs{v}_K+\bs{v}_J$, where $\bs{v}_K$ and $\bs{v}_J$ are respectively Lie$\K$- and Lie$\J$-valued, and in accordance $\bs{s}=\bs{s}_K+\bs{s}_J$. The defining $\K_\text{loc}$-transformation of the dressing field translates as $\bs{s}_K \u=-\bs{v}_K \u$, while its $\J_\text{loc}$-transformation assumed in Proposition \ref{Residual2} is encoded as $\bs{s}_J \u=[\u, \bs{v}_J]$. The dressed ghost field is thus 
$\bs{v}^\u=\u\-(\bs{v}_K+\bs{v}_J) \u + \u\- (\bs{s}_K + \bs{s}_J)\u = \u\-(\bs{v}_K+\bs{v}_J) \u + u\- (-\bs{v}_K \u + [\u, \bs{v}_J]) = \bs{v}_J$. Therefore, the modified (actually \emph{reduced}) BRST$^\u$ algebra is: $\bs{s}_J A^\u = -D^{A^\u} \bs{v}_J$ and $\bs{s}_J a^\u = -\rho_*(\bs{v}_J)a^\u$. As expected, it encodes the residual gauge transformations of the dressed fields.  
\smallskip

Consider the Lagrangian $L(A, \phi)$ of our initial $\H_\text{loc}$-gauge theory, and suppose a $\K_\text{loc}$-dressing field satisfying the above propositions is available. Due to the  $\H_\text{loc}$-invariance of the Lagrangian, which holds as a formal property of $L$ as a functional, and due to the formal similarity between a gauge transformation \eqref{PassiveGT}/\eqref{local-activeGT} and a dressing operation \eqref{local-dressed-fields}, we have that $L(A, \phi)=L(A^\u, \phi^\u)$. 
That is, the $\H_\text{loc}$-gauge theory can be rewritten in terms of $\K_\text{loc}$-invariant variables, which means that it becomes a $\J_\text{loc}$-gauge theory: the $\K_\text{loc}$-gauge symmetry sector has been neutralised. 
 In case a $\H_\text{loc}$-dressing field is available, the gauge symmetry of the theory $L$ is fully reduced.

Remark again that as $\u\notin \K_\text{loc}\subset \H_\text{loc}$, the dressed fields $\upchi^\u=(A^\u, \phi^\u)$ are not points in the gauge orbits $\O[\upchi]$ of the gauge variables $\upchi=(A, \phi)$. So, the dressed Lagrangian $L(A^\u, \phi^\u)$ is not a gauge-fixed version of $L(A, \phi)$. 


\subsubsection{Ambiguity in Choosing a Dressing Field}
\label{Ambiguity in choosing a dressing field}

The dressed fields may  exhibit residual transformations of another kind resulting from a potential ambiguity in choosing the dressing field.
A priori two dressings $u, u' \in \D r[G, K]$ may be related by $u'=u\xi$, where $\xi:\P \rarrow G$.
Since by definition $R^*_k u =k\-u$ and $R^*_k u' =k\-u'$, one has $R^*_k \xi = \xi$. Let us denote the group of such basic maps $\G \defeq \left\{\,   \xi:\P \rarrow G\, |\,  R^*_k \xi = \xi  \, \right\}$, and denote its action on a dressing field as $u^\xi=u\xi$. 
By definition $\G$ has no action on the space of connections $\C$ or on $\Omega^\bullet_\text{tens}(\P, V_i)$: note this $\omega^\xi=\omega$ and $\alpha^\xi=\alpha$. On the other hand, it is clear how $\G$ acts on  dressed  fields: 
\begin{align}
\label{ResidualGT-2Kind}
(\omega^u)^\xi &\defeq (\omega^\xi)^{u^\xi}=\omega^{u\xi}=\xi\- \omega^u \xi + \xi\-d\xi, \quad \text{and} \notag\\
 \quad (\alpha^u)^\xi &\defeq(\alpha^\xi)^{u^\xi}=\alpha^{u\xi}=\rho(\xi\-) \alpha^u.
\end{align}
In particular, $(\Omega^u)^\xi=\xi\- \Omega^u \xi$. 
The  new dressed field  $(\omega^u)^\xi$ and $(\alpha^u)^\xi$ are also $K$-basic, and therefore $\K$-invariant. It means that the bijective correspondance between the $K$-dressings $(\chi^u)^\xi$, for $\chi=(\omega, \alpha)$, and their gauge $\K$-orbits $\O_\K[\chi]$ holds $\forall \xi \in \G$. So there is a $1:1$ correspondance $\O_\K[\chi] \sim \O_{\G}[\chi^u]$.

The local counterpart of the above is clear: The group $\G_\text{loc}:=\big\{ \upxi=\s^*\xi : U\subset \M \rarrow G, \text{for } \xi \in \G \big\}$ parametrises the ambiguity in choosing the local dressing field, s.t. $\u'=\u\upxi$, and consequently acts on the local dressed fields as, 
\begin{align}
\label{ResidualGT-2Kind-local}
(A^\u)^\upxi =\upxi\- A^\u \upxi + \upxi\-d\upxi, \quad \text{and}
 \quad (a^\u)^\upxi =\rho(\upxi\-) a^\u.
\end{align}
In particular  $(F^\u)^\upxi=\upxi\- F^\u \upxi$ and $(D^{A^\u}\phi_i^\u)^\upxi=\rho(\upxi\-)D^{A^\u}\phi_i^\u$. 
And again, we have a correspondance between the gauge $\K_\text{loc}$-orbits of fields $\upchi=(A, a)$ and the $\G_\text{loc}$-orbits of their dressings $\upchi^\u$: $\O_{\K_\text{loc}}[\upchi] \sim \O_{\G_\text{loc}}[\upchi^\u]$.
\medskip

What it tells us is that owing to the ambiguity in the choice of dressing, the reduced gauge symmetry is replaced with a local symmetry which is (at least) as big. 
 A way in which a meaningful constraint on this arbitrariness could arise is if the dressing field is built from the  initial gauge variables,
 $\u: \A \times \Gamma(E_i) \rarrow \D r[K, G]_\text{loc}$, $(A, \phi) \mapsto \u(A, \phi)$. Then it may be that this constructive procedure is such that $\G$ (or $\G_\text{loc}$) is reduced to a "small", rigid/global, or perhaps even discrete subgroup. Even if it is not so, this $\G$-symmetry may be an interesting new symmetry. 
These situations are represented in most fruitful applications [\cite{Attard-Francois2016_I, Attard-Francois2016_II, Attard_et_al2017, Francois2018}].
Notably, 
in the context of the tetrad formulation of General Relativity (GR), $\G_\text{loc}=\GL(4, \RR)$ is the group of local coordinate changes.

\subsubsection{A Connection-Form  on $\mathcal{A}$}
Here we comment on another very useful parallel between dressings and a more standard geometric construction, on $\mathcal{A}$, that may shed some light on the arbitrariness of the dressing. 

The idea is that, just as the action of $H$ on $\mathcal{P}$ permits interesting geometrical constructions, having to do with parallel transport and horizontality, so does the action of  $\mathcal{H}$ on $\mathcal{A}$. In truth, the infinite-dimensional space $\mathcal{A}/\mathcal{H}$ is not perfectly analogous to $\mathcal{P}/H$. 
$\mathcal{A}/\mathcal{H}$ is not a bona-fide (infinite-dimensional) manifold due to the presence of stabilizers; instead it forms only a stratified manifold (cf. \cite{kondracki1983, Mitter:1979un}). 

Nonetheless, the subset of $\cal A$ without stabilizers is generic (open and dense), and there one can introduce \textit{a functional connection-form} \cite{Gomes-Riello2018, Gomes-et-al2018, GomesRiello_2020}. The main idea of the functional connection-form is to infinitesimally (i.e. perturbatively) define a right-equivariant horizontal distribution. 

Thus a functional connection-form is a one-form on the field-space $\cal A$ that is valued on $\mathcal H$, obeying (infinite-dimensional versions of the) equivariance and projection equations, as in \eqref{eq:omega_equi}-\eqref{eq:omega_proj}. 

One of the main advantages of the connection-form is that in theories that possess a kinematic term in a Lagrangian, one can whittle down the enormous space of possible connection-forms, by using this term to define a supermetric on field-space, and thereby define a connection-form by orthogonality with respect to the gauge orbits. This choice has several pragmatic advantages \cite[Sec. 2]{GomesRiello_2020}. 

Moreover, when the connection-form is integrable, i.e. when it possesses no associated curvature, it also provides a complete dressing. Namely, in the Abelian theory, if $A(s)$ is a path in $\cal A$, with $A(0)=0$ and $A(1)=A$, then we can integrate the connection-form $\varpi$ along this path to obtain the Dirac dressing as: $$u(A):=\int\varpi(\frac{d}{ds}A(s))\, ds .
$$
The integrability condition means that the resulting dressing is independent of the path chosen. 
We refer to \cite[Sec. 5]{GomesRiello_2020} and \cite{Gomes-Riello2018} for more detail. 

Lastly, one can also reproduce the BRST transformations (cf. Section  \ref{Residual gauge transformations}) geometrically in this formalism \cite{GomesRiello2016}. 

\subsubsection{On Substantial \emph{vs} Artificial Gauge Symmetries}
\label{On substantial vs artificial gauge symmetries}

As previous sections have explained, section \ref{AVS} in particular, there is a (usually) recognised trade-off between  \emph{locality}, as understood in a field-theoretic sense,\footnote{Related of course but not to be conflated with ``local" in the bundle-theoretic sense used up until now, meaning on an open subset $U\subset\M$ of the base spacetime manifold.} and 
\emph{gauge-invariance}: a gauge theory is either written in terms of local gauge-variant variables, or in terms of non-local gauge-invariant variables. 
A theory displaying such trade-off would be said to have a \emph{substantial} gauge symmetry, whereas a theory that does not and can be (re)-written in terms of local gauge-invariant variables would be said to have an \emph{artificial} gauge symmetry. 
From that viewpoint,  a true/substantial gauge symmetry  signals that physical degrees of freedom (d.o.f.) have a form of non-locality to them. ``Fake"/artificial gauge symmetries signal nothing of the sort, and can be dispensed with at no physical cost: they are genuine \emph{surplus} as defined in \ref{INT}.   The  interpretive issues surrounding gauge symmetries and their physical relevance then applies to the substantial kind only. 

Connecting to this discussion, the DFM may suggest a way to assess the nature of the gauge symmetry in a theory $L(A, \phi)$. If one is able to find, or build, a (field-theoretically) local dressing field $\u(A, \phi)$, then  the theory can be rewritten as $L(A^\u, \phi^\u)$ in terms of the variables $A^\u$ and $\phi^\u$ that are gauge-invariant and remain local, showing decisively that the gauge symmetry of the theory is artificial. In a theory with substantial gauge symmetry, any $\u(A, \phi)$ would be non-local, and so would then be the gauge-invariant variables $A^\u$ and $\phi^\u$: the rewriting $L(A, \phi)=L(A^\u, \phi^\u)$ would then be the formal expression of the trade-off alluded to above. 
Of course, the failure to find a local dressing field may be faulted on a failure of imagination on the part of the theorist, or on a less than thorough search. So the strategy is asymmetric: finding a local dressing field is conclusive, not finding is not. Yet, to all practical purpose it is rather effective. 
\smallskip

As a matter of interpretation, if $\u=\u(A)$ one could say that $\phi^\u$ represents the bare charged matter field shrouded in the gauge field it sources. Similarly, $A^\u$ would be a self-enveloping charged gauge field acting as a source for itself  (e.g. gluons in QCD). For abelian gauge fields, the latter interpretation if not  available. If on the other hand $\u=\u(\phi)$, one could see $A^\u$ as the gauge field embedded in a pervasive ``sea" generated by $\phi$,  idem for $\phi^\u$ itself (which is reminiscent of a Higgs-like interpretation, 
since, for this interpretation to hold, it must be nowhere vanishing
).

Let us consider some examples before coming to the main case of interest, the electroweak model.

\subsection{Examples}
\label{Examples}

An early example of (abelian) dressing field is the so-called \emph{Stueckelberg field}, introduced in [\cite{Stueckelberg1938-I, Stueckelberg1938-II-III}], see [\cite{Ruegg-Ruiz}] for a review. An abelian Stueckelberg-type model involves a potential  $A\in \Omega^1\left(U, \text{Lie}U(1)\right)$ and a Stueckelberg field $B \in \Omega^0(U, \RR)$, respectively transforming as $A^\upgamma=A - d\theta$ and  $B^\upgamma=B - \mu \theta$, with $\upgamma=e^{i\theta} \in \mathcal{U}(1)$ and $\mu$ some constant. 
A prototypical (minimal)  Stueckelberg $\mathcal{U}(1)$ model would be 
 \begin{align}
 \label{Stueck}
 L(A, B) = \tfrac{1}{2}F\,*\!F +   \mu^2 (A-\tfrac{1}{\mu}dB)\,*\!(A-\tfrac{1}{\mu}dB),
 \end{align}
 Here, $*:\Omega^\bullet(U) \rarrow \Omega^{n-\bullet}(U)$, with $n=$dim$\M$, is the Hodge dual operator.
As just said, $B$ is actually a \emph{local} abelian dressing field: defining $\u(B):=e^{\sfrac{i}{\mu}B}$, it is clear that $\u(B)^\upgamma\defeq \u(B^\upgamma)=e^{\sfrac{i}{\mu}(B-\mu \theta)}=\upgamma\-\u(B)$. The associated $\mathcal{U}(1)$-invariant local dressed field is then $A^\u:=A+i\u\-d\u=A-\tfrac{1}{\mu}dB$,  with  field strength  $F^\u=F$. So \eqref{Stueck} is manifestly rewritten as, 
\begin{align}
L(A, B)=L(A^\u)= \tfrac{1}{2}F^\u\,  *\!F^\u + \mu^2 A^\u\, *\!A^\u,
\end{align}
which is a Proca Lagrangian for $A^\u$ with no gauge symmetry. According to the DFM, as $\u$ and $A^\u$ are local, the original $\mathcal{U}1)$ symmetry of the model \eqref{Stueck} is artificial. 

\medskip

Theories $L(A, \phi)$ with an abelian gauge potential $A\in \Omega^1\left(U, \text{Lie}U(1)\right)$ coupled to a charged scalar field $\phi \in \Omega^0(U, \CC)$ provide another illustration. The gauge transformations of the  potential $A$ and the $\CC$-scalar field $\phi$ are $A^\gamma=A + \upgamma\-d\upgamma$ and $\phi^\upgamma=\upgamma\-\phi$, for $\upgamma\in \mathcal{U}(1)$. Now, one can extract a $\mathcal{U}(1)$-dressing field from the scalar field by the polar decomposition $\phi=\u\rho$ with $\rho=\sqrt{\phi^*\phi}$ its modulus and $\u$ its phase. Obviously $\rho$ is invariant while the phase carries the transformation $\u^\upgamma=\upgamma\-\u$. The latter is indeed a local dressing field whose associated gauge-invariant local fields are $A^\u=A+\u\-d\u$ and $\phi^\u=\u\- \phi=\rho$. Any such theory can be rewritten in terms of these variables, $L(A, \phi)=L(A^\u, \phi^\u)$, which shows  that the $\mathcal{U}(1)$-gauge symmetry is artificial. 

In particular, the Aharonov-Bohm (AB) effect - see section \ref{ABeffect}  - formulated in the framework of $\CC$-scalar EM looses  its puzzling edge, as identified by Wallace [\cite{Wallace2014}], since it can be interpreted as resulting from the local interaction of the gauge-invariant local fields $A^\u$ and $\phi^\u$ outside the cylinder.  

The abelian Higgs model also belongs to this framework. The Lagrangian of the theory would be, 
\begin{align}
\label{AbHiggs1}
L(A, \phi)&=\tfrac{1}{2}\ F \,*\!F + (D^A\phi)^* *\!D^A\phi - V(\phi) *\!\jone,\\[1mm] & \quad \text{with } \quad   V(\phi)= \mu^2 \phi^*\phi + \lambda (\phi^*\phi)^2, \notag
\end{align}
where  $*\jone=\vol_n$ is a volume $n$-form on $U$,  $\phi^*$ is the conjugate of $\phi$, and in the potential $V: \CC^2 \rarrow \RR$ we must have $\lambda >0$. 
Taking this formulation of the model at face value, one notices that for $\mu^2>0$ there is only one invariant vacuum solution $\phi_0=0$ minimizing $V$, but for $\mu^2<0$ there is a whole $\mathcal{U}(1)$-orbit of vacua with modulus $|\phi_0|=\sqrt{\frac{-\mu^2}{2\lambda}}$. 
If and when $\phi$ settles for one of these vacua - spontaneously breaking $\mathcal{U}(1)$ - and fluctuates around it, $\phi=\phi_0+H$, a mass term $m_A=g\phi_0$ for $A$ appears in the Lagrangian via the minimal coupling term $D^A\phi=d\phi+gA\phi$, with $g$ a coupling constant. A mass for the gauge potential $A$ is thus generated, it seems, via spontaneous gauge symmetry breaking (SSB). 

Yet as just seen, the model can be rewritten in a $\mathcal{U}(1)$-invariant way via dressing as, 
\begin{align}
\label{AbHiggs2}
L(A^\u, \rho)&=\tfrac{1}{2}\ F^\u \,*\!F^\u + (D^{A^\u}\rho)^* *\!D^{A^\u}\rho - V(\rho) *\!\jone,   \\[1mm] & \quad \text{with } \quad   V(\rho)= \mu^2 \rho^2 + \lambda \rho^4. \notag
\end{align}
Thus rewritten, there is no gauge symmetry to break. The potential is now $V:\RR^+ \rarrow \RR$ and has a \emph{unique} vacuum configuration for either sign of $\mu^2$, $\rho_0=0$ and $\rho_0=\sqrt{\sfrac{-\mu^2}{2\lambda}}$. Writing $\rho=\rho_0+H$, we see that the theory still has a massless ($\mu^2>0$, $m_A=0$) and a massive ($\mu^2<0$, $m_A=g\rho_0$) phase. The mass  is generated  via a vacuum phase transition, but it is not tied to a SSB. 
And indeed as said above, as $A^\u$ and $\rho$ are local gauge-invariant fields, the $\mathcal{U}(1)$-symmetry of the initial model is artificial and plays no physical role. 

In line with the remarks made earlier in the general setting, despite a formal resemblance the above should not be confused with the \emph{unitary} gauge fixing of the model: the dressed model \eqref{AbHiggs2} is not a gauge fixing of  \eqref{AbHiggs1}.\footnote{ Curiously, in the textbook [\cite{Rubakov1999}] the abelian Higgs model is given the dressing treatment, while the electroweak model is treated via the usual unitary gauge and SSB approach. }

\bigskip
For pure $\H_\text{loc}$-gauge theories $L(A)$, there are not many options to work with to build a dressing field $\u(A)$. 
One attempt that has been explored in relation with the 
 ``proton spin decomposition controversy'' [\cite{LorceGeomApproach, Leader-Lorce}] is to split the potential as $A=A_\text{phys}+ A_\text{pure}$. 
 By assumption,   only $A_\text{phys}$ contributes to the field strength $F=F_\text{phy}$ and it transforms as $A_\text{phys}^\upgamma=\upgamma\- A_\text{phys}\upgamma$ for $\upgamma \in \H_\text{loc}$. So, $A_\text{pure}$ is  pure gauge $F_\text{pure}=0$, which means that it  can be written as $A_\text{pure}=\u d\u\-$ for some $H$-valued function $\u:U \rarrow H$. Since it must also transforms as a connection, $A_\text{pure}^\upgamma=\upgamma\- A_\text{pure}\upgamma + \upgamma\- d\upgamma$, it means that $\u^\upgamma =\upgamma\- \u$. In other words $\u=\u(A)$ is a local $H$-dressing field. The dressed fields are then $A^\u=\u\- A_\text{phy} \u=:A_\text{phys}^\u$ and $F^\u=F_\text{phys}^\u=dA_\text{phy}^\u +\sfrac{1}{2}[A_\text{phy}^\u, A_\text{phy}^\u]$. So that the theory is rewritten $L(A)=L(A_\text{phy}^\u)$. The same can be done for a theory including spinors, i.e. fermions, $L(A, \psi)=L(A_\text{phy}^\u, \psi^u)$. 
 
 This however is unsatisfactory. Indeed  the ansatz decomposition $A=A_\text{phys}+ A_\text{pure}$ reflects the affine nature of the connection space $\C$, so  that $A_\text{phys}$ is the local representative of a tensorial form, and $A_\text{pure}$ that of a flat connection. But then it can be shown that the existence of a global flat connection means the underlying bundle $\P$ is trivial, which further imply that the ambiguity group in choosing $\u$ is isomorphic to the initial gauge group, $\G_\text{loc}\simeq\H_\text{loc}$. In view of \eqref{ResidualGT-2Kind-local}, $\G_\text{loc}$ is a (gauge) symmetry of $L(A_\text{phy}^\u, \psi^u)$. So nothing has been really achieved by the ansatz as the bare and dressed theories are entirely isomorphic. See [\cite{FLM2015_I}] for details. 
\medskip

As far as is known, in pure $\H_\text{loc}$-gauge theories $L(A)$ any  dressing field $\u(A)$ is non-local, so that according to the DFM the initial $\H_\text{loc}$ symmetry is substantial. The same seems likely for non-abelian gauge theories including spinor fields, $L(A, \psi)$, as there is no polar decomposition of a spinor $\psi$ from which one could extract a local dressing field $\u(\psi)$. 
%
Applications of the DFM in the context of such theories,  building non-local $\u(A)$'s, provide in particular a geometric basis for 
Dirac's gauge-invariant formulation of QED [\cite{Dirac55}] and [\cite{Dirac58}] (section 80) - see [\cite{Francois2018}] for a discussion - 
as well as for the construction of quark ($\psi^\u$) and gluons ($A^\u$) states in QCD such as [\cite{McMullan-Lavelle97, Lavelle-McMullan-Bagan2000}]. 

We notice that it is when formulated in the context of spinorial EM that the AB effect retains is physical significance by displaying how EM properties are encoded non-locally. Indeed, the effect cannot be explained by the local interaction of gauge invariant fields outside the cylinder (or so it seems), as the gauge-invariant fields $A^\u$ and $\psi^\u$ are non-local. 
\medskip
 
 Before turning to the main example, let us remark that the latest instance to date of application of the DFM concerns the 
 so-called ``edge modes''  invoked as a way to deal with the problem of boundaries in the study of the symplectic structure of gauge theories. About edge modes see [\cite{DonnellyFreidel2016, Geiller2017, Speranza2018, Teh-et-al2020, Teh2020}], about the problem of boundary and an alternative proposition to edge modes see [\cite{Gomes-et-al2018, Gomes-Riello2018, Gomes2019}].

 \subsection{Invariant Formulation of the Electroweak Model}
 \label{Invariant formulation of the electroweak model}
 
 The basic idea behind the DFM featured repeatedly in reformulations of  theories undergoing SSB. It is seen in the pioneering work of Higgs on abelian models [\cite{Higgs66}] and of Kibble on non-abelian models [\cite{ Kibble67}].  It resurfaced in the work of Banks \& Rabinovici on the abelian Higgs model 
 [\cite{Banks-Rabinovici1979}], and shortly after in the work of Frohlich, Morchio \&Strocchi [\cite{Frohlich:1980gj, Frohlich:1981yi}] on the invariant formulation of  the electroweak model, which is still today a point of reference (known as the FMS approach, see section \ref{Sec:FMS approach} next). Since then, the idea is found again in several works also concerned with invariant formulations of (aspects of) the electroweak model [\cite{GrosseKnetter-Kogerler1993, Buchmuller-1994, McMullan-Lavelle95, Chernodub2008, Faddeev2009, Ilderton-Lavelle-McMullan2010, Masson-Wallet, Kondo:2018qus,Rosten:2010vm,Morris:1999px,Morris:2000fs}]. The recent review on Higgs physics [\cite{Maas:2017wzi}] emphasizes  the importance of gauge-invariant formulations, and flavors of the DFM can be recognised there. 
 
In the past fifteen years, the fact that such reformulations may cast a new light on the electroweak physics, and gauge physics more generally,  has been appreciated by  philosophers of physics such as 
[\cite{Smeenk2006}], 
[\cite{Lyre2008}], 
[\cite{Struyve2011}], and 
[\cite{Friederich2013, Friederich2014}]. 

In the following we propose the most natural reformulation via DFM of a simplified electroweak model (considering only leptons and massless neutrinos). It may be compared to the  FMS approach for this case.

The space of field of the model is $\upchi=\{a,b, \vphi, \psi_L, \psi_R \}$. The gauge potentials are  $a \in \Omega^1(U, \text{Lie}(U(1)))$ and $b \in \Omega^1(U, \text{Lie}(SU(2)))$, with field strength $F$ and $G$. We have a scalar field in the fundamental representation of $SU(2)$, $\vphi=(\vphi_1, \vphi_2)^T \in \Omega^0(U, \CC^2)$, as well as a left-handed (Weyl) fermion doublet (leptons say) $\psi_L= (\nu_L , \ell_L )^T$, and a right-handed  fermion singlet $\psi_R=\ell_R$. 
The  the scalars and fermions couples minimally with the gauge potentials via the covariant derivatives
\begin{align*}
 D\vphi &= d\vphi +(gb+ g'a)\vphi, \\
 D\psi_L &= d\psi_L + (gb - g'a) \psi_L, \\
  D\psi_R &= d\psi_R  - g'2\,a \psi_R,
\end{align*}
 with $g$ and $g'$  coupling constants.
The gauge group  $\H_\text{loc}=\mathcal{U}(1)\times \SU(2)$ acts, for  $\alpha \in \mathcal{U}(1)$ and $\beta\in\SU(2)$,  as:
\begin{equation}
\begin{aligned}
\label{FieldGT-EW}
a^{\,\alpha}&=a +\tfrac{1}{g'}\alpha^{-1}d\alpha, \quad 
b^{\,\alpha}=b, \quad
\vphi^{\,\alpha} =\alpha^{-1}\vphi,\quad 
 \text{and} \quad  \psi_{L/R}^{\,\alpha} = \left\{  \begin{array}{c}  \alpha\, \psi_{L} \\ \alpha^2\, \psi_R  \end{array} \right. \\
a^{\,\beta}&=a, \quad
b^{\,\beta}=\beta^{-1}b\beta + \tfrac{1}{g}\,\beta^{-1}d\beta, \quad   
\vphi^{\,\beta}=\beta^{-1}\vphi, \quad   
 \text{and}\quad  \psi_{L/R}^{\,\beta} = \left\{  \begin{array}{c} \beta^{-1} \psi_{L} \\ \psi_R \end{array} \right.
\end{aligned}
\end{equation}
The $\H_\text{loc}$-invariant Lagrangian form of the theory is, 
\begin{align}
\label{EW-Lagrangian}
L(a, b, \vphi, \psi_{L/R})&=\tfrac{1}{2}\Tr(F\wedge *F) +\tfrac{1}{2}\Tr(G\wedge *G) + \langle D\vphi, \ *D\vphi\rangle - V(\vphi)  \notag\\
					&  + \langle \psi_{L/R}, \slashed D \psi_{L/R }\rangle + f_\ell\,\langle \psi_L, \vphi \rangle\, \psi_R + f_\ell\, \b\psi_R \langle  \vphi, \psi_L\rangle. 
\end{align}
The potential term is $ V(\vphi) = \mu^2 \langle\vphi, *\vphi\rangle +\lambda \langle\vphi, *\vphi\rangle^2$ with $\mu^2 \in \RR$, $\lambda > 0$  and $\langle\  , \,  \rangle$ is a Hermitian form on $\CC^2$. The Dirac operator is $\slashed D = \gamma \w *D$, with $\gamma=\gamma_\mu\, dx^{\,\mu}$ the Dirac matrices-valued 1-form. The constants $f_\ell \in \RR$ are Yukawa couplings specific of each type of leptons ($\ell= e, \mu, \tau$). 

As the usual narrative goes (see e.g [\cite{Becchi-Ridolfi}]) if $\mu^2<0$, the electroweak vacuum given by $V(\vphi)=0$ seems degenerate as it appears to be an $\SU(2)$-orbit of non-vanishing vacuum expectation values for $\vphi$. When the latter settles randomly, spontaneously, on one of them, this breaks $\SU(2)$ and generates mass terms for the  fields with which it couples (minimally of via Yukawa terms). 
\smallskip

The dressing field method suggests an alternative proposition.  
Indeed it is not hard to find a dressing field in the electroweak model.
 Considering the polar decomposition in $\CC^2$ of the scalar field $\vphi=\rho \u $ with
\begin{align}
\label{decomp-phi}
    \u(\vphi)&= \tfrac{1}{\rho} \begin{pmatrix} \vphi_2^* & \vphi_1 \\ -\vphi_1^*  & \vphi_2  \end{pmatrix} \in SU(2) \quad \text{ and }\quad \rho :=  \vect{0 \\ ||\vphi||} \in \RR^+ \subset \CC^2, \notag\\[1mm] &\text{one has} \quad \vphi^\beta  \quad    \Rightarrow  \quad   \u^\beta=\beta\- \u. 
\end{align}
Thus,   $\u$ is a  $\SU(2)$-dressing field that can be used to construct the $\SU(2)$-invariant composite fields:
\begin{align}
 b^\u&=\u\-b\u+\tfrac{1}{g}\u\-d\u\rdefeq B, \quad \text{and} \quad  G^\u = \u\-G\u = dB+\tfrac{g}{2}[B, B]  
 \label{eq:CompositeW}  \\
 \vphi^\u&=\u\-\vphi=\rho, \quad \text{and} \quad  (D\vphi)^\u=D^\u\rho=d\rho + (g B + g'a )\rho, 
 \label{eq:Radial} \\
 \psi_L^\u&=\u\- \psi_L\rdefeq \vect{ \nu_L^\u \\[1mm] \ell_L^\u } \quad \text{and} \quad (D\psi_L)^\u=D^\u\psi_L^\u=d\psi_L^\u + (g B + g'a )\psi_L^\u.
 \label{eq:CompositeLepton}
\end{align}
Since $\u$ is \emph{local}, so are the above composite fields. Therefore, we might suggest that the $\SU(2)$-gauge symmetry of the model is \emph{artificial}, so that the theory defined by the electroweak Lagrangian \eqref{EW-Lagrangian} is rewritten as the $\mathcal{U}(1)$-gauge theory\footnote{Note that the quantum version of the Lagrangian receives a further term ($3\log\rho$) from a Jacobian.},
\begin{align}
\label{EW-Lagrangian2}
L(a, B, \rho, \psi_L^\u, \psi_R)=&\ \tfrac{1}{2}\Tr(F\wedge *F) +\tfrac{1}{2}\Tr(G^\u \wedge *G^\u)   \notag\\[.5mm]	
& + \langle D^\u \rho, \ *D^\u \rho \rangle - V(\rho)    \notag \\
+ \langle\psi_L^\u, \slashed D^\u \psi_L^\u \rangle  & +   \langle \psi_R, \slashed D \psi_R \rangle  + f_\ell\,\langle \psi_L^\u, \rho \rangle\, \psi_R + f_\ell\, \b\psi_R \langle  \rho, \psi_L^\u \rangle. 
\end{align}
The interpretation of the model in terms of SSB is here superfluous, and indeed impossible when expressed in the form \eqref{EW-Lagrangian2}. 
Analyzing the residual substantial $\mathcal{U}(1)$-gauge symmetry of the model allows us to go a step further in exhibiting the physical d.o.f.
\smallskip

Let us remark that the dressed fields above essentially reproduces the invariant variables used in  [\cite{Banks-Rabinovici1979}] and the seminal FMS approach [\cite{Frohlich:1980gj, Frohlich:1981yi}]. In particular it is easy to compare $\rho \sim ( \vphi_1^* \vphi_1 + \vphi_2^*\vphi_2)^{\sfrac{1}{2}}$ and $ \vect{ \nu_L^\u \\[1mm] \ell_L^\u } =  \tfrac{1}{\rho} \vect{  \vphi_2 \nu_L - \vphi_1 \ell_L  \\[1mm]  \vphi_1^* \nu_L + \vphi_2^* \ell_L } $ to e.g. eq.(6.1) of [\cite{Frohlich:1981yi}]. 



 \subsubsection{Residual $\mathcal{U}(1)$ Symmetry}
  \label{Residual U1 symmetry}
 
By its very definition $\rho^{\,\beta}=\rho^\alpha=\rho$, so it is already a fully $\H_\text{loc}$-gauge invariant scalar field which  then qualifies as a potential  observable. 
As explained in section \ref{Reduction of gauge symmetries via dressing},  the  $\mathcal{U}(1)$-residual gauge transformations of the  $\SU(2)$-invariant composite fields depend on the $\mathcal{U}(1)$-gauge transformation of the dressing field $\u$:  $(\upchi^\u)^\alpha=(\upchi^\alpha)^{\u^\alpha}$. One finds that,
\begin{align*}
\u(\vphi)^\alpha \defeq \u(\vphi^\alpha)= \u(\vphi) \t\alpha, \qquad \text{where} \qquad  \t\alpha= \begin{pmatrix} \alpha & 0  \\ 0 & \alpha\- \end{pmatrix}.
\end{align*}
This is not the kind of residual transformation shown in Proposition \ref{Residual2}, yet the general logic applies and we get  $(\upchi^\u)^\alpha=(\upchi^\alpha)^{\u\t\alpha}$. So, using \eqref{FieldGT-EW} we easily find: 
\begin{align}
\label{U(1)-GT}
\begin{aligned}
B^\alpha &= \t\alpha\- B \t\alpha + \tfrac{1}{g}\t\alpha\-d\t\alpha, & \qquad (G^\u)^\alpha&=\t\alpha \- G^\u \t\alpha, \\
(\psi^\u_L)^\alpha &= \t\alpha\- \alpha\, \psi_L^\u,     &   \qquad (D^\u\psi^\u_L)^\alpha &= \t\alpha\- \alpha\, D^\u\psi_L^\u,  \\
\rho^\alpha &= (\alpha\t\alpha)\- \rho=\rho,            &        \qquad  (D^\u\rho)^\alpha &= (\alpha\t\alpha)\- D^\u\rho.  
\end{aligned}
\end{align}
By a simple inspection of the matrices $\alpha \t\alpha = \begin{pmatrix} \alpha^2 & 0 \\ 0 & 1 \end{pmatrix}$ and $\t\alpha\- \alpha = \begin{pmatrix} 1 & 0 \\ 0 & \alpha^2 \end{pmatrix}$, it is clear on the one hand that the top component $\nu_L^\u$ of $\psi_L^\u$ is $\U(1)$-invariant,\footnote{While $(\ell^\u_L)^\alpha=\alpha^2 \ell^\u_L$, is now  a match for the singlet $\psi^\alpha_R=\ell^\alpha_R=\alpha^2 \ell_R$, which is relevant to the final form of the Yukawa couplings (eq.\eqref{EW-Lagrangian3}).} and on the other hand that $\mathcal{U}(1)$-invariant combinations of $a$ and (components of) $B$ are to be found in the covariant derivatives. 
And indeed, given the decomposition $B=B_a \sigma^a$, where $\sigma^a$ are the hermitian Pauli matrices and $B_a \in i\RR$, we have explicitly:
\begin{align}
\label{B-field}
&B=\begin{pmatrix} B_3 & B_1-iB_2 \\ B_1+iB_2 & -B_3  \end{pmatrix}=:\begin{pmatrix} B_3 & W^- \\ W^+ & -B_3  \end{pmatrix}, \notag \\[1.5mm] 
\textrm{so that}  \qquad &B^\alpha=\begin{pmatrix}  B_3 +\frac{1}{g}\alpha\-d\alpha & \alpha^{-2}W^- \\[5pt] \alpha^2W^+ & -B_3 -\frac{1}{g}\alpha\-d\alpha \end{pmatrix}. 
\end{align}
 The linear combination $gB_3- g' a =: (g^2+{g'}^2)^{\sfrac{1}{2}}\,Z^0$, obviously $\mathcal{U}(1)$-invariant, appears  in both $D^\u\rho$ and $D^\u\psi^\u_L$. 
 One may observe that the combination $A := (g^2+{g'}^2)^{-\sfrac{1}{2}}(g'B_3 +g a)$, $\mathcal{U}(1)$-transforms as $A^\alpha=A+\tfrac{1}{e} \alpha\-d\alpha$ with $e=\sfrac{gg'}{\sqrt{g^2+g'^2}}$. It would be natural to expect it appearing, together with $Z^0$, in  the bottom component of $(D^\u\psi^\u_L)$. Explicitly
\begin{align}
D^u\rho&= d\rho+(g'a + gB) \rho =\vect{gW^-\rho \\ d\rho +(g'a- gB_3)\rho} \notag \\[1mm]
&=\vect{g W^-\rho \\ d\rho - (g^2+{g'}^2)^{\sfrac{1}{2}}\,Z^0 \rho} =\vect{g W^-\rho \\ d\rho - \tfrac{e}{\cos\theta_W \sin\theta_W}\,Z^0 \rho},  \label{coupling-Higgs-WeakFields}
\end{align}
so,
\begin{align*}
 (D^u\rho)^\alpha =\begin{pmatrix} \alpha^{-2} & 0 \\ 0 & 1 \end{pmatrix} \vect{gW^-\rho \\ d\rho - \tfrac{e}{\cos\theta_W \sin\theta_W}\,Z^0 \rho}, 
 \end{align*}
 by \eqref{U(1)-GT} or \eqref{B-field}. And, 
 \begin{align}
 \label{coupling-leptons-ElectroweakFields}
 D^\u\psi^\u_L&=  d\psi^\u_L+(gB-g'a) \psi^\u_L 
 			=\vect{ 	d\nu^\u_L +(gB_3-g' a) \nu^\u_L + gW^- \ell^\u_L    \\[1mm] d\ell^\u_L -(gB_3 + g' a) \ell^\u_L + gW^+ \nu^\u_L  } \notag \\[1mm]
 			&= \vect{ 	d\nu^\u_L + (g^2+{g'}^2)^{\sfrac{1}{2}} Z^0 \nu^\u_L + gW^- \ell^\u_L    \\[1mm] d\ell^\u_L -2e A  \ell^\u_L  - \tfrac{g^2-{g'}^2}{\sqrt{g^2+{g'}^2}} Z^0 \ell^\u_L + gW^+ \nu^\u_L  }  \\[1mm]
&=\vect{ 	d\nu^\u_L + \tfrac{e}{\cos\theta_W \sin\theta_W} Z^0 \nu^\u_L + gW^- \ell^\u_L    \\[1mm] d\ell^\u_L -2e A  \ell^\u_L  - e\left( \tfrac{1}{\cos\theta_W \sin\theta_W} - 2\tfrac{\sin\theta_W}{\cos\theta_W}\right) Z^0 \ell^\u_L + gW^+ \nu^\u_L  }, \notag	
\end{align}
so
\begin{align*}
  (D^\u&\psi^\u_L)^\alpha=  \\ &\begin{pmatrix} 1 & 0 \\ 0 & \alpha^{-2} \end{pmatrix}  \vect{ 	d\nu^\u_L + \tfrac{e}{\cos\theta_W \sin\theta_W} Z^0 \nu^\u_L + gW^- \ell^\u_L    \\[1mm] d\ell^\u_L -2e A  \ell^\u_L  - e\left( \tfrac{1}{\cos\theta_W \sin\theta_W} - 2\tfrac{\sin\theta_W}{\cos\theta_W}\right) Z^0 \ell^\u_L + gW^+ \nu^\u_L  },  
\end{align*}
by \eqref{U(1)-GT} or \eqref{B-field}.
Above is introduced the weak mixing (or Weinberg) angle variable  $\theta_W$ via  $\cos \theta_W \defeq \sfrac{g}{\sqrt{g^2+g'^2}}$ and $\sin\theta_W=\sfrac{g'}{\sqrt{g^2+g'^2}}$, so that the change of field variable $(a, B_3) \rarrow (Z^0, A)$ can be written as a rotation in field space,
\begin{align*}
\vect{A \\ Z^0} =\begin{pmatrix} \cos\theta_W &  \sin\theta_W \\ -\sin\theta_W & \cos\theta_W \end{pmatrix} \vect{a \\ B_3} = \vect{\cos\theta_W a + \sin\theta_W B_3\\  \cos\theta_W B_3 - \sin\theta_W a}.
\end{align*}

The electroweak theory \eqref{EW-Lagrangian2} is then expressed in terms of the $\H_\text{loc}$-invariant fields $\rho, Z^0, \nu_L^\u$ and  the $\mathcal{U}(1)$-gauge fields $W^\pm, A, \ell_L^\u, \ell_R$. Writing explicitly the parts of the Lagrangian relevant to the next point to be  discussed, we have:
\begin{align}
\label{EW-Lagrangian3}
& L(A, W^\pm, Z^0, \rho, e^\u_L, e_R, \nu^\u_L)= \   \tfrac{1}{2}\Tr(F\w * F)+ \tfrac{1}{2}\Tr(G^u\w * G^u)  \notag\\
		&+ d\rho \wedge*d\rho  -  g^2\rho^2\  W^+\wedge*W^-   -  (g^2+g'^2)\rho^2\  Z^0\wedge*Z^0  -\left( \mu^2\rho^2 +\lambda \rho^4  \right)  *\!\jone  \notag \\
		& + \langle \psi_L^\u, \slashed D^\u \psi_L^\u \rangle  +   \langle \psi_R, \slashed D \psi_R \rangle  + f_\ell\,\left( \b \ell^\u_L\, \rho\, \ell_R + \b \ell_R\, \rho\, \ell^\u_L \right). \end{align}

One can expand the $\RR^+$-valued scalar field $\rho$ around its \emph{unique} groundstate $\rho_0$, given by $V(\rho)=0$, as $\rho=\rho_0+H$, where $H$ is the gauge invariant Higgs field.
Then, in the phase $\mu^2<0$ of the theory, where $\rho_0=\sqrt{\sfrac{-\mu^2}{2\lambda}}$,  mass terms $m_{Z^0}=\rho_0\sqrt{(g^2+g'^2)}$ and  $m_{W^\pm}=\rho_0g$ for  $Z^0, W^\pm$ appear from the couplings of the electroweak fields with $\rho$,\footnote{Since $A$ does not couple to $\rho$ directly, eq.\eqref{coupling-Higgs-WeakFields},  it is massless. The two photons decay channel of the H-field involves intermediary fermions.}  and the latter's self interaction produces a mass $m_H=\rho_0\sqrt{2\lambda}$  for $H$, while mass terms $m_\ell=\rho_0 f_\ell$ for the Dirac spinor leptons $\ell = (\ell^\u_L, \ell_R)^T$ are produced by Yukawa couplings.

Masses for gauge fields and leptons are obtained through a phase transition of the unique electroweak vacuum, but it is not congruent with a spontaneous gauge symmetry breaking, as the model is $\SU(2)$-invariant -- and the physical d.o.f. are manifest -- in both phases.\footnote{According to 
[\cite{Westenholz}], the very meaning of the terminology ``\textit{spontaneous} symmetry breaking'' lies in the fact that the manifold of vacua is not reduced to a point.}
The DFM approach to the electroweak model is consistent with  Elitzur's theorem [\cite{Elitzur1975}] stating that in lattice gauge theory a gauge symmetry cannot be spontaneously broken.



 \subsubsection{Discussion}
  \label{Discussion}
  
 To reiterate again a general remark in this context, \eqref{EW-Lagrangian3}  formally looks like the electroweak Lagrangian in the unitary gauge, yet it is conceptually different as a dressing is not a gauge-fixing.
 
Another noteworthy difference, is that while the model \eqref{EW-Lagrangian}  is defined for $\vphi \in \CC^2$, the dressed version  \eqref{EW-Lagrangian2}/\eqref{EW-Lagrangian3}  is only for $\vphi \in \CC^2/\{0\}$ because the polar decomposition \eqref{decomp-phi} is not well-defined at $\vphi=0$. Thus, the standard and dressed versions have different scalar field configuration  topologies.
In the massive phase ($\mu^2<0$), this should be of little concern regarding the perturbative regime, and appears also to be irrelevant non-perturbatively [\cite{Fernandez:1992jh}].

This is more troubling, however, in the phase $\mu^2>0$, as this means that the absolute minimum $\rho_0=0$ is not an available configuration, so that the mass terms are not vanishing, but vanishingly small. One could be tempted to retreat behind the fact that this phase of the theory is not realised in nature at present and beyond experimental reach, yet electroweak phase transition is believed to have occurred in the early universe and contributed to baryogenesis. So one cannot evade the necessity to assess the consequences (cosmological and otherwise) of not having strictly zero masses in the $\mu^2>0$ phase. It turns out there are arguments as to why this may finally be irrelevant, or at worst lead to relic monopoles [\cite{Fernandez:1992jh}].

Another question worth pursuing is the quantization of the dressed model.  As it is formally similar to the unitary-gauge version of the theory, indications of in-principle possibility to quantize the model in the unitary gauge (instead of the usual $R_\xi$-gauge) [\cite{Irges-Koutroulis2017, Ross1973, Mainland-Oraifeartaigh1974, Woodhouse1974}] may speak in favor of the view that \eqref{EW-Lagrangian3} lends itself well to perturbation theory. 
It is not obvious that the quantized version of  \eqref{EW-Lagrangian3} is exactly equivalent to the standard one, so that it may be interesting to compare them to see if one has some theoretical edge over the other. This has been done in lattice simulations, and within systematic errors no deviations have been observed [\cite{Evertz:1985fc,Philipsen:1996af}].
\smallskip

The above mentioned problem in the $\mu^2>0$ phase is avoided in the alternative invariant FMS approach to the model, which is also the one for which serious perturbative and lattice calculations have been done [e.\ g.\ \cite{Maas:2018xxu,Maas:2017wzi, Dudal:2019pyg, Dudal:2020uwb, Sondenheimer:2019idq, Maas-et-al2020, Maas:2020kda}], so is most easily weighted against the standard literature. It has also the advantage of being easily generalisable to $\SU(n)$ gauge theories.

\section{The FMS Approach}
\label{Sec:FMS approach}

So far, we viewed and used the DFM as a direct reformulation of the degrees of freedom of the electroweak sector of the standard model to demonstrate that the local gauge structure is artificial and a change to a manifestly gauge-invariant formulation is possible. On practical grounds, in particular regarding quantization, an alternative viewpoint on the DFM is useful. Therefore, we keep the basic philosophy of the previous sections but slightly change the perspective. 

In the following, we quantize the actual gauge symmetry based on the gauge-dependent elementary degrees of freedom but consider only $n$-point functions of strictly gauge-invariant objects. From that perspective, we perform the analysis of physical observables in a gauge theory with BEH mechanism in a QCD-like fashion. In QCD, quarks and gluons are used to describe the microscopic degrees of freedom but observable quantities are only gauge-invariant bound states, e.g., the hadrons. For a general BEH theory, we can do the same. At first sight this seems to be at odds with the tremendous success of the common perturbative treatment to describe electroweak processes at current and past collider facilities. However, certain properties of these gauge-invariant objects can be mapped on properties of gauge-dependent objects within particular classes of gauges which was first observed by Fr\"ohlich, Morchio, and Strocchi (FMS) [\cite{Frohlich:1980gj,Frohlich:1981yi}].

\subsection{The FMS Approach for the Electroweak Model}

First of all note that we can already reinterpret the dressed fields defined in Eq.~\eqref{eq:CompositeW} and Eq.~\eqref{eq:CompositeLepton} as gauge-invariant composite bound state operators. Ignoring the common treatment of the BEH mechanism, these are precisely some of the simplest possible gauge-invariant operators one would construct as observables of an $\SU(2)$ gauge theory with fundamental scalar and fermion fields. For instance, one would construct an $\SU(2)$ gauge-invariant combination of  a left-handed fermion field with a scalar $\vphi^{\dagger}\psi_{L}$ (cf. $l_{L}^{u}$ in Eq.~\eqref{eq:CompositeLepton}) or the charge conjugate scalar $\tilde{\vphi}^{\dagger}\psi_{L} = (\epsilon \vphi^{*})^{\dagger}\psi_{L}$ (cf. $\nu_{L}^{u}$ in Eq.~\eqref{eq:CompositeLepton}) with $\epsilon$ the two dimensional Levi-Civita tensor of $\SU(2)$. Similarly, we can construct gauge-invariant vector operators (with respect to the Lorentz group), e.g., $\vphi^{\dagger} D \vphi - \tilde{\vphi}^{\dagger} D \tilde{\vphi}$,  $\tilde{\vphi}^{\dagger} D \vphi$, and  $\vphi^{\dagger} D \tilde{\vphi}$ (cf. $b^{u}$ in Eq.~\eqref{eq:CompositeW}).

Keeping this strategy, we can also define a strictly gauge-invariant scalar operator, $\vphi^{\dagger}\vphi$. Here, we do not rely on the polar decomposition of the complex scalar doublet $\vphi$ which factors out the $\SU(2)$ gauge dependent contribution as in Eq.~\eqref{eq:Radial} but construct a gauge-invariant scalar object by dressing the elementary scalar field with its hermitian conjugate.  An additional advantage of this viewpoint is the fact that these type of bound state operators can be investigated for all potential forms of the scalar potential $V(\vphi)$ independently as to whether it obeys only one minimum at vanishing field configuration or a multitude of different (possibly even gauge-inequivalent) minima. Thus, we have now a conceptually clean setup which can be used in all parameter regions of the model. However, the apparent disadvantage of these gauge-invariant formulations is given by the circumstance that we have to compute properties of composite objects instead of using perturbative techniques for the elementary degrees of freedom.  

In general, a bound state is a nontrivial object and the computation of its properties from first principles requires nonperturbative techniques. Nevertheless, the $n$-point functions of some potential bound state operators can be computed in a fairly simple way in a BEH model. In order to examine this, FMS proposed to gauge fix the field configurations, e.g., via 't Hooft gauge, such that the scalar field acquires a nontrivial VEV. In this case, we are able to perform the conventional split 
\begin{align}
 \vphi(x) = \frac{v}{\sqrt{2}}\vphi_{0} + \Delta\vphi(x),
 \label{eq:Split}
\end{align}
where $\vphi_{0}$ is a unit vector in gauge space denoting the direction of the VEV, e.g., $\vphi_{0} = (0,1)^{\mathrm{T}}$ is a common choice, $v$ is the modulus of the VEV and $\Delta\vphi$ denotes fluctuations around it. The latter contains the field that is usually identified with the Higgs boson as well as the three would-be Goldstone modes which mix with those gauge bosons that acquire a nonvanishing mass term due to the BEH mechanism. With the aid of $\vphi_{0}$, we can extract these fields in a covariant but obviously not in a gauge-invariant way. Therefore, they cannot belong to the physical spectrum of the model if the gauge structure is merely a redundancy in our description. The Higgs field $h = \sqrt{2} \re(\vphi_{0}^{\dagger}\Delta\vphi)$ is the radial component of the fluctuation field in the direction of the VEV while the Goldstone modes are excitations in the remaining orthogonal directions, $\Delta\breve{\vphi} = \Delta\vphi - \re(\vphi_{0}^{\dagger}\Delta\vphi)\vphi_{0}$. 

By using such a gauge with nonvanishing VEV, we are able to rewrite the $n$-point functions of the gauge-invariant bound state operator in terms of $n$-point functions of gauge-variant objects. For instance, we obtain for the connected part of the propagator,
\begin{align}
 \langle \big(\vphi^{\dagger}\vphi\big) (x)\, \big(\vphi^{\dagger}\vphi\big) (y) \rangle &= v^{2} \langle h(x)\, h(y) \rangle + 2v \langle h(x)\, \big(\Delta\vphi^{\dagger}\Delta\vphi\big) (y) \rangle \notag\\
 &\quad + \langle \big(\Delta\vphi^{\dagger}\Delta\vphi\big)(x)\, \big(\Delta\vphi^{\dagger}\Delta\vphi\big) (y) \rangle.
 \label{eq:FMSScalar}
\end{align}
We ordered the terms on the right-hand side according to the number of fluctuation fields $\Delta\vphi$ appearing in the $n$-point functions (note that $h$ is also a component of $\Delta\vphi$). However, this FMS expansion of the bound state $\vphi^{\dagger}\vphi$ should not merely be viewed as an expansion in small fluctuations around the VEV. The FMS expansion is finite by construction and rather an exact rewriting of the original gauge-invariant operator. Thus, Eq.~\eqref{eq:FMSScalar} holds for any field amplitude $\Delta\vphi$ even in the nonperturbative regime. Nevertheless, using the number of fluctuation fields as an ordering scheme is an efficient method to extract the main information of the FMS expansion, in particular in the weak coupling regime. The first term on the right-hand side, i.e., the leading order term with respect to the ordering parameter $\Delta\vphi/v$, is the propagator of the (gauge-variant) elementary Higgs field $h$. Therefore, certain properties of the gauge-invariant bound state propagator can already be extracted from $\langle h(x)\, h(y) \rangle$.

For instance, let us consider the mass and decay width of the state generated by $\vphi^{\dagger}\vphi$. These properties are encoded in the pole structure of its propagator. Ignoring for a moment the higher-order terms of the FMS expansion, we obtain that the pole of the gauge-invariant bound state propagator coincides with the pole structure of the elementary Higgs propagator. In addition, it can be shown to all orders in a perturbative expansion of the $n$-point functions that the higher-order terms of the FMS expansion do not alter the pole structure on the right-hand side [\cite{Maas:2020kda}]. Therefore, the on-shell properties of $\vphi^{\dagger}\vphi$ are well described by the propagator $\langle h(x)\, h(y) \rangle$. Of course, the pole of the bound state operator has to be gauge-invariant by construction. This translates at the level of the elementary $h$ field to the well-known Nielsen identities which show that the pole of $\langle h(x)\, h(y) \rangle$ is independent of the gauge-fixing parameter within 't Hooft gauges [\cite{Nielsen:1975fs,Grassi:2001bz}]. 

However, the latter fact does not mean that the elementary Higgs field can be associated with the experimental observed Higgs boson. The Nielsen identities merely show that certain gauge-invariant informations of the model can be extracted from the field $h$, but $h$ itself is still gauge dependent. In particular, every single term on the right-hand side of the FMS expansion is gauge dependent and can only be computed within the specifically chosen gauge. Without gauge fixing, any of these Green's functions will vanish since the action as well as the path integral measure are gauge invariant. The fact that they are nontrivial within the common treatment is merely due to the conventional gauge fixing procedure. Choosing a gauge implies automatically an explicit breaking of the gauge symmetry. However, this is done by hand and should not be confused with spontaneous symmetry breaking. For instance gauges can be constructed which induce a vanishing VEV of the scalar field even if the potential has a nontrivial global minimum. For these type of gauges, the mass parameters of the various elementary fields would be zero to any order in a perturbative expansion. Nonetheless, the properties of a gauge-invariant object as the scalar bound state operator $\vphi^{\dagger}\vphi$ are independent of the gauge. The FMS formulation basically reveals that in some gauges, namely those that are conventionally used in the particle physics community, some gauge-invariant informations of the system can be computed in a convenient way as they are stored in the $n$-point functions of elementary fields. Further, we have perturbative access to them in the weak coupling regime as all terms on the right-hand side of Eq.~\eqref{eq:FMSScalar} can be computed via perturbative techniques. Therefore, we have reduced the problem of calculating the properties of a complicated but strict gauge-invariant bound state operator on computing $n$-point functions of elementary fields and composites of elementary fields in a gauge-fixed setup. 

That this nontrivial relation is indeed realized has been validated by nonperturbative lattice simulations for an $\SU(2)$ Yangs-Mills-Higgs theory [\cite{Maas:2013aia,Maas:2014pba}]. The lattice formulation provides a clean setup for this check as no gauge fixing is required to compute the properties of a gauge-invariant bound state. Furthermore, gauge-fixed configurations can be generated that allow for a nonperturbative investigation of the elementary $n$-point functions. Thus, both sides of the relation can be investigated independently. By contrast, a perturbative analysis can only investigate the terms on the right-hand side due to the necessity to gauge fix. Investigating the spectrum in the scalar channel of the model, lattice simulations confirm that the mass of the gauge-invariant bound state operator coincides with the mass of the elementary Higgs field as dictated by the FMS relation. Considering the vector channel, one would expect three degenerate massive vector bosons due to the BEH mechanism from the conventional analysis. Constructing bound state operators, we are able to write down a gauge-invariant triplet of states that precisely map on the elementary triplet of vector bosons via the FMS formulation in this model. Also this relation has been confirmed by lattice investigations.

\subsubsection{Gauge-Invariant Description of the Electroweak Particles}
\label{Gauge-invariant description of the electroweak particles}

The remaining question is now if such a type of mapping between a gauge-invariant bound state operator and the elementary fields of the Lagrangian can be implemented for all fields of the electroweak sector. Before we discuss the FMS formulation of the full electroweak model, for the sake of simplicity we neglect the $\mathcal{U}(1)$ hypercharge gauge group and the Yukawa couplings  for a moment and focus on the non-Abelian $\SU(2)$ part. To be specific, we consider the Lagrangian~\eqref{EW-Lagrangian} in the limit $f_{l} \to 0$ and put the Abelian gauge field $a$ to zero. Besides the local $\SU(2)$ gauge structure given in the second line of Eq.~\eqref{FieldGT-EW}, the model also obeys a less obvious, additional global $\SU(2)_{R}$ symmetry. It solely acts on the scalar field but in a non-linear way as it relates $\vphi$ with $\t\vphi$, 
\begin{align}
 &\vphi^{\kappa} = \kappa_{1} \vphi + \kappa_{2} \t\vphi, \quad \t\vphi^{\kappa} = -\kappa_{2}^{*} \vphi + \kappa_{1}^{*} \t\vphi \notag \\ 
 &\text{where} \quad \kappa_{1/2}\in \mathbb{C} \quad \text{and} \quad |\kappa_{1}|^{2} +  |\kappa_{2}|^{2} = 1
\end{align}
($b^{\kappa} = b$, $\psi_{L/R}^{\kappa} = \psi_{L/R}$). Note, that this is a particularity of $\SU(2)$ as only for this group the dual field of the fundamental scalar $\vphi$ transforms under the fundamental representation as well.  
From the FMS perspective, we can now classify gauge-invariant bound state operators according to their transformation properties with respect to this global $\SU(2)_{R}$ symmetry, i.e., as $\SU(2)_{R}$ multiplets. Again, this is similar to pure QCD with $N_{f}$ fermion flavors where the physical spectrum is described in terms of $\SU(3)$-invariant meson, hadron, and more exotic bound states that form certain multiplets of the global flavor symmetry group. 
The additional global symmetry can be made more transparent by introducing a bi-doublet, $\Phi = \begin{pmatrix}\t\vphi & \vphi \end{pmatrix}$. The usual gauge transformations act on $\Phi$ by multiplication from the left, $\Phi^{\beta} = \beta\Phi$. The additional global (flavor-like) symmetry acts as multiplication from the right $\Phi^{\kappa} = \Phi\kappa$, $\kappa \in \SU(2)_{R}$. Note, that this bi-doublet is precisely used to construct the local dressing field $u = ||\vphi||^{-1} \Phi$. Further, we would like to emphasize that this global symmetry is broken via the BEH mechanism as well. Nonetheless, in case a gauge with nonvanishing VEV is chosen, a global diagonal subgroup of $\SU(2)\times\SU(2)_{R}$ remains  such that the precise breaking pattern of the model reads $\SU(2) \times \SU(2)_{R} \to \SU(2)_{\mathrm{diag}}$. This remaining symmetry also manifests in the elementary spectrum. For instance the weak vector bosons receive the same mass term due to the BEH mechanism and transform as an $\SU(2)_{\mathrm{diag}}$ triplet after gauge fixing.\footnote{Note that we still neglect the hypercharge sector which causes the splitting of the $W$ and $Z$ mass terms in the standard model.} 

In order to analyze the spectrum of the model, we characterize states due to their global quantum numbers. Therefore, we have an additional quantum number due to the global $\SU(2)_{R}$ symmetry group. First, we consider the scalar channel, i.e., operators that generate states that can be associated with scalar particles. The simplest gauge-invariant operators in this channel contain two elementary scalar fields. These can be combined such that we obtain two different irreducible $\SU(2)_{R}$ multiplets, a singlet or a triplet.  We already discussed the scalar $\SU(2)_{R}$ singlet $\Tr(\Phi^{\dagger}\Phi) = \vphi^{\dagger}\vphi + \t\vphi^{\dagger}\t\vphi = 2\vphi^{\dagger}\vphi$. In case a gauge is chosen such that $\vphi$ acquires a nonvanishing VEV, this operator can be mapped on the elementary Higgs field as discussed above. Technically, we can also construct a triplet state $\Tr(\tau_{i_{R}}\Phi^{\dagger}\Phi)$ where $\tau_{i_{R}}$ denotes the generators of $\SU(2)_{R}$.\footnote{In order to make transformation properties with respect to the different $\SU(2)$ groups directly transparent, we will use index notation in the following. Indices with a subscript $R$ or $D$ denote objects transforming with respect to $\SU(2)_{R}$ or $\SU(2)_{\mathrm{diag}}$, respectively. Furthermore, we will use Lorentz indices (greek letters) to highlight the spin quantum number of the operators.} However, this multiplet vanishes identically, $\Tr(\tau_{i_{R}}\Phi^{\dagger}\Phi) =0$, which becomes directly apparent in the $\vphi$-$\t\vphi$ notation where the triplet is given by $\big(\re(\t\vphi^{\dagger}\vphi),\im(\vphi^{\dagger}\t\vphi),\vphi^{\dagger}\vphi - \t\vphi^{\dagger}\t\vphi \big) = (0,0,0)$. 

In the vector channel, we perform the same analysis. The global $\SU(2)_{R}$ triplet, $\Tr(\tau_{i_{R}}\Phi^{\dagger}D^{\mu}\Phi)$, expands in leading order of the FMS mechanism to the triplet of massive $\SU(2)_{\mathrm{diag}}$ vector fields. Choosing $\vphi_{0} = (0, 1)^{\mathrm{T}}$, i.e., $\Phi = \frac{v}{\sqrt{2}}\mathbbm{1} + \mathcal{O}(\Delta\vphi)$, we have 
\begin{align}
\Tr(\tau_{i_{R}}\Phi^{\dagger}D^{\mu}\Phi) = \frac{gv^{2}}{2} \Tr(\tau_{i_{R}} b^{\mu} ) + \mathcal{O}(\Delta\vphi) 
= \frac{gv^{2}}{4} \delta_{i_{R}i_{D}} b_{i_{D}}^{\mu} + \mathcal{O}(\Delta\vphi). 
\end{align}
Therefore, we obtain a gauge-invariant vector operator transforming as an $\SU(2)_{R}$ triplet that can be mapped on the triplet of massive elementary vector fields similar to the $\vphi^{\dagger}\vphi\,$-$\,h$ mapping in the scalar sector. Of course, we are also able to construct a vector operator that transforms as an $\SU(2)_{R}$ singlet, $\Tr(\Phi^{\dagger}D_{\mu}\Phi)$. However, this operator does not provide a mapping on an elementary vector field as the $\mathcal{O}(\Delta\vphi^{0})$ term vanishes due to the properties of the $\SU(2)$ group. At $\mathcal{O}(\Delta\vphi^{1})$, we obtain a nontrivial term given by $\sqrt{2}vd_{\mu}h$. Thus, investigating the propagator of the vector singlet, we expect a pole at the mass of the elementary Higgs. Nevertheless, this pole structure does not give rise to a new vector particle with mass $\mh$ as it appears only in the longitudinal part of the correlator such that it does not exhibit the correct Lorentz structure of a vector particle. Note, that not only the pole structure but also the correct Lorentz structure is necessary for a proper particle interpretation. In case of the vector channel, one would expect a structure $(g^{\mu\nu}-p^\mu p^\nu / m_V^2)/(p^2-m_V^2)$ for a proper massive vector particle having Spin = 1. 
However, we only obtain $p^\mu p^\nu/(p^2-m_h^2)$ which has no meaningful particle interpretation because not only the analytic structure of the propagator is important but also its Lorentz tensor structure. It rather reflects the fact that a derivative acting on a scalar operator transforms as a vector and therefore mixes with operators in the vector channel. Therefore, we do not obtain a new vector particle from the gauge-invariant description, which thus remains consistent with the common perturbative treatment of the model. 

Furthermore, we have the fermionic sector of the model. Neglecting the hypercharge sector, the right-handed fermion fields are part of the physical spectrum as they are already gauge-invariant, see Eq.~\eqref{FieldGT-EW}.\footnote{Note that we neglect any influence of the strong force under which the right-handed quarks transform as fundamental objects.} However, the left-handed flavors of quarks and leptons within one generation are actually weak gauge charges and thus unobservable due to their non-Abelian nature. Due to the global $\SU(2)_{R}$ symmetry, we are able to construct $\SU(2)$ gauge-invariant fermionic operators that are $\SU(2)_{R}$ doublets. In leading order of the FMS expansion, these expand to the elementary left-handed fermionic fields,
\begin{align}
 \Phi^{\dagger} \psi_{L} = \begin{pmatrix} \t\vphi^{\dagger}\psi_{L} \\ \vphi^{\dagger}\psi_{L} \end{pmatrix} 
 = \frac{v}{\sqrt{2}} \begin{pmatrix} \t\vphi_{0}^{\dagger}\psi_{L} \\ \vphi_{0}^{\dagger}\psi_{L} \end{pmatrix} + \mathcal{O}(\Delta\vphi) 
 = \frac{v}{\sqrt{2}} \begin{pmatrix} \nu_{L} \\ \ell_{L} \end{pmatrix} + \mathcal{O}(\Delta\vphi).
\end{align}
Therefore, the different flavors of the left-handed components observed within one generation are actually not the weak gauge charges but rather the physically well defined $\SU(2)_{R}$ quantum numbers. 

So far, we only discussed the spectrum within our reduced electroweak model, i.e., neglecting Yukawa coupling and hypercharge contributions. Allowing for nonvanishing Yukawa and hypercharge couplings, we explicitly break the global $\SU(2)_{R}$ symmetry. Therefore, the $\SU(2)_{\mathrm{diag}}$ symmetry of the gauge fixed formulation is broken as well. Nonetheless, we are still able to investigate gauge-invariant operators that generate states of the aforementioned $\SU(2)_{R}$ multiplets which map on the corresponding $\SU(2)_{\mathrm{diag}}$ multiplets. The only difference to the previous discussion is a splitting of the multiplet levels in the various quantum number channels due to the explicit breaking terms in the Lagrangian which then results in the different observed mass terms for charged leptons and neutrinos or the mass splitting of the $W$ and $Z$ bosons. Furthermore, we have to incorporate 
a gauge-invariant treatment of the additional $\mathcal{U}(1)$ gauge structure. As the hypercharge sector is an Abelian gauge theory, we might use common dressings via a Dirac phase factor as in QED. For more details see the end of Sec.~\ref{Examples} or Ref.~[\cite{Maas:2017wzi}].

\subsubsection{Phenomenological Implications of the FMS Formulation}
\label{Phenomenological implications of the FMS formulation}

\begin{figure}
\centering
\includegraphics[width=0.8\textwidth]{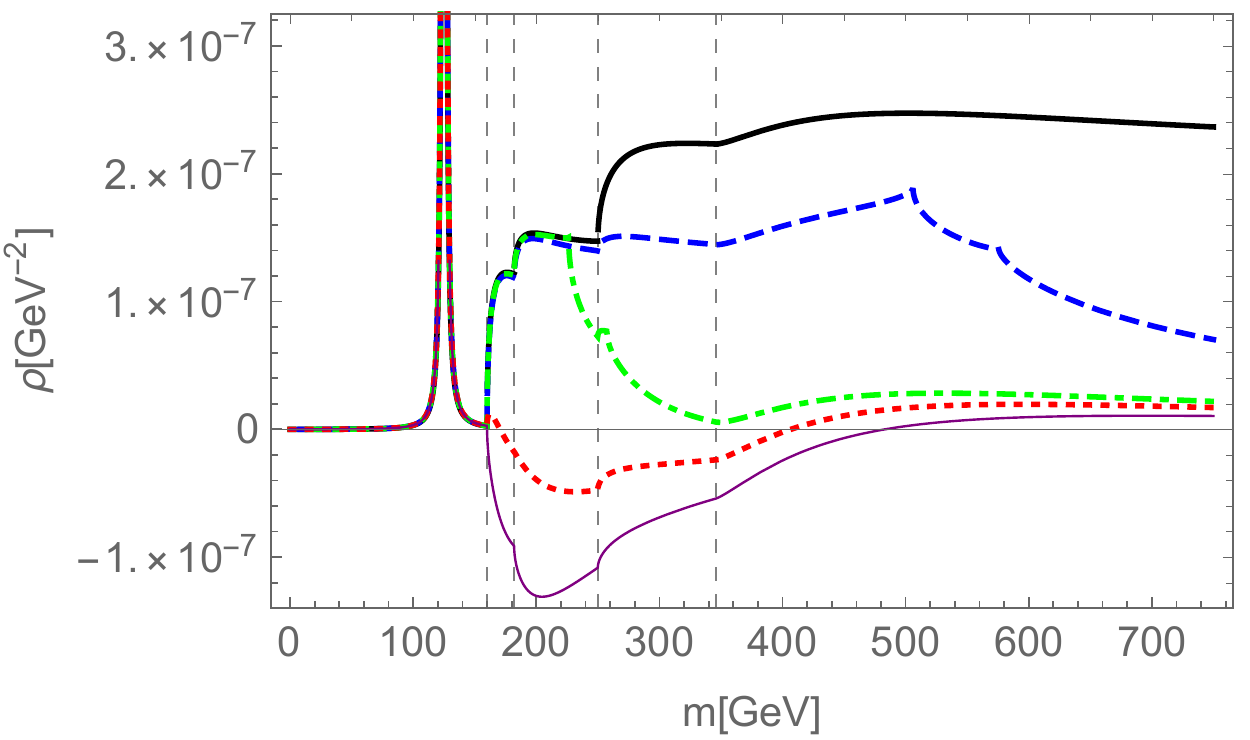}
\caption{Spectral density of the elementary Higgs field for different values of the gauge-fixing parameter $\xi$. We depict the spectral function for $\xi=1$ (red dotted line), $\xi=2$ (green dash-dotted line), and $\xi=10$ (blue dashed). Further, we depict the Higgs spectral function extracted from the pinch technique (purple line). The black solid line shows the spectral function of the gauge-invariant bound state $\vphi^{\dagger}\vphi$. The vertical gray dashed lines indicate the mass thresholds at $2m_{\mathrm{W}}$, $2m_{\mathrm{Z}}$, $2m_{\mathrm{h}}$, and $2m_{\mathrm{top}}$ from left to right. For further details see [\cite{Maas:2020kda}]}
\label{fig:spectral}
\end{figure}

Besides nontrivial lattice checks of the FMS relation, also perturbative investigations that include higher-order FMS terms shed a new light on the gauge-invariant definition of observables in a gauge theory with BEH mechanism. Note that conventional investigations of, e.g., the properties of the Higgs, merely focus on the first term of the FMS expansion in Eq.~\eqref{eq:FMSScalar}. Although the on-shell properties of $h$ do not depend on the gauge-fixing parameter, the off-shell properties do. In order to examine this further, let us extract the K\"all\'{e}n-Lehmann spectral representation $\rho_{\mathrm{h}}(\lambda)$ from the propagator $\langle h(x)\, h(y) \rangle$. In momentum space we have
\begin{align}
 \langle h(p)\, h(-p) \rangle = \int_{0}^{\infty} d\lambda \frac{\rho_{\mathrm{h}}(\lambda)}{p^{2}-\lambda}.
\end{align}
We depict the spectral function for the elementary Higgs field computed via a one-loop approximation of the propagator in Fig.~\ref{fig:spectral}, see [\cite{Maas:2020kda}] for further details. The analysis was performed for different gauge fixing parameters $\xi$. The red dotted curve denotes the result for $\xi=1$ (Feynman-'t Hooft gauge), the green dash-dotted curve represents the result for $\xi=2$, and the blue dashed line is $\xi=10$. First of all, we obtain a clear peak at $\mh = 125$ GeV independently of the chosen gauge. Also the width of this peak is the same for all gauges. This is expected due to the Nielsen identities as the peak position (mass) and width (decay width) are determined by the pole of the propagator.

Furthermore, we find several continuum thresholds that are associated with observed particle masses. These are indicated as vertical, thin grey dashed lines. The first such line is the $2m_{\mathrm{W}}$ threshold starting at twice the mass of the $W$ boson. Going to higher energies, we also find the thresholds at $2m_{\mathrm{Z}}$, where $m_{\mathrm{Z}}$ is the mass of the Z boson, $2\mh$, as well as $2m_{\mathrm{top}}$. However, we also find unphysical thresholds which are not related to physical particles. These thresholds depend on the gauge-fixing parameter $\xi$ which can easily be figured out by varying $\xi$. More precisely, these unphysical thresholds start at $2\sqrt{\xi}m_{\mathrm{W}}$ and $2\sqrt{\xi}m_{\mathrm{Z}}$. Indeed, for $\xi = 10$ (blue curve), we have two additional spikes at $\approx 505$ GeV and $\approx 575$ GeV while the additional thresholds appear at $\approx 226$ GeV and $\approx 257$ GeV for $\xi=2$ (green line). For $\xi = 1$ (red line), we don't find additional structures in the spectral density as the unphysical thresholds start at the masses of the physical $W$ and $Z$ mass scale which leads to a nontrivial modification of the latter physical thresholds. At this point we would also like to emphasize that the spectral function becomes negative for some gauge conditions. This is also a clear hint that the elementary Higgs field $h$ cannot be identified with a physical observable as the spectral function of such a quantity has to be nonnegative for a physical interpretation. Similar results can also be obtained for the Abelian Higgs model [\cite{Dudal:2019aew,Dudal:2019pyg,Dudal:2020uwb,Dudal:2021pvw}]

Of course, the fact that the elementary Higgs propagator depends on $\xi$ is known for a long time. However, the Higgs particle is unstable within the standard model and occurs only as an intermediate resonance in a physical process. Calculating physical $S$-matrix elements, e.g., scattering processes of stable particles, the gauge parameter dependence of the internal Higgs propagators will get cancelled by propagator-like pieces from triangle and box diagrams [\cite{Papavassiliou:1995gs,Papavassiliou:1996zn,Papavassiliou:1995fq}]. Taking these processes into account, a $\xi$-independent definition of the propagator and thus of the spectral function can be introduced via the so called pinch technique [\cite{Binosi:2009qm,Papavassiliou:1997pb}]. This pinch technique propagator is cured from unphysical thresholds by definition. However, its spectral function still violates positivity as can be seen in Fig.~\ref{fig:spectral} (purple solid line).

By contrast, let us investigate the spectral function of the bound state operator $\vphi^{\dagger}\vphi$. For that, we include the other two terms of the FMS expansion in Eq.~\eqref{eq:FMSScalar} on the same footage as the elementary Higgs propagator, i.e., we perform a one-loop approximation as the simplest possible nontrivial approximation for these terms as well. The first important result of this calculation is given by the fact that all $\xi$-dependent contributions to the leading order term $\langle h(x)\, h(y) \rangle$ get canceled by gauge-dependent contributions to the other two Green's functions $\langle h(x)\, \big(\Delta\vphi^{\dagger}\Delta\vphi\big) (y) \rangle$ and $\langle \big(\Delta\vphi^{\dagger}\Delta\vphi\big)(x)\, \big(\Delta\vphi^{\dagger}\Delta\vphi\big) (y) \rangle$. Of course, this is not a surprise as the sum of all terms on the right-hand side of the FMS expansion is gauge invariant by construction. Thus, the unphysical thresholds are absent. The second important result is the positivity of the spectral function such that a physical particle interpretation is possible for the bound state [\cite{Maas:2020kda}].

Apart from these more advanced analyses for the Higgs boson, also other interesting phenomenological implications have been investigated in first exploratory studies. For instance, the potential influence on anomalous couplings and the size of the gauge-invariant $W$-Higgs bound state has been studied in [\cite{Maas:2018ska}]. Furthermore, the bound state formulation of observables in the electroweak sector also influences high precision measurements of other sectors as QCD. The necessity to describe hadrons not only as gauge-invariant objects with respect to the strong interaction but also with respect to the weak interaction once they are embedded in the larger standard model context implies that some of them contain additional scalar fields as constituents [\cite{Egger:2017tkd}]. These effects can be addressed in a parton distribution function type of language [\cite{Fernbach:2020tpa}]. Also predictions for potential future lepton colliders should be investigated in light of the FMS formulation as off-shell properties of leptons will get altered similar to the case of the Higgs boson. If some of these effects are not properly accounted for, they could easily be missinterpreted as signals for new physics while being only nontrivial effects of standard model physics.

\subsection{The FMS Formulation for General Gauge Theories with a BEH Mechanism}
\label{The FMS formulation for general gauge theories with a BEH mechanism}

In the previous sections, we discussed different strategies to construct a gauge-invariant formulation of the electroweak sector of the standard model. One of the central advantages of the FMS formulation is its direct generalization to arbitrary gauge groups with scalar fields in arbitrary representations. This is of particular importance as also recent lattice investigations challenge the conventional interpretation of the spectrum of gauge theories with a BEH mechanism [\cite{Maas:2016ngo,Torek:2017czn,Maas:2018xxu,Afferrante:2020hqe}]. States that one would naively expect by the conventional analysis were not found by the nonperturbative lattice simulations of the models. This failure has far reaching consequences for potential model building [\cite{Maas:2017xzh,Sondenheimer:2019idq}]. Using the FMS approach provides a coherent picture for currently observed phenomena of the lattice spectra. 

The basic FMS ingredients, namely:
\begin{itemize}
 \item[(a)] construct strict gauge-invariant operators with respect to the original gauge group and classify them according to the global symmetries of the model and
 \item[(b)] chose a gauge with nonvanishing VEV of the scalar field and investigate the FMS expansion, 
\end{itemize}
can be used for any BEH model. In the following, let us consider a gauge theory with gauge group $\H$ that breaks in the conventional treatment of the BEH mechanism to a subgroup $\K \subset \H$. As discussed in detail in the previous sections, this viewpoint has various philosophical and field theoretical inconsistencies. From a field theoretical perspective, the BEH mechanism should rather be considered as a duality relation between the spectra of an $\H$ gauge theory and a $\K$ gauge theory with specific field content [\cite{Sondenheimer:2019idq}]. The FMS formalism reveals which of the potential states in both theories are related. This duality relation can be read in two ways. From a top down perspective, the FMS mechanism shows which $\H$-invariant operators can be computed by potential simpler objects in a $\K$ gauge theory. From a bottom up perspective, the FMS formalism explains which states of a $\K$ gauge theory can be embedded into the spectrum of an $\H$ gauge theory. 

At first sight, one may be tempted to conclude that the FMS strategy provides a gauge-invariant description of all quantities that are usually considered from the perspective of gauge symmetry breaking, similar to the standard model case. As a simple example, consider the elementary scalar field that is proportional to the direction of the VEV and thus always transforms as a singlet with respect to the unbroken remaining gauge group $\K$ of the BEH mechanism. We find always a strict $\H$-invariant operator that has precisely this particular gauge-dependent field as the nontrivial leading order term of the FMS expansion. We always have $(\phi^{\mathbf{a}})^{*}\phi^{\mathbf{a}} = v h + \cdots$ where $h$ is the elementary $\K$ singlet, $\phi$ a scalar field in an arbitrary representation of the gauge group $\H$ whose potential has nontrivial minima, and $\mathbf{a}$ is a multi-index characterizing the representation. Similar constructions of $\H$-invariant operators can also be done for all those elementary fields that transform as $\K$ singlets. This has been confirmed in all models that have been investigated in lattice calculations so far [\cite{Maas:2016ngo,Torek:2017czn,Maas:2018xxu,Afferrante:2019vsr,Afferrante:2020hqe}]. 

For instance, consider an $\H = \SU(3)$ gauge theory with fundamental scalar field. Any nontrivial minimum of the potential has a $\K=\SU(2)$ subgroup as stabilizer. Therefore, we would expect a breaking $\SU(3)\to\SU(2)$ due to the BEH mechanism from the conventional perspective. On the level of the particle spectrum this translates in a formulation of $\SU(2)$-invariant objects instead of strict $\SU(3)$-invariant composite bound states. The constituents of the former can be extracted from the $\SU(3)$ gauge and scalar field and arranged in multiplets of the remaining $\SU(2)$ group. For the considered example, we can decompose the $\SU(3)$ gauge field into three different $\SU(2)$ multiplets. Five gauge bosons acquire a nonvanishing mass term. These can be subdivided into a field $A_{\mathrm{s}}$ that transforms as a singlet with respect to the remaining $\SU(2)$ gauge transformations while the other four components form a fundamental multiplet $A_{\mathrm{f}}$. The remaining three (massless) gauge bosons, which we denote by $A_{\mathrm{a}}$, form the pure Yang-Mills sector of the $\SU(2)$ gauge theory. As the elementary $A_{\mathrm{s}}$ is already invariant with respect to the non-Abelian $\SU(2)$ gauge group, it belongs to the gauge-invariant spectrum of the $\SU(2)$ gauge theory. Similar to the elementary Higgs field $h$, we can also construct an $\SU(3)$-invariant vector operator that precisely maps on this particular vector field, $\phi^{\dagger}D^{\mu}\phi \sim A_{\mathrm{s}}^{\mu} + \mathcal{O}(\varphi/v)$ which has been confirmed via lattice investigations [\cite{Maas:2016ngo,Maas:2018xxu}]. 

By contrast, there is no $\SU(3)$-invariant operator that maps on any other elementary vector field. This is not a surprise. In general we start our investigation with a strict $\H$-invariant operator. All terms that can be extracted from such an object have to be invariant with respect to the remaining $\K$ gauge transformations by construction such that we can only obtain $\K$ singlets but we do not obtain a component of a $\K$ charged multiplet ($A_{\mathrm{f}}$ or $A_{\mathrm{a}}$ for our $\H = \SU(3)$ example). Of course, this does not imply that an object on the right-hand side cannot contain nontrivial $\K$ multiplets. $\K$-invariant combinations of $\K$ multiplets can be extracted from $\H$-invariant composite operators. For our current $\SU(3)$ example, the $\SU(3)$-invariant glueball operator $\Tr(F^{2})$ can be decomposed into an $\SU(2)$ glueball $\Tr(F_{\mathrm{a}}^{2})$ with $F_{\mathrm{a}} = d A_{\mathrm{a}} + \sfrac{1}{2}\left[A_{\mathrm{a}}, A_{\mathrm{a}}\right]$, an $\SU(2)$ bound state formed by the fundamental vector fields $A_{\mathrm{f}}^{\dagger}A_{\mathrm{f}}^{\phantom{\dagger}}$, as well as several other $\SU(2)$-invariant combinations of the elementary $\SU(2)$ multiplets. Although, we can extract these $\SU(2)$-invariant states from a strict $\SU(3)$-invariant operator in a gauge-fixed setup by decomposing the multiplets, the associated states have not been found on the lattice yet. Why this is the case is currently under investigation and an open problem. So far, we can identify two differences that distinguish, for instance, a $\K$ glueball operator from the the elementary Higgs field or the elementary singlet vector field for the $\SU(2)$ case. 

First, the latter operators can be obtained from $\SU(3)$-invariant operators not only via the standard multiplet decomposition but appear also in a unique way at nontrivial leading order of the FMS expansion of some $\SU(3)$-invariant operators. By contrast, no such mapping of an $\H$-invariant operator on an $\K$-invariant glueball operator exist via the split $\phi = \frac{v}{\sqrt{2}}\phi_{0} + \Delta\phi$ as the constituents of the $\K$ glueball operator are in a subspace orthogonal to the direction of the VEV $\phi_{0}$. 
Second, the $\K$ glueball as well as various other operators are composites of elementary $\K$ multiplets and form nontrivial bound states already from the perspective of the $\K$ gauge theory. As to whether the BEH duality extends to these objects and if the FMS mappings are able to explain the spectra on a pure group theoretical basis or dynamical effects of bound state formations from either the $\H$ or $\K$ perspective play an important role needs further detailed investigations. 

The fact that only $\K$-invariant operators can be extracted from $\H$-invariant ones has far reaching implications for model building beyond the standard model. As a simple toy model, let us consider an $\SU(2)$ gauge theory with a scalar field in the adjoint representation. Performing the conventional analysis, the breaking pattern reads $\SU(2) \to \mathcal{U}(1)$ and the particle spectrum consists of a scalar particle described by an elementary scalar field which is a $\mathcal{U}(1)$ singlet, a massive vector boson that is charged with respect to the remaining $\mathcal{U}(1)$ symmetry and its corresponding antiparticle with opposite $\mathcal{U}(1)$ charge, as well as a massless gauge boson being the force carrier of the $\mathcal{U}(1)$ gauge group. Thus, a variety of potential states can be described by elementary fields from the conventional perspective of gauge symmetry breaking. From the FMS perspective, we have to construct $\SU(2)$-invariant states and investigate their FMS expansions. Indeed, it is straightforward to find strict $\SU(2)$-invariant operators that map on the elementary scalar boson as well as on the massless vector particle [\cite{Maas:2017xzh}]. However, no $\SU(2)$-invariant operator exist that maps on an operator generating a $\mathcal{U}(1)$ charged state [\cite{Sondenheimer:2019idq}] which is in accordance with lattice investigations [\cite{Lee:1985yi,Afferrante:2019vsr,Afferrante:2020hqe}]. Again, we can only extract a $\K$ $(=\mathcal{U}(1))$-invariant operator from any $\H$-invariant operator. Thus, it is not possible to embed the $\mathcal{U}(1)$ charged states from the perspective of the $\mathcal{U}(1)$ gauge theory into the spectrum of the $\SU(2)$ gauge theory. This is a generic problem for any physical theory beyond that standard model (BSM) that tries to embed the $\mathcal{U}(1)$ gauge group of the standard model into a larger gauge symmetry. 

From that perspective, the standard model electroweak gauge group is special. First, it explicitly contains a $\mathcal{U}(1)$ (hypercharge) group whose properties translate into the properties of the remaining $\mathcal{U}(1)$ (electromagnetism) gauge group via the BEH mechanism. Second, the non-Abelian $\SU(2)$ weak gauge sector has a global counterpart which is also described by an $\SU(2)$ structure which purely acts on the scalar fields. Therefore, a sufficient large number of $\SU(2)$ gauge invariant operators can be constructed and classified according to global $\SU(2)_{R}$ multiplets. Due to these particular reasons, we are able to construct the spectrum in a strict gauge-invariant way and the FMS mapping provides a convenient description of it in terms of the conventional analysis via the elementary fields of the gauge-fixed Lagrangian. Similar constructions can be done for BSM models that fulfill the same requirements as the standard model, e.g., two-Higgs-doublet models [\cite{Maas:2016qpu}] as well as general $N$-Higgs-doublet models, such that these models provide reliable BSM models that pass all FMS constraints.

\section{Critical Assessment, Reflections, and Challenges}\label{Sec:Reflection}

The success of gauge theories in particle physics opened the door to optimism concerning unification in physics based on the concept of symmetry \citep{Yang1980}. Understanding gauge symmetries as descriptive redundancies seems adequate in the light of reasonable conceptual desiderata such as determinism, parsimony of the posited unobservable ontology and elimination of superfluous structure (Section \ref{Sec:Symmetry}).  Yet, the fundamental significance of this success seems to be challenged by the apparent indispensability of gauge fixing and spontaneous symmetry breaking. In the context of the BEH mechanism, the very possibility of providing a gauge invariant account (presented already in the unitary-gauge-fixed formulations by \citealp{Higgs66} and \citealp{Kibble67}) can appear as providing a viewpoint benefiting from the best of both worlds \citep{Struyve2011}, reconciling gauge invariance and accounting for massive vector bosons at the same time. However, the foundational importance of such attempts remains questionable as long as they stand as mere reformulations of existing theories, achieving certain theoretical virtues at the price of sacrificing others. While gauge invariance may further resolve some technical issues that arise in the context of lattice theories (Section \ref{Sec:Breaking}), it may seem far from being clear whether these advantages can compete with those of the established framework of spontaneous symmetry breaking. 

The DFM (Section \ref{The dressing field method}) and the FMS approach (Section \ref{Sec:FMS approach})  may each suggest that the  gauge invariant approaches can, in fact, open the door to a wider heuristic and conceptual frameworks. Both of these methods identify  gauge invariant field variables, thus achieving reduction of gauge symmetries without compromising on the theoretical virtues that motivate gauge invariance (see Section \ref{Sec:Why}). Applied to the electroweak model, they converge on the conclusion that the spontaneous breaking of gauge symmetry is not a physical phenomenon in this case, and furthermore, at the classical level the results they provide coincide \citep{Maas:2017wzi}, giving rise to a local gauge invariant description of the massive gauge bosons, that renders the $SU(2)$ symmetry an artificial one. However, while neither  the DFM nor the FMS are yet in a state of full maturity, we can already point out that  despite the aforementioned similarities  the two approaches seem to entail different research programs that face different challenges. 

For example the question of quantizing a dressed theory, with its invariant field variables, is still a programmatic endeavor.
The FMS approach is more developed in that regard, as it treats invariant fields as composites of gauge-variant fields and borrows technics from QCD, and it already shows great promises as it appears to mitigate problems appearing in the standard formulation based on SSB (as detailed by the end of Section 6). 

If future research confirms that such local invariant formulations of the electroweak model have a theoretical edge over the usual approach (gauge-fixing and SSB), then a web of interconnected questions arise: 
If  $\SU(2)$  is indeed artificial, and given that the gauge principle applied to the sole $\mathcal{U}(1)$ substantial gauge symmetry is not enough to explain the structure of the model, presumably this re-opens the question of its conceptual and theoretical foundations -- or at least provide a new angle to reassess those foundations. This question itself is then nested within that of the underlying principle(s) explaining the structure of the full Standard Model, whose substantial gauge symmetry group would then be $\mathcal{U}(1) \times \SU(3)$:
A more fundamental theory giving the SM in the effective low energy regime should then explain why it presents this mix of substantial and artificial symmetries. Following that thread could be another avenue toward beyond the SM physics.

Constraining the formalism to gauge invariant field variables might come at the price of increased complexity and/or loss of manifest locality. 
This shows that while gauge symmetries are convenient, they are not always necessary in order to formulate the relevant physical theories. We conclude with some reflections on implications of these findings, and a (possible) future role of gauge-invariant approaches in physical practice.

Concerning physical practice, there is overwhelming agreement among physicists that gauge dependent quantities are not empirical.  
On the other hand, as far as day-to-day physical practice is concerned, this statement is often applied only within the narrow window of perturbative or effective treatments. Only a few, mainly mathematical, physicists have consistently pointed out that the perturbative approach is viable only for very specific theories. In matters of practice, only within the community of lattice theoreticians has non-perturbative gauge invariance become, by necessity, mandatory. Nonetheless, the subtleties in relation to the perturbative treatment, which were emphasized especially by Fr\"ohlich, Morchio and Strocchi, have not been widely appreciated [see \cite{Maas:2017wzi} for an overview of the developments]. Assuming that, in accordance with the recommendations presented here, formulating theories in gauge-invariant ways from the start becomes accepted as a methodological guideline in the future, this leaves us with a number of puzzling insights and challenges.

On the one hand, it appears that many, or perhaps even all, gauge theories can be reformulated as theories without gauge symmetries. However, these reformulations come with non-trivial features, like non-local contributions, non-power-countable Lagrangians, involved target spaces, or an infinite number of fields. Moreover, provided that dualities between different gauge theories hold, there could be multiple different gauge theories associated with the same set of gauge-invariant quantities.  

On the other hand, peaceful co-existence between theoretical practice and gauge symmetries is definitely possible, as long as we maintain a commitment to express observable quantities  in terms of gauge-invariant observables. That is, even if gauge symmetries are still a part of the theoretical framework, we assign physical relevance only to quantities that are gauge-independent; this should be contrasted with eliminative approaches that are strictly formulated using gauge invariant variables. Though, as the example of QCD shows, if this commitment to gauge-invariant observable quantities is manifested by e.g. lattice QCD, this resolution may require only marginally less effort than eliminating the gauge symmetry altogether, as manifested by e.g. a reformulation of QCD in terms of Wilson lines. The enormous amount of computing time and person-years in development of algorithms for lattice QCD needs to be compared to the conceptual and technical complications of a reformulation of QCD in terms of Wilson lines.

But a number of conceptual challenges emerge in the eliminative approach, from the physics point of view: Can indeed every gauge theory be written in terms of a formalism without gauge-dependent quantities and hence without gauge symmetries? Do there exist gauge theories whose non-gauge version is genuinely local, without the gauge symmetry being trivial? Is any such theory relevant to experiment? 
Do theories which are dual to experimentally relevant ones exist, which have  different gauge symmetries? Answering these questions would tell us a lot about to which extent gauge symmetries are uniquely tied to the observables. 

Even if the ease of use implies that  gauge symmetry will find continued employment in actual calculations, consistently adopting the stance that gauge symmetries are conceptually redundant would have far-reaching implications. For this would imply that, as they are represented in the Lagrangian, each and every elementary particle in the current standard model of particle physics\footnote{With the exception of right-handed neutrinos, if they exist.} is not physical, as the fields corresponding to these particles are all gauge-dependent! The only physical degrees of freedom would be those that correspond to hadrons, the electroweak objects of the FMS approach, and photon-cloud dressed QED states. The conventional notions of quarks, electrons etc.\ would need to be regarded as mere auxiliaries that are technically useful but do not have any physical reality. Given the role these objects play even at the level of school textbooks, this would be a fundamental shift of what are widely, and popularly, regarded as the furniture of reality and the fundamental building blocks of nature\footnote{Such a picture is anyhow imprecise due to the identification of particles with fields, rather than speaking of particles as localized field excitations. However, the issue of particle-field duality is unrelated to the issue of gauge symmetry, and thus glossed over here.}.

Thus, eliminating gauge-dependent objects as physical objects might well be the most consequential shift in the way in which we portray nature since the advent of quantum field theory.

However, this leaves one stark observation: Every experimentally relevant theory can be written either in a local form using gauge symmetries, or in a non-local form without gauge symmetries. This raises several questions, namely: If we wanted to preserve locality, is the preservation of gauge symmetries our only option? Is a description of experiments without gauge symmetry only possible non-locally? As such, are we guided correctly in assuming that the gauge principle is essential in more fundamental theories? Or does this unnecessarily narrow our perspective? Even when thinking about approaches like loop quantum gravity, gauge-invariant variables originally derive from a local formulation. Could and should a general non-local (or non-gaugeable) approach be searched for? Without experimental guidance, this appears challenging at least. So, as a more pragmatic benchmark, we can ask: does eliminating gauge-dependent objects as an element of reality  create progress?

These questions need to be answered. And it needs to be understood whether abandoning our current view in terms of quarks, electrons, etc.\ is necessary, or at the very least advantageous.

As a concluding comment,  we discussed here primarily the situation in ordinary relativistic quantum field theories in flat space-time. But the problem extends beyond those. Most notably, similar problems arise in (quantum) gravity theories, which can be considered to be gauge theories of translations and, in presence of torsion, Lorentz symmetry [\cite{hehl1995gauge,Hehl:1976kj}]. This does not even touch upon the possibilities in more extensive settings, e.\ g.\ string theory.

The problems encountered with gauge symmetries are then amplified in gauge theories of gravity, as the space-time structure itself becomes gauge-dependent, including the notion of time. Likewise, similar approaches have been advocated to eliminate the problem of gauge dependence. Most notably, loop-quantum gravity [\cite{Ashtekar:2011ni}] seeks an alternate quantization procedure by quantizing manifestly gauge-invariant quantities. More particle-physics-like approaches are also discussed, where the quantum theory remains a gauge theory. This leads to ideas similar to the dressing-field method [see e.\ g.\ \cite{Donnelly:2015hta,Giddings:2019wmj}] or the FMS approach [\cite{Maas2020}]. However,  the concept of locality in particular becomes far more involved. In this context, the concept of local observables is far less developed, and important questions, e.\ g.\ the role of affine parameters instead of space-time coordinates, are far from understood. This issue has also been observed in the philosophy of physics [\cite{Lyre:2004,Healey:2007}]. However, without a clear understanding of the role of gauge symmetries in particle physics a full clarification in the quantum gravity setting appears unlikely.

\appendix  

\bibliography{refs}

\section*{Acknowledgements}

P.B. acknowledges support by the Austrian Science Fund (FWF) [P 31758]. During the completion of this work, J.F. has been supported by the Fonds de la Recherche Scientifique -- FNRS under grant PDR No. T.0022.19 (``Fundamental issues in extended
gravitational theories''), and by the FNRS grant MIS No.\ F.4503.20
(``HighSpinSymm''). 
R.S. acknowledges support by the DFG under Grant No. SO1777/1-1

\end{document}